\documentclass[a4paper,11pt]{article}
\pdfoutput=1 

\usepackage{jheppub} 
\usepackage{multirow}
\usepackage[T1]{fontenc} 
\usepackage[section]{placeins}
\usepackage{bbold}
\usepackage[normalem]{ulem}
\usepackage{jheppub}
\usepackage{hyperref}
\newcommand\identity{1\kern-0.25em\text{l}}
\newcommand{\qsqmin}{q^2_{\rm min}}
\newcommand{\qsqmax}{q^2_{\rm max}}
\def\mpole{m_\mathrm{pole}}
\def\gev{\,\mathrm{Ge\kern-0.1em V}}
\def\mev{\,\mathrm{Me\kern-0.1em V}}
\newcommand{\abs}[1]{\left| #1 \right|}   

\title{Bayesian inference for form-factor fits regulated by
unitarity and analyticity}

\preprint{CERN-TH-2023-047}

\author[a,b]{J.M. Flynn,}
\author[a,b,c]{A. Jüttner,}
\author[c]{J.T. Tsang}

\affiliation[a]{School of Physics and Astronomy, University of Southampton, Southampton, SO17 1BJ, UK}
\affiliation[b]{STAG Research Centre, University of Southampton, Southampton, SO17 1BJ, UK}
\affiliation[c]{Theoretical Physics Department, CERN, Geneva, Switzerland}

\emailAdd{j.m.flynn@soton.ac.uk}
\emailAdd{Andreas.Juttner@cern.ch}
\emailAdd{j.t.tsang@cern.ch}

\abstract{We propose a model-independent framework for fitting hadronic form-factor data, which is often only available at discrete kinematical points, using parameterisations based on to unitarity and analyticity. In this novel approach the latter two properties of quantum-field theory regulate the ill-posed fitting problem and allow model-independent predictions over the entire physical range. Kinematical constraints, for example for the vector and scalar form factors in semileptonic meson decays, can be imposed exactly. The  core formulae are straight-forward to implement with standard math libraries. We take account of a generalisation of the original Boyd~Grinstein~Lebed (BGL) unitarity constraint for form factors and demonstrate our method for the exclusive semileptonic decay $B_s\to K \ell \nu$, for which we make a number of phenomenologically relevant predictions, including  the CKM matrix element $|V_{ub}|$.}

\begin{document} 
\maketitle
\flushbottom
\section{Introduction}\label{sec:Introduction}
Hadronic  form factors  are a crucial ingredient in precision tests of the Standard Model (SM)~\cite{Workman:2022ynf}. They allow us to better understand the structure of hadrons at different length scales, and to study their constituents. In the case of flavour-changing hadron decays, they enable the determination of CKM matrix elements from experiment. This motivates ongoing experimental effort to improve decay-rate measurements. On the theory side, lattice QCD is one of the main tools for computing form factors from first principles~\cite{FlavourLatticeAveragingGroupFLAG:2021npn}, allowing us to predict their overall normalisation and momentum dependence. QCD sum rules play a similar role and are often complementary to lattice QCD in their kinematical range of applicability~\cite{Colangelo:2000dp, Khodjamirian:2020btr}. In order to match experimental efforts it is crucial to reduce errors in the theory computations.

One often finds, however, that neither experimental nor theoretical approaches are able to cover the entire physical kinematical range of the decay process. For instance, differential decay rates for flavour-changing exclusive semileptonic decays as measured in experiment for, \emph{e.g.}, heavy-light mesons, are kinematically suppressed when the momentum transfer between the initial and final meson, $q$, approaches the zero-recoil point $q^2_{\rm max}$. Lattice simulations of the same process on the other hand, which compute the corresponding hadronic form factors as a function of momentum transfer, have difficulties in controlling systematic effects for small $q^2$. Thus, very often one finds oneself in a situation where results for a small number $N_{\rm data}$ of $q^2$ values (or bins) are available in one particular kinematical regime, and one would like to make predictions about the entire physically allowed range. Or, one has data points or bins in two distinct kinematical regimes and would like to combine the data for a global analysis.

To this end, model independent form-factor parameterisations based on the quantum-field theory principles of unitarity and analyticity~\cite{Boyd:1994tt, Caprini:1997mu, Bourrely:2008za} have been devised in order to relate and combine results for different kinematical regimes. 
In this paper we propose a method to determine the parameters of one such parameterisation, the one by Boyd-Grinstein-Lebed (BGL)~\cite{Boyd:1994tt}, with \textcolor{black}{controlled} truncation errors. In order to make BGL and therefore our approach applicable to a wider range of decay channels, we also adopt a generalisation~\cite{Gubernari:2020eft,Gubernari:2022hxn,Blake:2022vfl} of the BGL unitarity constraint. \textcolor{black}{Furthermore we study a modified BGL expansion for which the asymptotic behaviour of the form factors for large values of the momentum transfer $|q^2|$, as found in perturbative QCD, provides further relations between the expansion coefficients~\cite{Buck:1998kp,Becher:2005bg}}.

To illustrate the problem further, let us consider a frequentist fit, where the number of parameters $K$ that can be determined is  primarily limited by the number $N_{\rm data}$ of input-data points. A further common limitation is poor statistical quality of the data, which can further reduce the effective usable number $N_{\rm data}$ of data points, indicated by a badly conditioned correlation matrix of the input data. In any case, the constraint on the number of degrees of freedom $N_{\rm dof}=N_{\rm data}-K\ge 1$ for a meaningful frequentist fit imposes a strict upper bound on the truncation $K$. This may not be too much of a limitation in situations where abundant independent data is available, allowing one to observe how final results depend
on the choice of the order $K$. All too often, however, data is scarce, leaving little room for estimating truncation errors reliably.

The problem of finding a model- and truncation-independent parameterisation of a finite set of data is hence ill-defined and some form of regulator is required to keep the parameters not well constrained by the data under control. We propose to address the problem starting from Bayes' theorem. As we will argue and demonstrate, a \textcolor{black}{form-factor parameterisation with controlled truncation errors can be determined, relying merely on analyticity and unitarity as regulators.} The resulting
form-factor parameterisation is free from systematic effects besides those potentially afflicting the underlying experimental, lattice or sum-rule data.

The proposal made here is similar in spirit to the determination of model-independent parameterisations based on the recently revived dispersive-\-matrix~(DM) 
method~\cite{Bourrely:1980gp,Lellouch:1995yv,DiCarlo:2021dzg}. Both approaches use the same physical information  and should produce consistent results, but the proposal presented here is considerably simpler to implement and multiple, potentially correlated, data sets from both experiment and theory can  be included straightforwardly in the fitting problem. We make our own implementation available in the form of a Python code~\cite{fittingPaperCode}. 

We note that a number of novel ideas applying Bayesian inference in the context of quantum-field theory have recently been put forward in a variety of contexts: fitting of parton-distribution functions~\cite{DelDebbio:2021whr}, analysis of fits to lattice data~\cite{Neil:2022joj,Jay:2020jkz,Frison:2023lwb}, or the estimation of missing higher-order terms in perturbation theory~\cite{Duhr:2021mfd}. 

{
\color{black}
Starting from a set of 
 reference data points for the form factor that we assume to follow Gaussian statistics, we show how the 
parameterisation, for which we 
assume uniform (flat) parameter priors, 
is defined in terms of a multivariate normal distribution. We explain how representative samples for observables based on the 
parameterisation can then be computed by drawing Gaussian random numbers in a way that takes the unitarity constraint into account. Moreover, kinematical constraints like the equality of vector and scalar form factor in pseudoscalar-to-pseudoscalar-transition form-factors at zero momentum transfer can be imposed exactly. We therefore hope that the approach presented here will be attractive to, and adopted by, a wide user community in theory, phenomenology and experiment.}

The main results of this paper are:
\begin{itemize}
 \item using a generalisation of the BGL unitarity constraint
 \item a simple method for 
 form-factor fits subject to 
 unitarity and kinematical constraints
 \item an algorithm and its implementation in a Python code~\cite{fittingPaperCode}
 \item a demonstration of the method for individual and combined fits of 
 lattice~\cite{Bouchard:2014ypa,Bazavov:2019aom,Flynn:2023nhi}, sum-rule~\cite{Khodjamirian:2017fxg} and experimental~\cite{LHCb:2020ist,LHCb:2020cyw} data for semileptonic $B_s\to K\ell\nu$ decay, making predictions for a number of phenomenologically relevant observables
 \item a comparison to fits based on the dispersive-matrix method~\cite{Bourrely:1980gp,Lellouch:1995yv,DiCarlo:2021dzg}
\end{itemize}
The paper is structured as follows. In Sec.~\ref{sec:basics and BGL} we first introduce some basic notation for semileptonic decays in the Standard Model and then discuss the BGL parameterisation and the generalised unitarity constraint. In Sec.~\ref{sec:fitting} we first revisit the theory of frequentist form-factor fits to introduce basic notation, followed by the discussion of Bayesian inference and an algorithm to solve it in practice. In Sec.~\ref{sec:BstoK} we apply the new method for $B_s\to K \ell\nu$ exclusive semileptonic decay using lattice data from HPQCD 14~\cite{Bouchard:2014ypa}, FNAL/MILC 19~\cite{Bazavov:2019aom} and RBC/UKQCD 23~\cite{Flynn:2023nhi} and compare to frequentist fits and the dispersive-matrix method. Finally we make predictions that can be used for phenomenology in Sec.~\ref{sec:Phenomenological analysis}.

\clearpage
\section{Form factors, unitarity and analyticity}\label{sec:basics and BGL}
While the ideas presented here are universally applicable to parameterisations of hadronic form factors, we find it instructive to base the presentation on a particular example, the semileptonic meson decay $B_s\to K \ell\nu$. \textcolor{black}{The application} to other decay channels is
straightforward. In this section we introduce basic definitions and recall the details of the unitarity- and analyticity-based model-independent form-factor parameterisation by Boyd-Grinstein-Lebed (BGL)~\cite{Boyd:1994tt}. The case of $B_s\to K \ell\nu$ is particularly interesting since its kinematics and the analytical properties of the corresponding form factors motivated us to use a modified BGL unitarity constraint.

\subsection{Decay rate and form factors}

The differential decay rate for $B_s\to K \ell\nu$ in the $B_{s}$ rest frame is given by
\begin{equation}
\label{eq:B_semileptonic_rate}
  \begin{split}
  \frac{d\Gamma(B_{s}{\to} K\ell\nu)}{dq^2} &=
  \eta_\text{EW} \frac{G_F^2 |V_{ub}|^2}{24\pi^3} \,
  \frac{(q^2{-}m_\ell^2)^2 |{\bf  p}_K|}{(q^2)^2}
    \bigg[ \Big(1{+}\frac{m_\ell^2}{2q^2}\Big)
           |{\bf p}|^{2}|f_+(q^2)|^2\\
  &\mathrel{\phantom{=}} {}
      +\frac{3m_\ell^2}{8q^2}
      \frac{(M_{B_s}^2-M_K^2)^2}{M_{B_s}^2}|f_0(q^2)|^2
    \bigg]\,.
  \end{split}
\end{equation}
The kaon three momentum is $|{\bf  p}_K| = (E_K^2-M_K^2)^{1/2}$, where $E_K$ is the kaon energy. 
The momentum transfer between the $B_s$  meson and the kaon is $q = p_{B_s}-p_K$, $m_\ell$ is the lepton mass and $\eta_\text{EW}$ is an
electroweak correction factor.\footnote{We follow Ref.~\cite{Na:2015kha}
and take $\eta_\text{EW}=1.011(5)$ by combining the factor computed by
Sirlin~\cite{Sirlin:1981ie} with an estimate of final-state
electromagnetic corrections using the ratio of signal yields from
charged and neutral decay channels.} 
The form factors $f_+$ and
$f_0$ arise in the decomposition of the QCD matrix element
\begin{align}
  \langle K(p_K) |\mathcal{V}^\mu| B_{s}(p_{B_s}) \rangle
    &= f_+(q^2)\big(p_{B_s}^\mu +p_K^\mu \big)+f_-(q^2)\big(p_{B_s}^\mu-p_K^\mu\big)\nonumber\\
    &= 2f_+(q^2)\bigg(p_{B_s}^\mu -\frac{p_{B_s}\cdot q}{q^2}q^\mu\bigg)
    + f_0(q^2)\frac{M_{B_s}^2-M_K^2}{q^2}q^\mu\,,
\end{align}
where the kinematical constraint 
\begin{equation}\label{eq:kinematical constraint}
f_+(0)=f_0(0)
\end{equation}
can be deduced from 
$f_0(q^2)=f_+(q^2)+q^2/(M_{B_s}^2-M_K^2)f_-(q^2)$. We will use this constraint 
in the later discussion.
In the SM, $\mathcal V^\mu = \bar u\gamma^\mu b$
is the continuum charged current operator. Lattice computations
of the matrix element are by now standard and can be computed 
with per-cent-level 
precision~\cite{FlavourLatticeAveragingGroupFLAG:2021npn,Bouchard:2014ypa,Flynn:2015mha,Bazavov:2019aom,Flynn:2023nhi}.
\subsection{BGL parameterisation with generalised unitarity constraint}\label{Sec:BGL modified}
Unitarity- and analyticity-based form-factor parameterisations have in common that they map 
the complex $q^2$ plane with a cut for $q^2\ge t_\ast$ onto the unit-disc of a new complex kinematical variable $z$~\cite{Bourrely:1980gp,Boyd:1994tt,Boyd:1995sq,Lellouch:1995yv,Boyd:1997qw,Caprini:1997mu,Arnesen:2005ez,Bourrely:2008za}
using the map
\begin{equation}
\label{eq:z-fn-defn}
  z(q^2;t_*,t_0) = \frac{\sqrt{t_*-q^2} - \sqrt{t_*-t_0}}%
                      {\sqrt{t_*-q^2} + \sqrt{t_*-t_0}}\,.
\end{equation}
For use below we set $t_\pm =
(M_{B_s}\pm M_K)^2$, with $t_- = q^2_\text{max}$ the upper end of the kinematical
range for physical semileptonic decay. The opening of the cut at $t_*$ is fixed by the lowest appropriate two-particle production threshold $t_*=(M_B+M_\pi)^2$, which is determined by the flavour content of the electroweak current $\mathcal{V}$. 
The value of $t_0$ can be chosen to fix the range in
$z$ corresponding to a given range in $q^2$. We choose $t_0$ to
symmetrise the range of $z$ about zero for $q^2$ in the range $0\leq
q^2\leq q^2_\text{max} = t_-$:
\begin{equation}
  \label{eq:t-opt}
  t_0 = t_\mathrm{opt} = t_* - \sqrt{t_*(t_*-t_-)}.
\end{equation}

Boyd, Grinstein and Lebed (BGL)~\cite{Boyd:1994tt} write the form
factor as
\begin{equation}
  \label{eq:BGLparametrisation}
  f_X(q^2) = \frac1{B_X(q^2)\phi_X(q^2,t_0)} \sum_{n\geq0}
  a_{X,n} (t_0)z^n,
\end{equation}
where $X=+,0$, \textcolor{black}{$\phi_X(q^2,t_0)$ is a known ``outer function'' and} the Blaschke factor $B_X(q^2)$ is chosen to vanish at the
positions of sub-threshold poles $M^X_{i}$,
\begin{equation}
B_X(q^2)=\prod\limits_{i\in X\, {\rm poles}}z\left(q^2;t_\ast, \left(M^X_{i}\right)^2\right)\,.
\end{equation}
From now on we drop the explicit dependence of the BGL coefficients $a(t_0)$ on the parameter $t_0$.
For the vector form factor $f_+$ of the $B_s\to K\ell\nu$ decay 
the theoretically predicted $1^-$ vector-meson with  
$M^+_{B^\ast(1^-)}=5.32471\gev$~\cite{Workman:2022ynf}
sits above the physical semileptonic region $0\le q^2\le q^2_{\rm max}$,
but also below the $B\pi$ threshold at $(M_B+M_\pi)^2$
(specifically, 
$q^2_{\rm max}\le (M^+_{B^\ast(1^-)})^2\le t_\ast \to 23.73\gev^2\le 28.35\gev^2\le29.35\gev^2$). 
The pole is cancelled by the Blaschke factor $B_+(q^2)$.
For $f_0$ the theoretically predicted pole mass 
$M^+_{B^\ast(0^+)}=5.63\gev$~\cite{Bardeen:2003kt} sits above
the $B\pi$ threshold and no pole needs to be cancelled.
The outer functions $\phi_X(q^2,t_0)$ in Eq.~\eqref{eq:BGLparametrisation} \textcolor{black}{are given in}
Appendix~\ref{app:The BGL parameterisation}. 
What differentiates the $B_s\to K\ell\nu$ semileptonic decay from $B\to\pi \ell\nu$ is the observation that 
in the former the two-particle $B\pi$ production threshold lies below
the one of $B_sK$, \emph{i.e.}~$t_\ast<t_+$. This has recently been discussed in Refs.~\cite{Berns:2018vpl,Gubernari:2020eft,Gubernari:2022hxn,Blake:2022vfl}\footnote{Note some differences in notation to those 
papers, in particular our use of $t_*$ and $t_+$ for the locations of 
the $B\pi$ and $B_sK$ production thresholds, respectively.},
where it was pointed out that 
when inserting the BGL expansion~\eqref{eq:BGLparametrisation} into the unitarity constraint
\begin{equation}\label{eq:vanilla unitarity constraint}
\frac{1}{2\pi i}\oint_C\frac {dz}{z} |B_X(q^2)\phi_X(q^2,t_0)f_X(q^2)|^2\le 1 \,,
\end{equation}
the integration around the unit-circle includes contributions from below 
$t_+=(M_{B_s}+M_K)^2$, \emph{i.e.}~from below the $B_s K$ production threshold. The 
unitarity bound for $B_s\to K \ell \nu$ can in this way become too strong. 
The authors of Ref.~\cite{Gubernari:2020eft,Gubernari:2022hxn} propose to modify the BGL expansion, replacing the
monomials $z^i$, which are orthogonal on the unit circle,
\begin{equation}
\langle z^i|z^j\rangle=\frac 1{2\pi}\int\limits_{-\pi}^{+\pi} d\alpha (z^i)^\ast z^j|_{z=e^{i\alpha}}=\delta_{ij}\,,
\end{equation}
by polynomials $p_i(z)$ which are orthogonal with respect to  an inner product with the integral restricted to the relevant part of the unit circle, \emph{i.e.},
\begin{equation}\label{eq:Blake inner product}
\langle p_j(z)|p_j(z)\rangle_{\alpha_{B_s K}}=\frac 1{2\pi}\int\limits_{-\alpha_{B_s K}}^{+\alpha_{B_s K}} d\alpha (p_i(z))^\ast p_j(z)|_{z=e^{i\alpha}}=\delta_{ij}\,,
\end{equation}
with $\alpha_{B_s K}={\rm arg}\left[z(t_+;t_\ast,t_0)\right]$. An algorithm for constructing the $p_i(z)$ is
provided in Refs.~\cite{Szego:1939,Simon:2004,Gubernari:2022hxn,Blake:2022vfl}. Here we propose to modify just the unitarity constraint Eq.~\eqref{eq:vanilla unitarity constraint}
and leave the BGL expansion Eq.~\eqref{eq:BGLparametrisation} untouched. This has the benefit that existing analysis
codes barely have to be modified. In particular, we write the unitarity constraint as
\begin{equation}
\frac{1}{2\pi i}\oint_C\frac {dz}{z}\theta_z |B_X(q^2)\phi_X(q^2,t_0)f_X(q^2)|^2\le 1 \,,
\end{equation}
where the step function $\theta_z=\theta(\alpha_{B_s K}-|{\rm arg}[z]|)$ restricts
the integration over the unit circle to the relevant segment, \emph{i.e.}~the one corresponding to the branch cut above the $B_s K$ threshold $t_+$. Inserting the BGL expansion Eq.~\eqref{eq:BGLparametrisation}, the unitarity constraint takes the compact form
\begin{equation}
\label{eq:modified unitarity constraint}
\sum\limits_{i,j\ge 0}a_{X,i}^\ast \langle z^i|z^j\rangle_{\alpha_{B_sK}} a_{X,j} \equiv
    |{\bf a}_X|^2_{\alpha_{B_sK}}\le 1\,,
\end{equation}
where the inner product is known analytically,
\begin{equation}
  \langle z^i|z^j\rangle_\alpha =
  \frac 1{2\pi}\int\limits_{-\alpha}^\alpha d\phi (z^i)^\ast
  z^j|_{z=e^{i\phi}}
  = \begin{cases}
    \displaystyle
    \frac{\sin(\alpha(i-j))}{\pi(i-j)} & i\neq j\,,\\
    \displaystyle\frac\alpha\pi & i=j\,.
  \end{cases}
  \label{eq:metric}
\end{equation}
The proposal made here is equivalent to the one in Refs.~\cite{Gubernari:2020eft,Gubernari:2022hxn,Blake:2022vfl},
but technically much simpler to implement. We provide more details on the
relation to Refs.~\cite{Gubernari:2020eft,Gubernari:2022hxn,Blake:2022vfl} and the underlying work of
Refs.~\cite{Szego:1939,Simon:2004} in App.~\ref{app:Modified unitarity
  constraint}.  Note, that for decays where $t_\ast=t_+$, \emph{e.g.}~$B\to\pi\ell\nu$,
the original BGL unitarity constraint is recovered, since in this case
$\alpha_{B\pi}=\pi$.

\textcolor{black}{
We close this section with a comment regarding the large-$q^2$ (or $t$) behaviour of the vector form factor.
    In perturbation theory the large-$t$ behaviour is expected to be $f_+(t)\sim 1/t$~\cite{Lepage:1980fj,Akhoury:1994tnu}. The expression in Eq.~\eqref{eq:BGLparametrisation}
    in principle allows for terms that decay slower or even diverge in this limit.
    These terms are not controlled by the unitarity constraint. In particular, in the vicinity of $z=1$
    the leading contributions to Eq.~\eqref{eq:BGLparametrisation} are $t^{1/4},\, t^{-1/4}$ and $t^{-3/4}$. A set of sum-rules that constrain these unphysical terms was first proposed in Ref.~\cite{Buck:1998kp,Becher:2005bg}. In appendix~\ref{app:modified BGL} we work out a modified BGL expansion based on these
    constraints. It can be used to check whether the large-$t$ behaviour of the BGL ansatz affects the fit results in any way. Given that the constraints are only relevant far above threshold they are not expected to be of much relevance for the form factor in the semileptonic region (\emph{cf.} discussion in Ref.~\cite{Becher:2005bg}). All our numerical results indeed confirm this picture.
}
\clearpage
\section{The fitting problem}\label{sec:fitting}
In this section we discuss our proposed method for  determining the coefficients of the BGL expansion Eq.~\eqref{eq:BGLparametrisation} 
from a finite set of input data. 
In particular, we assume to have results
for the form factors 
$f_+(q_i^2)$ for $N_+$  $q^2$ values ($i=0,1,2,\dots,N_+-1$) 
and 
$f_0(q_j^2)$ for $N_0$ $q^2$ values ($j=0,1,2,\dots,N_0-1$),
respectively.
We find it convenient to combine  all 
data into a data vector
\begin{equation}
{\bf f}^{\,T}=({\bf f}_+^T,{\bf f}_0^T)=(f_+(q^2_0),f_+(q^2_1),\dots,f_+(q^2_{N_+-1}),
f_0(q^2_0),f_0(q^2_1),\dots,f_0(q^2_{N_0-1}))\,.
\end{equation}
The data is assumed to be correlated with  known covariance matrix $C_{\bf f}$.

While the fitting problem within the Bayesian framework is formally well defined with
infinitely many fit parameters, truncating the expansion will be necessary in practice, and is, for a finite number of input data a requirement for a meaningful 
frequentist fit. As we will discuss below, the model and truncation independence can
then still be demonstrated by showing the independence of the results of the chosen truncation as the truncation is gradually removed.
For the following discussion we therefore truncate the BGL expansion after $K_X$ terms,
\begin{equation}
  \label{eq:BGLparametrisation modified}
  f_X(q^2) = \frac1{B_X(q^2)\phi_X(q^2,t_0)} \sum\limits_{n=0}^{K_X-1}
  a_{X,n} z^n\,.
\end{equation}

\subsection{Frequentist fit}
Frequentist fits to form-factor data are common practice. We will discuss the 
method here, on the one hand to introduce our notation, on the other hand so we 
can later compare to it.
Due to the discrete nature of ${\bf f}$, we can express the BGL parameterisation in terms of a matrix-vector notation.
The combined frequentist fitting problem for $f_+$ and $f_0$ is defined by the sum of squares
\begin{equation}\label{eq:chisq_BGL}
\chi^2({\bf a},{\bf f})=
\left[{\bf f}-Z{\bf a}\right]^T
C_{\bf f}^{-1}
\left[{\bf f}-Z{\bf a}\right]\,,
\end{equation}
where 
\begin{equation}\label{eq:avec definition}
{\bf a}^T=({\bf a}_+^T,{\bf a}_0^T)=(a_{+,0},a_{+,1},a_{+,2},...,a_{+,K_+-1},a_{0,1},...,a_{0,K_0-1})\,,
\end{equation}
and where we defined the matrix
\begin{equation}
Z=\left(\begin{array}{rr}Z_{++}&Z_{+0}\\Z_{0+}&Z_{00}\end{array}\right)\,,
\end{equation}
with diagonal blocks
\begin{align}
  (Z_{++})_{ij}=& 
  \frac1{B_+(q_i^2)\phi_+(q_i^2,t_0)}
                    z^j(q^2_i)\,, \nonumber\\
  (Z_{00})_{ij}=&
  \frac1{B_0(q_i^2)\phi_0(q_i^2,t_0)}z^j(q^2_i) \,.
\end{align}
For reasons to be explained shortly we deliberately omitted the component $a_{0,0}$ in the definition of the vector ${\bf a}$
in Eq.~\eqref{eq:avec definition}.
The off-diagonal blocks $Z_{+0}$ and $Z_{0+}$ are determined as follows: We use the kinematical constraint $f_+(0)=f_0(0)$ 
to eliminate one parameter
 in the BGL expansion. For instance, the 
 constraint can be solved for
\begin{align}\label{eq:kinematical constraint}
    a_{0,0}=B_0(0)\phi_0(0,t_0)f_+(0)-\sum\limits_{k=1}^{K_0-1}a_{0,k} z^k(0)\,.
\end{align}
 In terms of the above matrix notation the 
 constraint then corresponds to
\begin{align}
(Z_{+0})_{ij}=&0\,,\nonumber\\
\\[-4mm]
(Z_{0+})_{ij}=&\frac{1}{B_+(0)\phi_+(0,t_0)}\frac{\phi_0(0,t_0)}{\phi_0(q_i^2,t_0)}z^j(0)\,.\nonumber
\end{align}

The solution of the fitting problem is 
given by the minimisation of the $\chi^2$ in 
Eq.~\eqref{eq:chisq_BGL}. Given the linear
parameter dependence the solution is
\begin{equation}\label{eq:frequentist_sln_a}
    {\bf a}=\left(Z^T C_{\bf f}^{-1} Z\right)^{-1}ZC_{\bf f}^{-1}{\bf f}\,,
\end{equation}
with covariance matrix for the parameters ${\bf a}$,
\begin{equation}\label{eq:frequentist_sln_Ca}
    C_{\bf a}=\left(Z^T C_{\bf f}^{-1}Z\right)^{-1}\,.
\end{equation}
A few comments are in order:
\begin{itemize}
 \item For the frequentist fit with the kinematical constraint to be meaningful
   requires $N_{\rm dof}=N_++N_0-K_+-K_0\ge 0$ for the number of degrees of
   freedom $N_{\rm dof}$. This constraint very often makes studying the
   dependence of results on the truncation difficult due to limited number of
   input data.
 \item A frequentist fit allows for a measure of `quality of fit' in terms 
 of the $p$-value, which is well-defined assuming Gaussian statistics of 
 the input data. The quality of fit can be helpful in assessing how well
 a particular fit ansatz is compatible with the data. Given that the finite
 number of data points always requires us to truncate the fit ansatz, having
 such a measure is crucial in assessing the validity of the fit.
\item The fit carried out in the way described in this section does not impose
the unitarity constraint in Eq.~\eqref{eq:modified unitarity constraint}. 
While an a-posteriori check of the unitarity of the central fit result is possible,
it can be difficult to make consistent statements on whether the fit is
more generally compatible with unitarity given the Gaussian nature of the
error estimate. In the following we will provide a solution to this problem
by consistently embedding the unitarity constraint in the fitting strategy.

\end{itemize}
\subsection{Bayesian inference }
In Bayesian inference the fitting problem is formulated in terms of probability distributions
encoding prior knowledge not only about the
fit function and data, but, for instance, also about fundamental properties of quantum-field theory like
unitarity and analyticity. 
Here we  consider the unitarity constraint in~\eqref{eq:modified unitarity constraint} as prior knowledge. Other knowledge, like previous results for parameters of the BGL expansion
could also qualify as prior knowledge. However, in order to maintain model-independence and to
avoid any bias, care has to be 
taken when choosing priors.

\subsubsection{Theoretical setup}
Bayes' theorem states that 
\begin{equation}\label{eq:Bayes}
    \pi(A|B)=\frac{\pi(B|A)\pi(A)}{\pi(B)}\,,
\end{equation}
where 
\begin{itemize}
    \item $\pi(A|B)$ is the conditional probability density of $A$ happening given $B$, 
    \item $\pi(B|A)$ is the conditional probability density of $B$ happening given $A$,
    \item $\pi(A)$ and $\pi(B)$ are the probability densities for $A$ and $B$ happening without any conditions.
\end{itemize}
Assuming one knows the probabilities on the r.h.s. of Eq.~\eqref{eq:Bayes}, expectation values for functions $g(A)$ of parameters $A$ can be computed as
\begin{equation}\label{eq:Bayes expectation}
    \langle g(A)\rangle = \frac 1{\mathcal{Z}}\int dA \,g(A)\,\pi(A|B)\,,
\end{equation}
where $\mathcal{Z}=\int dA \,\pi(A|B)$ is a normalisation.

We consider the following prior probability distributions:
\begin{itemize}
\item The unitarity constraint Eq.~\eqref{eq:modified unitarity constraint} and any 
    prior knowledge (subscript $p$) about the fit parameters ${\bf a}_p$ assumed to be following Gaussian statistics with metric $M$ are encoded in the conditional probability distribution
    \begin{equation}\label{eq:prior}
    \pi_{\bf a}({\bf a}|{\bf a}_p,M)\propto 
         \theta({\bf a})\,
        {\rm exp}\left(-\frac 12 ({\bf a}-{\bf  a}_p)^T M({\bf  a}-{\bf  a}_p)\right)\,,
    \end{equation}
    where $\theta({\bf a})=\theta(1-|{\bf a}_{+}|_{\alpha}^2)\theta(1-|{\bf a}_{0}|_{\alpha}^2)$.
    The step functions $\theta$ impose the unitarity constraint for both the vector and scalar form 
    factors.
    The Gaussian term with metric $M$ allows the inclusion of prior knowledge, if available, about the fit parameters. In order
    to avoid introducing bias we will not add any such prior knowledge to the fits below, \textcolor{black}{\emph{i.e.}, the coefficients $a_i$ are drawn from a uniform distribution.} We will
    only make use of the Gaussian term in an intermediate step when formulating an efficient
    algorithm for integrating Eq.~\eqref{eq:Bayes expectation}. The final results in this paper
    will however be independent of it.
\item The input data ${\bf f}_p$ with covariance $C_{{\bf f}_p}$ to which the BGL ansatz is fitted is assumed to follow Gaussian 
    statistics and is represented by the probability distribution
    \begin{equation}\label{eq:fprior}
    \pi_{\bf f}({\bf f}|{\bf f}_p,C_{{\bf f}_p})\propto
    {\rm exp}\left(-\frac 12 ({\bf f}-{\bf  f}_p)^T C^{-1}_{{\bf f}_p}({\bf  f}-{\bf  f}_p)\right)\,.
    \end{equation}
\item We consider the BGL ansatz prior knowledge, represented by the distribution
   \begin{equation}\label{eq:BGLprior}
    \Theta({\bf  f}, {\bf  a}|Z)\propto \delta \left(|{\bf f}- Z{\bf a}|\right)\,.
\end{equation}
\end{itemize}
Marginalising Eqs.~\eqref{eq:fprior} and~\eqref{eq:BGLprior} over ${\bf f}$, leads to
\begin{equation}
    \pi_{\bf a}({\bf a}|{\bf f}_p,C_{{\bf f}_p})\propto{\rm exp}\left(-\frac 12\chi^2({\bf a},{\bf f}_p)\right)\,,
\end{equation}
where $\chi^2({\bf a},{\bf f}_p)$ is as defined in Eq.~\eqref{eq:chisq_BGL}.

Combining the above into a single probability distribution we get
\begin{align}
    \pi_{\bf a}({\bf a}|{\bf f}_p,C_{{\bf f}_p})\pi_{\bf a}({\bf a}|{\bf a}_p,M)
    &\nonumber\\
       &\hspace{-20ex}\propto \theta({\bf a}){\rm exp}\left(-\frac 12 ({\bf  f}_p-Z{\bf  a})^TC_{{\bf f}_p}^{-1}({\bf  f}_p-Z{\bf  a})-\frac 12 ({\bf a}-{\bf  a}_p)^TM({\bf  a}-{\bf  a}_p)\right)\nonumber\\
    &\hspace{-20ex}=\,\theta({\bf a})
        {\rm exp}\left(-\frac 12 ({\bf  a}-{\bf  \tilde a})^T  C_{{\bf \tilde a}}^{-1}({\bf  a}-{\bf \tilde a})\right)\,,\label{eq:full_PD}
\end{align}
where in the last line
\begin{equation}
    C_{{\bf \tilde a}}^{-1}=Z^T C_{{\bf f}_p}^{-1}Z+M\label{eq:tildeCa}\,,
\end{equation}
and
\begin{equation}
{\bf \tilde a}=C_{\tilde {\bf a}}\left(Z^TC_{{\bf f}_p}^{-1} {\bf  f}_p+M {\bf a}_p\right)\,.\label{eq:tildea}
\end{equation}
In analogy to the expectation value $\langle g(A)\rangle$ in Eq.~\eqref{eq:Bayes expectation},
expectation values $\langle g({\bf a})\rangle$ can now be computed in terms of 
Monte-Carlo integration by drawing from a multivariate normal distribution
\(\sim\mathcal{N}({\bf \tilde a}, C_{{\bf \tilde a}})\), restricting to those samples
that are compatible with the unitarity constraint~\eqref{eq:modified unitarity constraint}, 
which in the probability distribution Eq.~\eqref{eq:full_PD} is imposed in terms of the 
step functions $\theta({\bf a})$.
Note that in the absence of priors the maximum of $\pi_{\bf a}({\bf a}|{\bf f}_p,C_{{\bf f}_p})\pi_{\bf a}({\bf a}|{\bf a}_p,M)$ is reached for
${\bf a}$ as in Eq.~\eqref{eq:frequentist_sln_a}. In cases where unitarity imposes
only mild constraints on the fit result, for a given choice
of truncation $(K_+,K_0)$ we therefore expect central values and covariances of
${\bf a}$ from both approaches to agree.
\subsubsection{Proposed algorithm}
The unitarity constraint $\theta({\bf a})$ restricts the vectors ${\bf a}_+$ and ${\bf a}_0$, respectively, to lie within $K_{+,0}$-dimensional ellipsoids. Drawing
random numbers $\sim\mathcal{N}({\bf \tilde a}, C_{{\bf \tilde a}})$ may therefore
become inefficient for higher truncations due to the large number of samples
that have to be dropped where they are incompatible with unitarity.
To mitigate this problem we propose, as an intermediate step, to 
start with a choice of priors ${\bf a}_p={\bf 0}$ and with
metric $M/\sigma^2$, where $\sigma$ is a parameter that 
can be used to tune the width of the prior.  
In order to ensure that final results are independent of this intermediate
prior we propose to correct the sampling by means of an accept-reject step:
\begin{itemize}
    \item[1)] Draw a vector of multivariate random numbers ${\bf a}$ following
    $\mathcal{N}(\tilde {\bf a},C_{\tilde {\bf a} })$, with
    ${\bf a}_p={\bf 0}$ and metric $M/\sigma^2$.
    \item[2)] Continue with 3) if $|{\bf a}_+|_{\alpha_{Bs K}}^2\le 1$ and $|{\bf a}_0|_{\alpha_{Bs K}}^2\le 1$, otherwise restart at 1) -- this ensures that the 
    		parameters satisfy the unitarity condition in Eq.~\eqref{eq:modified unitarity constraint}.
    \item[3)] Draw a single uniform random number $p\in[0,1]$ and accept the proposal for $\bf a$ from step 1) only if 
    	\begin{equation}
		p\leq \frac{c}{{\rm exp}(-{\bf a}^TM{\bf a}/2\sigma^2)}\,\label{eq:accept reject p}\,,
	\end{equation}
		where $c={\rm exp}(-1/\sigma^2)$ is a normalisation factor ensuring $p\in [0,1]$, which assumes that 
        $|{\bf a}_+|_{\alpha_{Bs K}}^2\le 1$ and $|{\bf a}_0|_{\alpha_{Bs K}}^2\le 1$ hold.  In practice,
        in order to ensure
        that Eq.~\eqref{eq:accept reject p} constitutes
        a normalised acceptance probability, the metric has to be 
        chosen such that 
        that ${\bf a}^TM{\bf a}\le 2$. 
        How this can be achieved is detailed in App.~\ref{app:prior metric}.
  \item[4)] Restart at 1) until the desired number of samples has been generated.
\end{itemize}

\subsection{A combined frequentist and Bayesian perspective}
The frequentist and Bayesian approach, respectively, provide complementary information. A frequentist fit can make probabilistic
statements about the compatibility of the fit-function and data in terms of the $p$-value as derived from the 
$\chi^2$ distribution. Within the Bayesian framework only relative statements, \emph{i.e.}~a preference of one fit over another,
can be made. For instance, the ratio of marginalised probabilities of one model over another gives the Bayes factor, 
which in terms of the Jeffrey scale~\cite{Jeffreys:1939xee} can be used for model selection~\cite{Cossu:2020yeg,Jay:2020jkz,Neil:2022joj}. 
While we propose Bayesian inference as the preferred ansatz for fitting parameterisations to
form-factor data, frequentist fits, as we will demonstrate below, can still be a useful tool for testing compatibility of fit function and data.
\textcolor{black}{
\subsection{Truncation dependence}
Any practical implementation in a computer program 
requires one to restrict the BGL ansatz to a finite number of terms. 
The fit is truncation independent once the results for fit coefficients and errors have converged to stable values as $K$ is further increased, 
and it can be shown that contributions from above the truncation are  sufficiently suppressed to any order. 
We now discuss the two cases $\alpha=\pi$ and $\alpha <\pi$ separately:}

\textcolor{black}{
For $\alpha=\pi$ the unitarity constraint Eq.~\eqref{eq:modified unitarity constraint} is defined in terms of the metric $\langle z^i|z^j\rangle=\delta_{ij}$. It therefore corresponds to a sum of positive semi-definite terms. Contributions from higher orders are suppressed by powers of $z^i$ with coefficients $|a_i|\le 1$ that can strengthen but not weaken the unitarity constraint.
}

\textcolor{black}{
For $0<\alpha<\pi$ the metric $\langle z^i|z^j\rangle$ in the unitarity constraint
Eq.~\eqref{eq:modified unitarity constraint} mixes the BGL coefficients of all orders and the weak unitarity constraint in the form { $|a_i|<~1$} no longer holds. The unknown coefficients with $i\geq K$ 
above the
truncation could  in principle modify the contribution to the unitarity sum for
a given coefficient $a_i$ with $i<K$, thereby accidentally weakening or strengthening the unitarity constraint. While a weakening of the unitarity constraint would lead to larger errors within the Bayesian-inference approach, a constraint accidentally strengthened through the truncation could lead to underestimated errors. We can protect ourselves against underestimated errors as follows: The contributions 
to the dispersion integrals  Eqs.~\eqref{eq:chi-defn_T} and \eqref{eq:chi-defn_L} over the range $ t\in [0,(M_{B_s}+M_K)^2)$ are positive semi-definite and neglecting them on the r.h.s., respectively, turns both equations into inequalities. The integrals restricted 
to the range $t\in [(M_{B_s}+M_K)^2,\infty)$ can 
then be mapped to the unit disk with the prescription in Eq.~\eqref{eq:z-fn-defn} setting $t_\ast =t_+=(M_{B_s}+M_K)^2$, which corresopnds to  $\alpha=\pi$. This then allows for truncated
BGL fits with well-defined truncation as discussed in the previous paragraph. Repeating each fit in this paper following this prescription and comparing results, we confirm that the truncated BGL fit with unitarity constraint Eq.~\eqref{eq:modified unitarity constraint} is not accidentally over-constraining. 
}

\textcolor{black}{
Note however, that global and combined fits over
data from, \emph{e.g.}, $B\to\pi \ell\nu$ and $B_s\to K \ell\nu$ with simultaneous unitarity constraint $|{\bf a}_{X,B\to\pi}|^2+|{\bf a}_{X,B_s\to K}|^2\le 1$ may require the BGL ansatz  for both channels to be based on the same $z$-expansion (in particular the same choice of $t_\ast$).
}
\clearpage
\section{An example: Semileptonic $B_s\to K\ell\nu$ decay}\label{sec:BstoK}
In this section we demonstrate how Bayesian inference works in practice. We study
 as an example  the case of semileptonic 
$B_s\to K\ell \nu$ decay. The data sets we consider are 
HPQCD 14~\cite{Bouchard:2014ypa}, FNAL/MILC 19~\cite{Bazavov:2019aom} and RBC/UKQCD 23~\cite{Flynn:2023nhi} from lattice QCD, and 
Khodjamirian 17~\cite{Khodjamirian:2017fxg} from sum rules. 
In the following  sections we will first briefly discuss the individual 
data sets, then analyse them individually with
Bayesian inference and, following that, present combined fits over the data sets.
Besides presenting results for a number of phenomenologically relevant 
observables, this study will emphasise the benefit of combining insights 
from both Bayesian and frequentist analyses.
\subsection{Data preparation}
\begin{figure}
	\begin{center}
	\includegraphics[width=14cm]{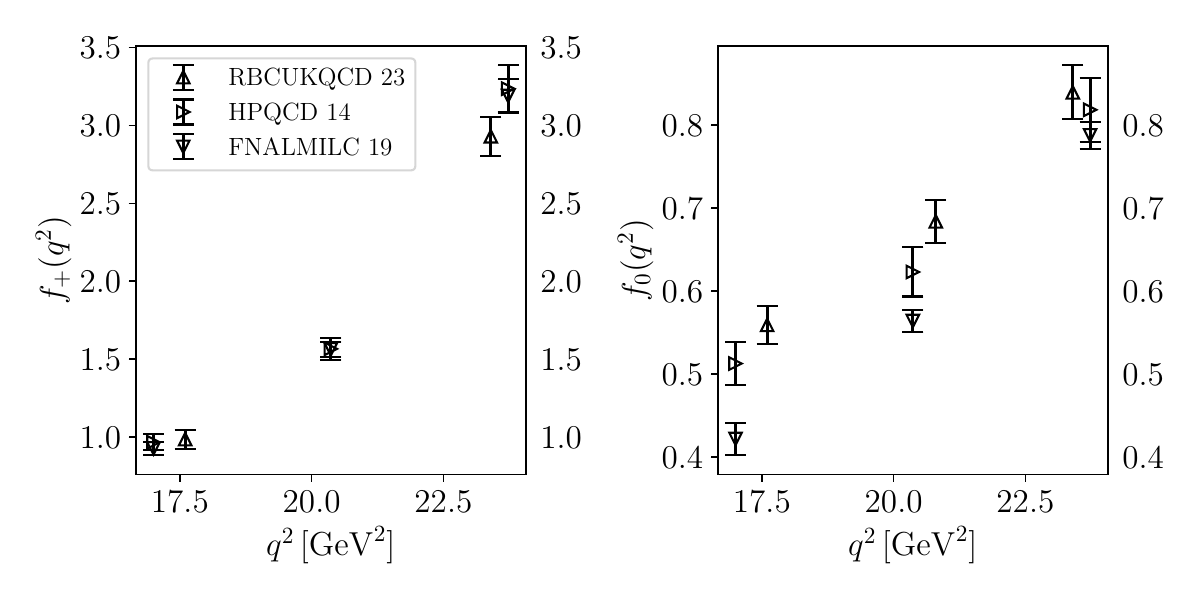}
	\end{center}
\caption{Summary of lattice data used in this study. The data was generated from BCL parameterisations provided in  by HPQCD 14~\cite{Bouchard:2014ypa} and 
FNAL/MILC 19~\cite{Bazavov:2019aom}. The values for RBC/UKQCD~23~\cite{Flynn:2023nhi} are from Tabs.~VII and VIII in their paper.}\label{fig:data_summary}
\end{figure}

\begin{itemize}
\item HPQCD 14~\cite{Bouchard:2014ypa} provide results in terms of central values, errors and correlation matrix
    for the coefficients of a BCL parameterisation~\cite{Bourrely:2008za} with truncation at order $K_+=3$ and $K_0=4$. The correlation matrix in Tab.~III of Ref.~\cite{Bouchard:2014ypa} is however only $6 \times 6$, since the kinematical constraint
    $f_+(0)=f_0(0)$ was imposed by eliminating one parameter in the expansion. We generate 
    central values and the covariance matrix for $f_+$($f_0$) at 3(3) reference-$q^2$ values in the region $17\,{\rm GeV}^2\le q^2\le q^2_{\rm max}$ by sampling BCL parameters from a multivariate normal distribution. The region of $q^2$ values
    corresponds to the kinematical region that is covered by lattice data in HPQCD 14.
\item FNAL/MILC 19~\cite{Bazavov:2019aom} provide results in terms of central values, errors and correlation matrix
    for the coefficients of a BCL parameterisation~\cite{Bourrely:2008za} with truncation at order $K_+=K_0=4$. The
    kinematical constraint $f_+(0)=f_0(0)$ is imposed via a Gaussian prior with a very narrow width $\epsilon=10^{-10}$. This constraint effectively
    eliminates one parameter (\emph{cf.} Eq.~\eqref{eq:kinematical constraint}). Formally
    the full $8 \times 8$ correlation matrix is
    therefore  singular (see discussion in  App. B of Ref.~\cite{Bazavov:2019aom}). 
    We generated
    synthetic data points by resampling
    3(4) reference-$q^2$ values for $f_+(f_0)$
    in the range  $17\,{\rm GeV}^2\le q^2\le q^2_{\rm max}$.
    We found the resulting correlation matrix to be poorly conditioned and
    therefore decided to produce synthetic data for only 3(3) 
    reference-$q^2$ values.
\item RBC/UKQCD 23~\cite{Flynn:2023nhi} provide results and a full 
    error budget for the form factors after their
    chiral and continuum extrapolation, \emph{i.e.}~before
    further analysing the data with a $z$ expansion
    and unitarity constraint. From Tabs.~VII and~VIII of their paper we 
    obtain values, errors and
    statistical and systematic covariances for form factors $f_+$ at $q^2=\{17.6, 23.4\}{\rm GeV}^2$ and for $f_0$ for $q^2=\{17.6,20.8, 23.4\}{\rm GeV}^2$. 
\item Khodjamirian 17~\cite{Khodjamirian:2017fxg} computed the result $f_+(0)=0.336(23)$ with QCD sum rules. For completeness we 
	note the earlier sum-rule results~\cite{Duplancic:2008tk,Faustov:2013ima,Wang:2012ab} for the form factor at $f_+(0)$.
\end{itemize}
We provide a summary of all lattice data in Tabs.~\ref{tab:HPQCD 14 input}-\ref{tab:RBCUKQCD 23 input} and in Fig.~\ref{fig:data_summary}. While all lattice data for $f_+$ are nicely compatible, there is a tension between
RBC/UKQCD 23 and HPQCD 14 on the one side, and FNAL/MILC 19 on the other side. A possible explanation for this tension was given in Ref.~\cite{Flynn:2023nhi},
but further studies will be required to understand and eventually resolve this tension.
\begin{table}
\begin{center}
\begin{center}
\tiny\begin{tabular}{l|cl|ccc|ccc}
\hline \hline 
\multicolumn{3}{c|}{}&\multicolumn{3}{c|}{$f_+$}&\multicolumn{3}{c}{$f_0$}\\
\cline{2-9}
\multicolumn{1}{c|}{}&$q^2\,[{\rm GeV}^2]$&&$17.00$&$20.36$&$23.73$&$17.00$&$20.36$&$23.73$\\
\cline{4-9}
\multicolumn{1}{c|}{}&&\multicolumn{1}{c|}{$f_{+/0}$}&0.968(51)&1.567(71)&3.24(15)&0.513(25)&0.623(30)&0.819(38)\\
\hline 
\multirow{3}{*}{$f_+$}&\multicolumn{1}{c|}{$17.00$}&0.968(51)&1.0000&0.9276&0.5854&0.4293&0.3864&0.3486\\
&\multicolumn{1}{c|}{$20.36$}&1.567(71)&0.9276&1.0000&0.8047&0.4346&0.4136&0.3645\\
&\multicolumn{1}{c|}{$23.73$}&3.24(15)&0.5854&0.8047&1.0000&0.4033&0.3707&0.3129\\
\hline 
\multirow{3}{*}{$f_0$}&\multicolumn{1}{c|}{$17.00$}&0.513(25)&0.4293&0.4346&0.4033&1.0000&0.9646&0.8713\\
&\multicolumn{1}{c|}{$20.36$}&0.623(30)&0.3864&0.4136&0.3707&0.9646&1.0000&0.9552\\
&\multicolumn{1}{c|}{$23.73$}&0.819(38)&0.3486&0.3645&0.3129&0.8713&0.9552&1.0000\\
\hline 
\hline 
\end{tabular}
\end{center}

\end{center}
\caption{Form factor data from HPQCD 14~\cite{Bouchard:2014ypa}. The table shows form-factor reference values
and errors for given $q^2$, and the corresponding the correlation matrix.}\label{tab:HPQCD 14 input}
\end{table}
\begin{table}
\begin{center}
\begin{center}
\tiny\begin{tabular}{l|cl|ccc|ccc}
\hline \hline 
\multicolumn{3}{c|}{}&\multicolumn{3}{c|}{$f_+$}&\multicolumn{3}{c}{$f_0$}\\
\cline{2-9}
\multicolumn{1}{c|}{}&$q^2\,[{\rm GeV}^2]$&&$17.00$&$20.36$&$23.73$&$17.00$&$20.36$&$23.73$\\
\cline{4-9}
\multicolumn{1}{c|}{}&&\multicolumn{1}{c|}{$f_{+/0}$}&0.928(43)&1.564(48)&3.19(11)&0.422(19)&0.564(14)&0.788(16)\\
\hline 
\multirow{3}{*}{$f_+$}&\multicolumn{1}{c|}{$17.00$}&0.928(43)&1.0000&0.8447&0.2180&0.6910&0.5889&0.3707\\
&\multicolumn{1}{c|}{$20.36$}&1.564(48)&0.8447&1.0000&0.6654&0.4604&0.5864&0.5070\\
&\multicolumn{1}{c|}{$23.73$}&3.19(11)&0.2180&0.6654&1.0000&0.1310&0.3447&0.3901\\
\hline 
\multirow{3}{*}{$f_0$}&\multicolumn{1}{c|}{$17.00$}&0.422(19)&0.6910&0.4604&0.1310&1.0000&0.8025&0.3754\\
&\multicolumn{1}{c|}{$20.36$}&0.564(14)&0.5889&0.5864&0.3447&0.8025&1.0000&0.7727\\
&\multicolumn{1}{c|}{$23.73$}&0.788(16)&0.3707&0.5070&0.3901&0.3754&0.7727&1.0000\\
\hline 
\hline 
\end{tabular}
\end{center}

\end{center}
\caption{Form factor data from FNAL/MILC 19~\cite{Bazavov:2019aom}. The table shows form-factor reference values and errors
for given reference-$q^2$ values, and  the correlation matrix.}
\label{tab:FNALMILC 19 input}
\end{table}
\begin{table}
\begin{center}
\begin{center}
\tiny\begin{tabular}{l|cl|cc|ccc}
\hline \hline 
\multicolumn{3}{c|}{}&\multicolumn{2}{c|}{$f_+$}&\multicolumn{3}{c}{$f_0$}\\
\cline{2-8}
\multicolumn{1}{c|}{}&$q^2\,[{\rm GeV}^2]$&&$17.60$&$23.40$&$17.60$&$20.80$&$23.40$\\
\cline{4-8}
\multicolumn{1}{c|}{}&&\multicolumn{1}{c|}{$f_{+/0}$}&0.988(60)&2.93(12)&0.559(23)&0.684(26)&0.840(33)\\
\hline 
\multirow{2}{*}{$f_+$}&\multicolumn{1}{c|}{$17.60$}&0.988(60)&1.0000&0.8473&0.7322&0.7654&0.7439\\
&\multicolumn{1}{c|}{$23.40$}&2.93(12)&0.8473&1.0000&0.6544&0.8146&0.8356\\
\hline 
\multirow{3}{*}{$f_0$}&\multicolumn{1}{c|}{$17.60$}&0.559(23)&0.7322&0.6544&1.0000&0.8816&0.8206\\
&\multicolumn{1}{c|}{$20.80$}&0.684(26)&0.7654&0.8146&0.8816&1.0000&0.9828\\
&\multicolumn{1}{c|}{$23.40$}&0.840(33)&0.7439&0.8356&0.8206&0.9828&1.0000\\
\hline 
\hline 
\end{tabular}
\end{center}

\end{center}
\caption{Form factor data from RBC/UKQCD 23~\cite{Flynn:2023nhi}. The table shows form-factor reference values and errors
for given $q^2$, and the corresponding 
the correlation matrix.}\label{tab:RBCUKQCD 23 input}
\end{table}
%
\subsection{Fits to individual data sets}
In this section we will apply both the frequentist and our new Bayesian-inference
fit strategies individually to the three lattice-data sets. We will first
discuss the BGL-fit results and then in Sec.~\ref{sec:Phenomenological analysis} discuss
a number of phenomenological predictions.
\begin{table}
\begin{center}
\tiny
{\normalsize HPQCD 14 -- ${\bf a}_+$}\\
\begin{tabular}{l@{\hspace{1mm}}llllllllllllllllllllllllllllllllllllllllllllllllll}
\hline\hline
$K_+$&$K_0$&\multicolumn{1}{c}{$a_{+,0}$}&\multicolumn{1}{c}{$a_{+,1}$}&\multicolumn{1}{c}{$a_{+,2}$}&$p$&$\chi^2/N_{\rm dof}$&$N_{\rm dof}$\\
\hline
2&2&0.0270(13)&-0.0792(50)&-& 0.03& 2.93&3&\\
2&3&0.0273(13)&-0.0760(63)&-& 0.02& 4.06&2&\\
3&2&0.0257(14)&-0.0805(50)&0.068(31)& 0.15& 1.89&2&\\
3&3&0.0262(14)&-0.0727(64)&0.096(34)& 0.97& 0.00&1&\\
\hline\hline\\
\end{tabular}
\\
{\normalsize FNAL/MILC 19 -- ${\bf a}_+$}\\
\begin{tabular}{l@{\hspace{1mm}}llllllllllllllllllllllllllllllllllllllllllllllllll}
\hline\hline
$K_+$&$K_0$&\multicolumn{1}{c}{$a_{+,0}$}&\multicolumn{1}{c}{$a_{+,1}$}&\multicolumn{1}{c}{$a_{+,2}$}&$p$&$\chi^2/N_{\rm dof}$&$N_{\rm dof}$\\
\hline
2&2&0.02489(94)&-0.0915(47)&-& 0.00& 6.52&3&\\
2&3&0.0263(10)&-0.0827(52)&-& 0.12& 2.12&2&\\
3&2&0.0239(10)&-0.0953(50)&0.044(19)& 0.00& 7.23&2&\\
3&3&0.0255(11)&-0.0858(57)&0.027(20)& 0.12& 2.38&1&\\
\hline\hline\\
\end{tabular}
\\
{\normalsize RBC/UKQCD 23 -- ${\bf a}_+$}\\
\begin{tabular}{l@{\hspace{1mm}}llllllllllllllllllllllllllllllllllllllllllllllllll}
\hline\hline
$K_+$&$K_0$&\multicolumn{1}{c}{$a_{+,0}$}&\multicolumn{1}{c}{$a_{+,1}$}&\multicolumn{1}{c}{$a_{+,2}$}&$p$&$\chi^2/N_{\rm dof}$&$N_{\rm dof}$\\
\hline
2&2&0.0293(11)&-0.0871(47)&-& 0.00& 9.52&2&\\
2&3&0.0249(16)&-0.0999(57)&-& 0.04& 4.33&1&\\
3&2&0.0245(16)&-0.0798(50)&0.093(21)& 0.84& 0.04&1&\\
\hline\hline\\
\end{tabular}
\\
{\normalsize HPQCD 14 -- ${\bf a}_0$}\\
\begin{tabular}{l@{\hspace{1mm}}llllllllllllllllllllllllllllllllllllllllllllllllll}
\hline\hline
$K_+$&$K_0$&\multicolumn{1}{c}{$a_{0,0}$}&\multicolumn{1}{c}{$a_{0,1}$}&\multicolumn{1}{c}{$a_{0,2}$}&$p$&$\chi^2/N_{\rm dof}$&$N_{\rm dof}$\\
\hline
2&2&0.0883(44)&-0.250(17)&-& 0.03& 2.93&3&\\
2&3&0.0880(44)&-0.242(19)&0.053(65)& 0.02& 4.06&2&\\
3&2&0.0906(45)&-0.240(17)&-& 0.15& 1.89&2&\\
3&3&0.0908(46)&-0.215(22)&0.138(71)& 0.97& 0.00&1&\\
\hline\hline\\
\end{tabular}
\\
{\normalsize FNAL/MILC 19 -- ${\bf a}_0$}\\
\begin{tabular}{l@{\hspace{1mm}}llllllllllllllllllllllllllllllllllllllllllllllllll}
\hline\hline
$K_+$&$K_0$&\multicolumn{1}{c}{$a_{0,0}$}&\multicolumn{1}{c}{$a_{0,1}$}&\multicolumn{1}{c}{$a_{0,2}$}&$p$&$\chi^2/N_{\rm dof}$&$N_{\rm dof}$\\
\hline
2&2&0.0775(28)&-0.275(13)&-& 0.00& 6.52&3&\\
2&3&0.0775(28)&-0.252(15)&0.153(39)& 0.12& 2.12&2&\\
3&2&0.0774(28)&-0.274(13)&-& 0.00& 7.23&2&\\
3&3&0.0774(28)&-0.254(15)&0.140(40)& 0.12& 2.38&1&\\
\hline\hline\\
\end{tabular}
\\
{\normalsize RBC/UKQCD 23 -- ${\bf a}_0$}\\
\begin{tabular}{l@{\hspace{1mm}}llllllllllllllllllllllllllllllllllllllllllllllllll}
\hline\hline
$K_+$&$K_0$&\multicolumn{1}{c}{$a_{0,0}$}&\multicolumn{1}{c}{$a_{0,1}$}&\multicolumn{1}{c}{$a_{0,2}$}&$p$&$\chi^2/N_{\rm dof}$&$N_{\rm dof}$\\
\hline
2&2&0.0981(36)&-0.287(15)&-& 0.00& 9.52&2&\\
2&3&0.0917(40)&-0.331(19)&-0.210(55)& 0.04& 4.33&1&\\
3&2&0.0950(37)&-0.262(16)&-& 0.84& 0.04&1&\\
\hline\hline\\
\end{tabular}

\end{center}
\caption{Results for the frequentist BGL fit to HPQCD 14, FNAL/MILC 19 and  RBC/UKQCD~23. The tables show the results for BGL coefficients for different
orders of the fit.}\label{tab:RBCUKQCD 23 frequ}
\end{table}
\subsubsection{Results for frequentist fits}
Tab.~\ref{tab:RBCUKQCD 23 frequ} summarises the results
of a frequentist analysis for all three data sets, 
where in each case we performed a simultaneous correlated fit to 
$f_+$ and $f_0$, subject to the 
constraints $f_+(0)=f_0(0)$ and $N_{\rm dof}\ge 1$. 
We make the following observations:
\begin{itemize}
    \item Judging by the $p$-value  fits with $(K_+,K_0)=(2,2),\, (2,3)$ are excluded by HPQCD 14 and RBC/UKQCD 23, while
    fits with $K_+\ge 3$ and $K_0\ge 2$ lead to acceptable fits for all data sets. Note that this is a data-dependent observation since one expects higher-order terms
    to be important for acceptable fits once results for form factors 
    with higher precision become available.
    \item For HPQCD 14 we find some variation of the ${\bf a}_{+,1}$ 
    coefficients at the 
    1$\sigma$ level between $(K_+,K_0)=(3,2)$ and (3,3). 
    For ${\bf a}_0$ we  see a similar variation in $a_{0,1}$, 
    and we obtain only one fit with acceptable $p$-value that is able to determine $a_{0,2}$.
    \item For FNAL/MILC 19 we obtain acceptable fits only for $(K_+,K_0)=(2,3$) and (3,3). We find the coefficients that are common
    to both truncations to agree within one standard deviation.
    \item For RBC/UKQCD 23 only fits with $(K_+,K_0)=(2,2),\,(2,3)$ and $(3,2)$ are 
    possible. There is essentially only one acceptable fit, the one with $(3,2)$. Consequently no statements about convergence  of the fit parameters
    are possible.
    \item HPQCD 14 and RBC/UKQCD 23 obtain compatible results, which are however in tension with FNAL/MILC 19 -- this is in line with the observation in Fig.~\ref{fig:data_summary}, that the respective
    data sets appear to be under tension.
\end{itemize}
For frequentist fits the constraint $N_{\rm dof}\ge 1$ severely limits 
the ability to probe the truncation dependence of the fit, and an
irreducible systematic error remains. After the above considerations one
could choose the results with truncations $(3,3)$ for HPQCD 14 and FNAL/MILC 19, respectively, and $(3,2)$ for RBC/UKQCD 23.
Whether higher-order coefficients could still
significantly modify these results has to be delegated to a 
systematic error budget, for which in our opinion no satisfactory 
procedure exists.
\subsubsection{Results for Bayesian inference}
\begin{figure}
    \centering
    \includegraphics[width=14cm]{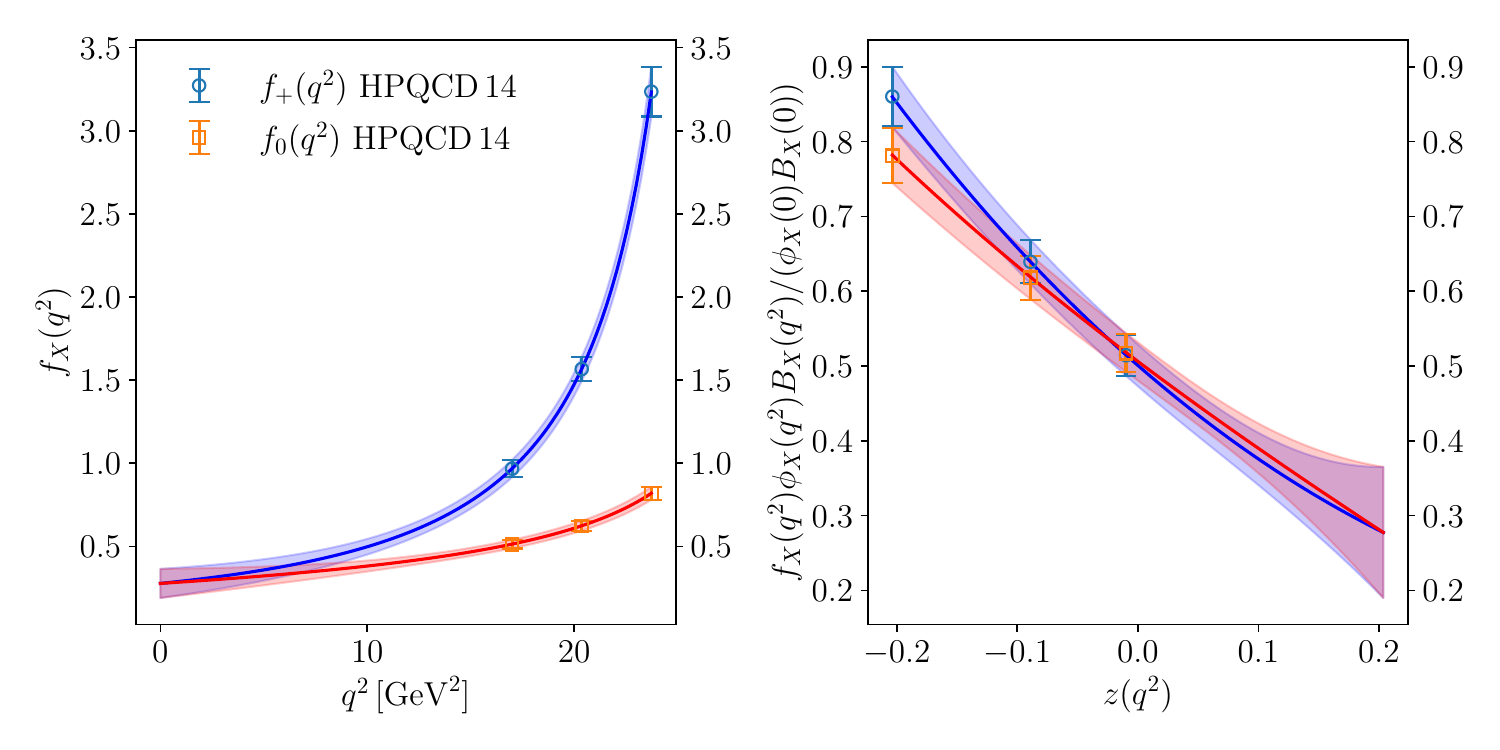}
    \caption{Illustration of the Bayesian-inference fit to the HPQCD-14 data~\cite{Bouchard:2014ypa} with $(K_+,K_0)=(5,5)$. Left: plot of the form
    factor vs. the squared momentum transfer; right: plot of the form factor
    after removing Blaschke and outer function, normalised such that the 
    kinematical constraint $f_0(0)=f_+(0)$ becomes apparent.}
    \label{fig:exemplary BGL fit HPQCD 14}
\end{figure}

%
\begin{table*}[hbt!]
\begin{center}
\tiny
{\normalsize HPQCD 14 -- ${\bf a}_+$}\\
\begin{tabular}{l@{\hspace{1mm}}llllllllllllllllllllllllllllllllllllllllllllllllll}
\hline\hline
$K_+$&$K_0$&\multicolumn{1}{c}{$a_{+,0}$}&\multicolumn{1}{c}{$a_{+,1}$}&\multicolumn{1}{c}{$a_{+,2}$}&\multicolumn{1}{c}{$a_{+,3}$}&\multicolumn{1}{c}{$a_{+,4}$}&\multicolumn{1}{c}{$a_{+,5}$}&\multicolumn{1}{c}{$a_{+,6}$}&\multicolumn{1}{c}{$a_{+,7}$}&\multicolumn{1}{c}{$a_{+,8}$}&\multicolumn{1}{c}{$a_{+,9}$}&\\
\hline
2&2&0.0270(12)&-0.0792(49)&- &- &- &- &- &- &- &-&\\
2&3&0.0273(13)&-0.0761(63)&- &- &- &- &- &- &- &-&\\
3&2&0.0257(14)&-0.0805(49)&0.069(30)&- &- &- &- &- &- &-&\\
3&3&0.0261(14)&-0.0728(64)&0.096(34)&- &- &- &- &- &- &-&\\
3&4&0.0261(14)&-0.0728(76)&0.096(39)&- &- &- &- &- &- &-&\\
4&3&0.0261(14)&-0.0729(68)&0.096(35)&0.008(90)&- &- &- &- &- &-&\\
4&4&0.0261(14)&-0.0730(77)&0.091(62)&-0.02(20)&- &- &- &- &- &-&\\
5&5&0.0262(15)&-0.0735(79)&0.084(67)&-0.03(19)&0.03(68)&- &- &- &- &-&\\
6&6&0.0261(14)&-0.0735(79)&0.086(69)&-0.03(19)&-0.00(64)&0.01(65)&- &- &- &-&\\
7&7&0.0262(14)&-0.0732(84)&0.088(69)&-0.02(18)&0.01(65)&0.02(73)&-0.03(70)&- &- &-&\\
8&8&0.0261(14)&-0.0732(80)&0.089(72)&-0.02(18)&-0.00(66)&0.03(86)&-0.04(90)&0.03(73)&- &-&\\
9&9&0.0261(14)&-0.0729(84)&0.095(75)&-0.02(19)&-0.04(68)&0.1(1.0)&-0.1(1.2)&0.1(1.1)&-0.06(79)&-&\\
10&10&0.0261(14)&-0.0726(89)&0.101(79)&-0.01(20)&-0.09(73)&0.2(1.3)&-0.3(1.7)&0.2(1.8)&-0.2(1.4)&0.08(87)&\\
\hline\hline\\
\end{tabular}
\\
{\normalsize FNAL/MILC 19 -- ${\bf a}_+$}\\
\begin{tabular}{l@{\hspace{1mm}}llllllllllllllllllllllllllllllllllllllllllllllllll}
\hline\hline
$K_+$&$K_0$&\multicolumn{1}{c}{$a_{+,0}$}&\multicolumn{1}{c}{$a_{+,1}$}&\multicolumn{1}{c}{$a_{+,2}$}&\multicolumn{1}{c}{$a_{+,3}$}&\multicolumn{1}{c}{$a_{+,4}$}&\multicolumn{1}{c}{$a_{+,5}$}&\multicolumn{1}{c}{$a_{+,6}$}&\multicolumn{1}{c}{$a_{+,7}$}&\multicolumn{1}{c}{$a_{+,8}$}&\multicolumn{1}{c}{$a_{+,9}$}&\\
\hline
2&2&0.02489(92)&-0.0916(46)&- &- &- &- &- &- &- &-&\\
2&3&0.02626(99)&-0.0827(51)&- &- &- &- &- &- &- &-&\\
3&2&0.0239(10)&-0.0955(49)&0.044(19)&- &- &- &- &- &- &-&\\
3&3&0.0255(11)&-0.0856(56)&0.027(20)&- &- &- &- &- &- &-&\\
3&4&0.0248(12)&-0.0949(80)&0.003(25)&- &- &- &- &- &- &-&\\
4&3&0.0248(12)&-0.0972(92)&-0.026(40)&-0.094(60)&- &- &- &- &- &-&\\
4&4&0.0248(12)&-0.0967(96)&-0.026(64)&-0.09(18)&- &- &- &- &- &-&\\
5&5&0.0248(12)&-0.0968(98)&-0.026(67)&-0.08(18)&0.05(67)&- &- &- &- &-&\\
6&6&0.0249(12)&-0.0964(98)&-0.021(68)&-0.07(17)&0.02(64)&-0.01(67)&- &- &- &-&\\
7&7&0.0248(12)&-0.0961(96)&-0.017(69)&-0.06(17)&0.03(63)&-0.03(73)&0.00(68)&- &- &-&\\
8&8&0.0248(12)&-0.096(10)&-0.012(73)&-0.05(17)&0.02(66)&-0.01(87)&-0.02(89)&0.01(72)&- &-&\\
9&9&0.0249(13)&-0.095(10)&-0.004(73)&-0.03(18)&-0.02(69)&0.0(1.1)&-0.0(1.2)&0.0(1.1)&-0.01(78)&-&\\
10&10&0.0249(12)&-0.094(10)&0.003(78)&-0.01(19)&-0.04(73)&0.1(1.3)&-0.1(1.7)&0.1(1.7)&-0.1(1.4)&0.03(85)&\\
\hline\hline\\
\end{tabular}
\\
{\normalsize RBC/UKQCD 23 -- ${\bf a}_+$}\\
\begin{tabular}{l@{\hspace{1mm}}llllllllllllllllllllllllllllllllllllllllllllllllll}
\hline\hline
$K_+$&$K_0$&\multicolumn{1}{c}{$a_{+,0}$}&\multicolumn{1}{c}{$a_{+,1}$}&\multicolumn{1}{c}{$a_{+,2}$}&\multicolumn{1}{c}{$a_{+,3}$}&\multicolumn{1}{c}{$a_{+,4}$}&\multicolumn{1}{c}{$a_{+,5}$}&\multicolumn{1}{c}{$a_{+,6}$}&\multicolumn{1}{c}{$a_{+,7}$}&\multicolumn{1}{c}{$a_{+,8}$}&\multicolumn{1}{c}{$a_{+,9}$}&\\
\hline
2&2&0.0293(11)&-0.0871(46)&- &- &- &- &- &- &- &-&\\
2&3&0.0249(16)&-0.0999(57)&- &- &- &- &- &- &- &-&\\
3&2&0.0245(16)&-0.0799(50)&0.093(21)&- &- &- &- &- &- &-&\\
3&3&0.0245(15)&-0.078(12)&0.101(49)&- &- &- &- &- &- &-&\\
3&4&0.0246(16)&-0.078(16)&0.100(70)&- &- &- &- &- &- &-&\\
4&3&0.0246(17)&-0.075(31)&0.102(49)&-0.07(72)&- &- &- &- &- &-&\\
4&4&0.0246(17)&-0.077(32)&0.100(68)&-0.03(70)&- &- &- &- &- &-&\\
5&5&0.0246(17)&-0.074(31)&0.099(70)&-0.08(67)&0.05(70)&- &- &- &- &-&\\
6&6&0.0247(16)&-0.073(32)&0.101(69)&-0.10(69)&0.09(74)&-0.05(71)&- &- &- &-&\\
7&7&0.0247(17)&-0.071(33)&0.107(70)&-0.11(72)&0.08(89)&-0.04(89)&0.03(73)&- &- &-&\\
8&8&0.0248(17)&-0.068(35)&0.102(74)&-0.18(77)&0.2(1.1)&-0.2(1.3)&0.1(1.2)&-0.06(82)&- &-&\\
9&9&0.0248(18)&-0.068(38)&0.107(85)&-0.16(82)&0.2(1.4)&-0.2(1.9)&0.1(1.9)&-0.1(1.5)&0.03(89)&-&\\
10&10&0.0247(18)&-0.067(43)&0.112(95)&-0.15(90)&0.2(1.8)&-0.2(2.6)&0.1(2.9)&-0.1(2.7)&-0.0(1.9)&0.02(98)&\\
\hline\hline\\
\end{tabular}
\\
\end{center}
\caption{Results for the individual Bayesian-inference BGL fits to HPQCD 14, FNAL/MILC 19 and  RBC/UKQCD 23, respectively. 
The tables show the results for BGL coefficients ${\bf a_+}$ for different orders of the fit.}\label{tab:BI results for BGL coefficients a+}
\end{table*}
\begin{table*}[hbt!]
\begin{center}
\tiny
{\normalsize HPQCD 14 -- ${\bf a}_0$}\\
\begin{tabular}{l@{\hspace{1mm}}llllllllllllllllllllllllllllllllllllllllllllllllll}
\hline\hline
$K_+$&$K_0$&\multicolumn{1}{c}{$a_{0,0}$}&\multicolumn{1}{c}{$a_{0,1}$}&\multicolumn{1}{c}{$a_{0,2}$}&\multicolumn{1}{c}{$a_{0,3}$}&\multicolumn{1}{c}{$a_{0,4}$}&\multicolumn{1}{c}{$a_{0,5}$}&\multicolumn{1}{c}{$a_{0,6}$}&\multicolumn{1}{c}{$a_{0,7}$}&\multicolumn{1}{c}{$a_{0,8}$}&\multicolumn{1}{c}{$a_{0,9}$}&\\
\hline
2&2&0.0883(44)&-0.250(17)&- &- &- &- &- &- &- &-&\\
2&3&0.0880(44)&-0.243(19)&0.052(65)&- &- &- &- &- &- &-&\\
3&2&0.0907(46)&-0.240(17)&- &- &- &- &- &- &- &-&\\
3&3&0.0906(44)&-0.215(22)&0.137(73)&- &- &- &- &- &- &-&\\
3&4&0.0907(47)&-0.215(22)&0.14(11)&-0.01(31)&- &- &- &- &- &-&\\
4&3&0.0907(45)&-0.214(22)&0.139(72)&- &- &- &- &- &- &-&\\
4&4&0.0907(46)&-0.215(25)&0.12(19)&-0.08(60)&- &- &- &- &- &-&\\
5&5&0.0909(46)&-0.218(25)&0.10(19)&-0.12(55)&0.04(63)&- &- &- &- &-&\\
6&6&0.0907(45)&-0.217(25)&0.10(19)&-0.11(53)&0.06(66)&-0.02(66)&- &- &- &-&\\
7&7&0.0907(46)&-0.217(26)&0.11(20)&-0.08(51)&0.03(73)&0.03(81)&-0.04(70)&- &- &-&\\
8&8&0.0908(46)&-0.217(25)&0.11(20)&-0.08(50)&-0.01(84)&0.1(1.0)&-0.09(96)&0.08(74)&- &-&\\
9&9&0.0907(46)&-0.215(25)&0.13(22)&-0.05(50)&-0.06(95)&0.2(1.4)&-0.2(1.5)&0.1(1.2)&-0.05(82)&-&\\
10&10&0.0907(46)&-0.214(27)&0.15(24)&-0.03(49)&-0.2(1.1)&0.4(1.8)&-0.5(2.2)&0.4(2.1)&-0.3(1.6)&0.13(90)&\\
\hline\hline\\
\end{tabular}
\\
{\normalsize FNAL/MILC 19 -- ${\bf a}_0$}\\
\begin{tabular}{l@{\hspace{1mm}}llllllllllllllllllllllllllllllllllllllllllllllllll}
\hline\hline
$K_+$&$K_0$&\multicolumn{1}{c}{$a_{0,0}$}&\multicolumn{1}{c}{$a_{0,1}$}&\multicolumn{1}{c}{$a_{0,2}$}&\multicolumn{1}{c}{$a_{0,3}$}&\multicolumn{1}{c}{$a_{0,4}$}&\multicolumn{1}{c}{$a_{0,5}$}&\multicolumn{1}{c}{$a_{0,6}$}&\multicolumn{1}{c}{$a_{0,7}$}&\multicolumn{1}{c}{$a_{0,8}$}&\multicolumn{1}{c}{$a_{0,9}$}&\\
\hline
2&2&0.0775(27)&-0.275(13)&- &- &- &- &- &- &- &-&\\
2&3&0.0775(27)&-0.253(15)&0.153(39)&- &- &- &- &- &- &-&\\
3&2&0.0773(28)&-0.274(13)&- &- &- &- &- &- &- &-&\\
3&3&0.0775(28)&-0.253(15)&0.141(40)&- &- &- &- &- &- &-&\\
3&4&0.0735(36)&-0.297(31)&0.088(51)&0.32(20)&- &- &- &- &- &-&\\
4&3&0.0734(38)&-0.305(36)&-0.01(10)&- &- &- &- &- &- &-&\\
4&4&0.0736(38)&-0.304(37)&-0.01(20)&-0.00(61)&- &- &- &- &- &-&\\
5&5&0.0735(38)&-0.303(36)&-0.00(20)&0.01(55)&-0.05(62)&- &- &- &- &-&\\
6&6&0.0736(37)&-0.301(36)&0.01(20)&0.04(52)&-0.07(64)&0.07(63)&- &- &- &-&\\
7&7&0.0735(38)&-0.300(36)&0.03(20)&0.07(51)&-0.18(73)&0.19(78)&-0.14(69)&- &- &-&\\
8&8&0.0737(38)&-0.298(36)&0.05(21)&0.09(51)&-0.25(85)&0.3(1.1)&-0.28(99)&0.15(74)&- &-&\\
9&9&0.0736(40)&-0.296(36)&0.08(22)&0.15(50)&-0.41(97)&0.6(1.4)&-0.6(1.5)&0.4(1.2)&-0.19(80)&-&\\
10&10&0.0738(36)&-0.292(35)&0.11(24)&0.17(49)&-0.6(1.1)&0.9(1.8)&-1.0(2.2)&0.8(2.1)&-0.5(1.6)&0.18(90)&\\
\hline\hline\\
\end{tabular}
\\
{\normalsize RBC/UKQCD 23 -- ${\bf a}_0$}\\
\begin{tabular}{l@{\hspace{1mm}}llllllllllllllllllllllllllllllllllllllllllllllllll}
\hline\hline
$K_+$&$K_0$&\multicolumn{1}{c}{$a_{0,0}$}&\multicolumn{1}{c}{$a_{0,1}$}&\multicolumn{1}{c}{$a_{0,2}$}&\multicolumn{1}{c}{$a_{0,3}$}&\multicolumn{1}{c}{$a_{0,4}$}&\multicolumn{1}{c}{$a_{0,5}$}&\multicolumn{1}{c}{$a_{0,6}$}&\multicolumn{1}{c}{$a_{0,7}$}&\multicolumn{1}{c}{$a_{0,8}$}&\multicolumn{1}{c}{$a_{0,9}$}&\\
\hline
2&2&0.0981(36)&-0.286(14)&- &- &- &- &- &- &- &-&\\
2&3&0.0917(39)&-0.331(19)&-0.211(53)&- &- &- &- &- &- &-&\\
3&2&0.0950(37)&-0.263(15)&- &- &- &- &- &- &- &-&\\
3&3&0.0953(43)&-0.254(41)&0.02(13)&- &- &- &- &- &- &-&\\
3&4&0.0955(44)&-0.254(42)&0.02(22)&-0.02(60)&- &- &- &- &- &-&\\
4&3&0.0954(43)&-0.254(40)&0.03(12)&- &- &- &- &- &- &-&\\
4&4&0.0953(42)&-0.254(42)&0.02(21)&-0.02(60)&- &- &- &- &- &-&\\
5&5&0.0954(44)&-0.254(41)&0.02(21)&-0.01(55)&-0.00(62)&- &- &- &- &-&\\
6&6&0.0957(42)&-0.251(41)&0.04(21)&-0.01(52)&-0.06(65)&0.07(65)&- &- &- &-&\\
7&7&0.0955(44)&-0.250(40)&0.06(20)&0.05(50)&-0.13(72)&0.17(79)&-0.12(69)&- &- &-&\\
8&8&0.0954(43)&-0.250(41)&0.06(22)&0.06(50)&-0.18(84)&0.2(1.0)&-0.21(99)&0.10(74)&- &-&\\
9&9&0.0956(44)&-0.247(41)&0.08(23)&0.06(50)&-0.27(96)&0.4(1.4)&-0.4(1.5)&0.3(1.2)&-0.15(80)&-&\\
10&10&0.0956(42)&-0.245(42)&0.11(24)&0.11(49)&-0.4(1.1)&0.7(1.8)&-0.8(2.2)&0.7(2.1)&-0.4(1.5)&0.16(87)&\\
\hline\hline\\
\end{tabular}

\end{center}
\caption{Results for the individual Bayesian-inference BGL fits to HPQCD 14, FNAL/MILC 19 and  RBC/UKQCD 23, respectively. 
The tables show the results for BGL coefficients ${\bf a_0}$  for different orders of the fit.}\label{tab:BI results for BGL coefficients a0}
\end{table*}
Here we repeat the same fits as in the previous section but now using the new Bayesian-inference approach,
which allows us to analyse the data with higher truncation $(K_+,K_0)$ than possible
in the frequentist case.
Tabs.~\ref{tab:BI results for BGL coefficients a+} 
and~\ref{tab:BI results for BGL coefficients a0}
show the results for Bayesian inference, 
and in Fig.~\ref{fig:exemplary BGL fit HPQCD 14} we exemplarily show the result of the 
Bayesian-inference fit to the HPQCD 14 data.
We make the following observations:
\begin{itemize}
    \item Frequentist fits and Bayesian inference, where possible at the same $(K_+,K_0$)
    agree.  This is 
    the expected behaviour: The fit results for the $B_s\to K\ell\nu$ decay considered
    here do not saturate the
    unitarity constraint Eq.~\eqref{eq:modified unitarity constraint}. In this
    situation  the maximum and width of the 
    probability distribution Eq.~\eqref{eq:full_PD} in Bayesian inference are
    described by the
    results obtained for central values ${\bf a}$ and covariance $C_{\bf a}$
    in the frequentist fit,
    as given in Eq.~\eqref{eq:frequentist_sln_a} and~\eqref{eq:frequentist_sln_Ca}.
    \item The power of Bayesian inference lies in the fact that  
    the order of the $z$ expansion can be extended beyond the frequentist constraint 
    $N_{\rm dof}\ge 1$: The data in Tabs.~\ref{tab:BI results for BGL coefficients a+} and~\ref{tab:BI results for BGL coefficients a0} shows that the central values for the BGL coefficients converge to stable central values.
    The unitarity constraint in
    Eq.~\eqref{eq:modified unitarity constraint} efficiently regulates the
    fluctuations of higher-order coefficients. By making use of unitarity and
    analyticity the hard-to-estimate
    truncation errors in the frequentist fit have been replaced by well-motivated and
    model-independent 
    statistical noise originating from the undetermined higher-order coefficients. 
    \item Note that in particular the samples of higher-order coefficients may not necessarily follow a normal
    distribution, in particular if they are determined mainly through the unitarity constraint.
    The errors given in the data tables have to be interpreted with this in mind. 
    It may in this context also at first be surprising, that some higher-order coefficients
    in the tables have central values, which apparently saturate the unitarity constraint. 
    Similarly, some coefficients have at first sight rather large `1$\sigma$' errors, which don't appear consistent
    with the unitarity constraint.
    However, such fluctuations are allowed and compatible with the modified unitarity constraint in 
    Eq.~\eqref{eq:modified unitarity constraint}.
    We check in our algorithm that this unitarity constraint is fulfilled at each step of the
    analysis. 
    \item The maximum truncation shown, $(K_+,K_0)=(10,10)$ is only for 
    demonstration purposes -- we see no significant changes in the
    fit coefficients and errors for $(K_+,K_0)\ge(5,5)$ and therefore
    choose this truncation for the main results of our study.

\end{itemize}
Bayesian inference regulated by unitarity and analyticity proves to be a 
powerful tool for truncation-independent fits to form-factor data. 
\subsubsection{Combined Bayesian and frequentist analysis}
The Bayesian-inference framework makes no statements about 
the quality of the BGL fit for a given truncation. Its power lies in its ability to fit the BGL ansatz without truncation error.
The frequentist fit on the other hand only provides meaningful results 
for $N_{\rm dof}\ge 1$, \emph{i.e.}~for a finite truncation. For this finite
truncation, however, quality measures like the $p$-value do make statements
about how well data and fit function are compatible. It is therefore
always advisable to consider both for a comprehensive data analysis.
Consider the case where a wrong assumption was made in the fit function,
or where the input data is erroneous -- apart from a visual inspection 
of a Bayesian-inference fit clearly indicating that something is wrong, only the frequentist fit provides a quantitative measure for the quality of the fit
that could indicate a problem. 
\subsection{Combined fits}
It is straight forward to combine results from different sources into one
global Bayesian-inference analysis. Essentially, this amounts to extending
the data vector ${\bf f}$ and covariance $C_{\bf f}$ by the additional 
data sets. Correlations between data set can be included by adding
the corresponding entries to the off-diagonal blocks of the enlarged
covariance matrix.
\subsubsection{Combined fits to lattice data}
We combine the results for HPQCD 14, FNAL/MILC 19 and RBC/UKQCD~23 in 
Tab.~\ref{tab:HPQCD 14 input}-\ref{tab:RBCUKQCD 23 input}, assuming the results
and errors from these three data sources to be independent.
We find that the FNAL/MILC 19 data on the one hand and the HPQCD 14 and RBC/UKQCD 23 on
the other are incompatible, as indicated by
visual inspection of Fig.~\ref{fig:data_summary}, and by unacceptably  small $p$-values of 
such a fit as summarised in Tab.~\ref{tab:all lattice data frequentist} in App.~\ref{sec:all lattice data results}.
We note that a Bayesian-inference analysis would nevertheless be
possible. This just underlines the importance of making best 
use of the complementary information one gains from 
frequentist and Bayesian fitting, respectively. 

\begin{figure}[bt!]
    \centering
    \includegraphics[width=14cm]{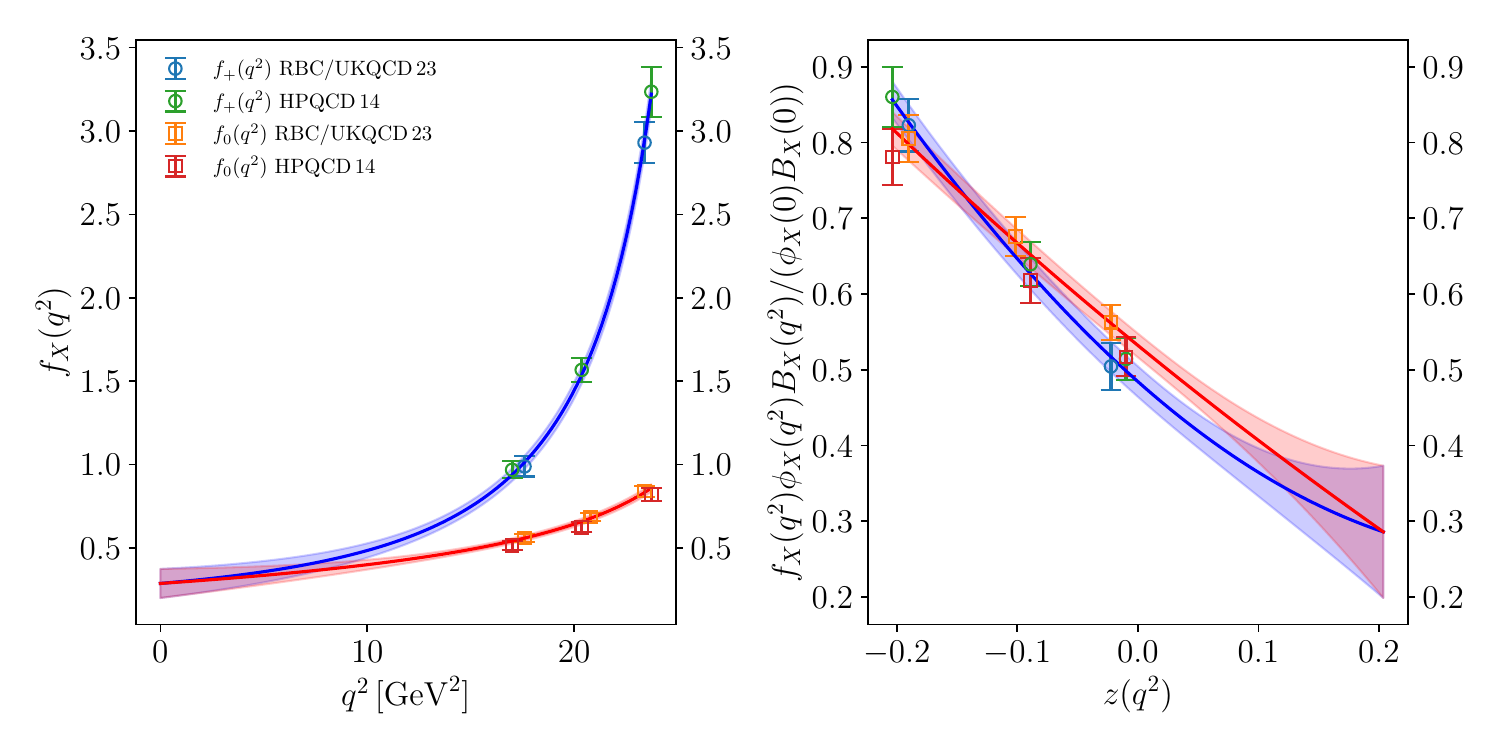}
    \caption{Illustration of the joint Bayesian-inference fit to the HPQCD 14~\cite{Bouchard:2014ypa} and RBC/UKQCD 23~\cite{Flynn:2023nhi} data sets
    with $(K_+,K_0)=(5,5)$. Left: plot of the form
    factor vs. the squared momentum transfer; right: plot of the form factor
    after removing Blaschke and outer function, normalised such that the 
    kinematical constraint $f_0(0)=f_+(0)$ becomes apparent.}
    \label{fig:exemplary BGL fit RBCUKQCD 23 HPQCD 14}
\end{figure}

We proceed considering only the combined fit over the data sets by
HPQCD 14 and RBC/UKQCD 23. The results for the frequentist and Bayesian
BGL fits are presented in Tabs.~\ref{tab:combined BGL frequ} 
and~\ref{tab:combined BGL Bayesian}, respectively.
\begin{table*}[bt!]
\begin{center}
\tiny
\begin{tabular}{l@{\hspace{1mm}}llllllllllllllllllllllllllllllllllllllllllllllllll}
\hline\hline
$K_+$&$K_0$&\multicolumn{1}{c}{$a_{+,0}$}&\multicolumn{1}{c}{$a_{+,1}$}&\multicolumn{1}{c}{$a_{+,2}$}&\multicolumn{1}{c}{$a_{+,3}$}&\multicolumn{1}{c}{$a_{+,4}$}&$p$&$\chi^2/N_{\rm dof}$&$N_{\rm dof}$\\
\hline
2&2&0.02805(81)&-0.0822(33)&- &- &-& 0.00& 4.02&8&\\
2&3&0.0266(10)&-0.0881(40)&- &- &-& 0.00& 3.69&7&\\
3&2&0.0250(10)&-0.0794(34)&0.083(16)&- &-& 0.47& 0.95&7&\\
3&3&0.0253(10)&-0.0731(52)&0.110(24)&- &-& 0.67& 0.67&6&\\
3&4&0.0253(11)&-0.0742(68)&0.105(32)&- &-& 0.56& 0.79&5&\\
4&3&0.0253(11)&-0.0738(58)&0.111(24)&0.024(89)&-& 0.56& 0.79&5&\\
4&4&0.0257(13)&-0.038(54)&0.61(74)&1.7(2.5)&-& 0.48& 0.87&4&\\
5&5&0.0261(14)&-0.002(77)&1.2(1.1)&5.3(6.3)&6.7(18.1)& 0.23& 1.46&2&\\
\hline\hline\\
\end{tabular}
\\
\begin{tabular}{l@{\hspace{1mm}}llllllllllllllllllllllllllllllllllllllllllllllllll}
\hline\hline
$K_+$&$K_0$&\multicolumn{1}{c}{$a_{0,0}$}&\multicolumn{1}{c}{$a_{0,1}$}&\multicolumn{1}{c}{$a_{0,2}$}&\multicolumn{1}{c}{$a_{0,3}$}&\multicolumn{1}{c}{$a_{0,4}$}&$p$&$\chi^2/N_{\rm dof}$&$N_{\rm dof}$\\
\hline
2&2&0.0938(27)&-0.270(11)&- &- &-& 0.00& 4.02&8&\\
2&3&0.0926(28)&-0.289(13)&-0.098(39)&- &-& 0.00& 3.69&7&\\
3&2&0.0942(27)&-0.256(11)&- &- &-& 0.47& 0.95&7&\\
3&3&0.0955(29)&-0.234(17)&0.091(56)&- &-& 0.67& 0.67&6&\\
3&4&0.0955(29)&-0.235(18)&0.07(10)&-0.08(30)&-& 0.56& 0.79&5&\\
4&3&0.0956(29)&-0.234(18)&0.093(57)&- &-& 0.56& 0.79&5&\\
4&4&0.0968(34)&-0.11(19)&1.8(2.6)&5.6(8.5)&-& 0.48& 0.87&4&\\
5&5&0.0967(35)&-0.07(22)&3.2(3.5)&19.7(21.6)&40.7(54.6)& 0.23& 1.46&2&\\
\hline\hline\\
\end{tabular}

\end{center}
\caption{Results for the frequentist BGL fit to HPQCD 14 and  RBC/UKQCD 23. The tables show the results for BGL coefficients for different orders of the fit.}\label{tab:combined BGL frequ}
\end{table*}
\begin{table*}[bt!]
\begin{center}
\tiny
\begin{tabular}{l@{\hspace{1mm}}llllllllllllllllllllllllllllllllllllllllllllllllll}
\hline\hline
$K_+$&$K_0$&\multicolumn{1}{c}{$a_{+,0}$}&\multicolumn{1}{c}{$a_{+,1}$}&\multicolumn{1}{c}{$a_{+,2}$}&\multicolumn{1}{c}{$a_{+,3}$}&\multicolumn{1}{c}{$a_{+,4}$}&\multicolumn{1}{c}{$a_{+,5}$}&\multicolumn{1}{c}{$a_{+,6}$}&\multicolumn{1}{c}{$a_{+,7}$}&\multicolumn{1}{c}{$a_{+,8}$}&\multicolumn{1}{c}{$a_{+,9}$}&\\
\hline
2&2&0.02805(80)&-0.0821(33)&- &- &- &- &- &- &- &-&\\
2&3&0.02659(99)&-0.0881(39)&- &- &- &- &- &- &- &-&\\
3&2&0.0250(10)&-0.0793(33)&0.083(16)&- &- &- &- &- &- &-&\\
3&3&0.0253(10)&-0.0733(50)&0.110(24)&- &- &- &- &- &- &-&\\
3&4&0.0252(11)&-0.0743(68)&0.105(32)&- &- &- &- &- &- &-&\\
4&3&0.0253(10)&-0.0740(58)&0.112(24)&0.028(89)&- &- &- &- &- &-&\\
4&4&0.0253(11)&-0.0738(66)&0.110(58)&0.02(20)&- &- &- &- &- &-&\\
5&5&0.0253(11)&-0.0738(74)&0.111(64)&0.02(19)&-0.04(68)&- &- &- &- &-&\\
6&6&0.0253(11)&-0.0739(74)&0.107(61)&0.01(19)&-0.01(63)&0.01(66)&- &- &- &-&\\
7&7&0.0253(10)&-0.0734(74)&0.113(64)&0.01(18)&-0.06(64)&0.05(72)&-0.07(69)&- &- &-&\\
8&8&0.0252(11)&-0.0732(78)&0.116(66)&0.01(19)&-0.09(65)&0.12(84)&-0.12(86)&0.10(72)&- &-&\\
9&9&0.0253(10)&-0.0727(75)&0.121(69)&0.01(19)&-0.12(69)&0.2(1.1)&-0.3(1.2)&0.2(1.1)&-0.10(78)&-&\\
10&10&0.0253(11)&-0.0720(85)&0.127(74)&0.00(20)&-0.20(75)&0.4(1.3)&-0.5(1.7)&0.5(1.8)&-0.3(1.4)&0.14(86)&\\
\hline\hline\\
\end{tabular}
\\
\begin{tabular}{l@{\hspace{1mm}}llllllllllllllllllllllllllllllllllllllllllllllllll}
\hline\hline
$K_+$&$K_0$&\multicolumn{1}{c}{$a_{0,0}$}&\multicolumn{1}{c}{$a_{0,1}$}&\multicolumn{1}{c}{$a_{0,2}$}&\multicolumn{1}{c}{$a_{0,3}$}&\multicolumn{1}{c}{$a_{0,4}$}&\multicolumn{1}{c}{$a_{0,5}$}&\multicolumn{1}{c}{$a_{0,6}$}&\multicolumn{1}{c}{$a_{0,7}$}&\multicolumn{1}{c}{$a_{0,8}$}&\multicolumn{1}{c}{$a_{0,9}$}&\\
\hline
2&2&0.0938(27)&-0.269(10)&- &- &- &- &- &- &- &-&\\
2&3&0.0927(28)&-0.289(13)&-0.097(38)&- &- &- &- &- &- &-&\\
3&2&0.0942(27)&-0.256(11)&- &- &- &- &- &- &- &-&\\
3&3&0.0955(28)&-0.235(17)&0.090(55)&- &- &- &- &- &- &-&\\
3&4&0.0954(29)&-0.235(18)&0.07(10)&-0.07(31)&- &- &- &- &- &-&\\
4&3&0.0956(29)&-0.234(18)&0.093(57)&- &- &- &- &- &- &-&\\
4&4&0.0955(29)&-0.234(21)&0.09(19)&-0.02(62)&- &- &- &- &- &-&\\
5&5&0.0956(28)&-0.234(22)&0.09(19)&-0.03(55)&-0.01(64)&- &- &- &- &-&\\
6&6&0.0955(28)&-0.234(22)&0.08(19)&-0.05(51)&0.00(66)&0.03(64)&- &- &- &-&\\
7&7&0.0956(28)&-0.233(22)&0.09(19)&-0.02(50)&-0.06(72)&0.09(79)&-0.07(69)&- &- &-&\\
8&8&0.0955(29)&-0.233(23)&0.10(21)&0.00(50)&-0.09(82)&0.2(1.0)&-0.16(98)&0.11(71)&- &-&\\
9&9&0.0956(29)&-0.231(23)&0.12(22)&0.02(49)&-0.18(98)&0.3(1.4)&-0.3(1.5)&0.2(1.2)&-0.12(79)&-&\\
10&10&0.0956(29)&-0.230(25)&0.13(23)&0.02(48)&-0.3(1.1)&0.5(1.8)&-0.5(2.2)&0.5(2.1)&-0.3(1.6)&0.14(88)&\\
\hline\hline\\
\end{tabular}

\end{center}
\caption{Results for the Bayesian-inference BGL fit to HPQCD 14 and  RBC/UKQCD 23. The tables show the results for BGL coefficients for different orders of the fit.}\label{tab:combined BGL Bayesian}
\end{table*}
Fig.~\ref{fig:exemplary BGL fit RBCUKQCD 23 HPQCD 14} shows the result of the combined
Bayesian-inference fit to the RBC/UKQCD 23 and the HPQCD 14 data.
A look at both tables clarifies that the fit-function is capable of describing
the joint data set for $(K_+,K_0)\ge(3,2)$ with an acceptable $p$-value but 
central values and errors for the higher-order coefficients still vary as
the values $(K_+,K_0)$ are further increased. While the higher-order coefficients 
fluctuate wildly due to the lack of unitarity constraint in the frequentist ansatz,
the results of the Bayesian-inference remain stable when increasing $(K_+,K_0)$.
The higher-order coefficients remain well controlled.
\subsubsection{Combined fits to lattice and sum-rule data}
\begin{figure}[bt!]
    \centering
    \includegraphics[width=14cm]{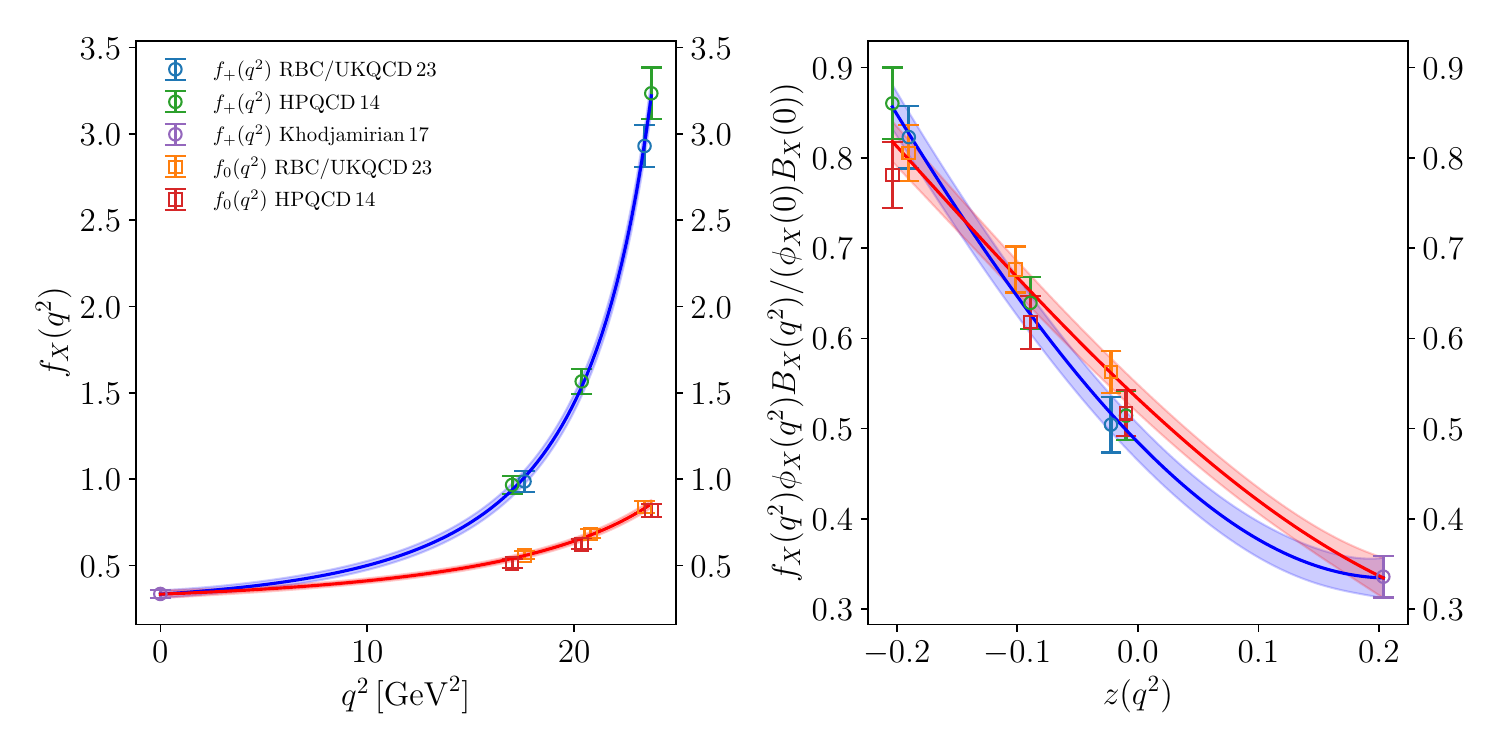}
    \caption{Illustration of the joint Bayesian-inference fit to the HPQCD-14~\cite{Bouchard:2014ypa}, RBC/UKQCD 23~\cite{Flynn:2023nhi} and Khodjamirian 17 data sets
    with $(K_+,K_0)=(5,5)$. Left: plot of the form
    factor vs. the squared momentum transfer; right: plot of the form factor
    after removing Blaschke and outer function, normalised such that the 
    kinematical constraint $f_0(0)=f_+(0)$ becomes apparent.}
    \label{fig:exemplary BGL fit RBCUKQCD 23 HPQCD 14 with sum rules}
\end{figure}

Repeating the fits of the previous section after including the 
sum-rule result Khodjamirian 17~\cite{Khodjamirian:2017fxg} leads 
to the results in Fig.~\ref{fig:exemplary BGL fit RBCUKQCD 23 HPQCD 14 with sum rules} (numerical results can be found in 
App.~\ref{sec:all lattice data with sum rules} in Tabs.~\ref{tab:combined BGL frequ with sum rule} and 
\ref{tab:combined BGL Bayesian with sum rules}).
While the frequentist fit achieves good $p$-values starting with $(K_+,K_0)=(3,3)$, the results of Bayesian inference converge towards
stable central values and errors starting with $(K_+,K_0)=(4,4)$.
Comparing Figs.~\ref{fig:exemplary BGL fit RBCUKQCD 23 HPQCD 14} 
and~\ref{fig:exemplary BGL fit RBCUKQCD 23 HPQCD 14 with sum rules},
highlights the importance that SM predictions at lower $q^2$ values can
have in stabilising the overall parameterisation of the form factor. This is then also reflected in the smaller error of
the respective BGL expansion coefficients listed in 
Tabs.~\ref{tab:combined BGL Bayesian} and~\ref{tab:combined BGL Bayesian with sum rules}.

\subsection{Comparison with dispersive-matrix method}\label{sec:Comparison to dispersive-matrix method}
\begin{figure}[bt!]
    \centering
    \includegraphics[width=14cm]{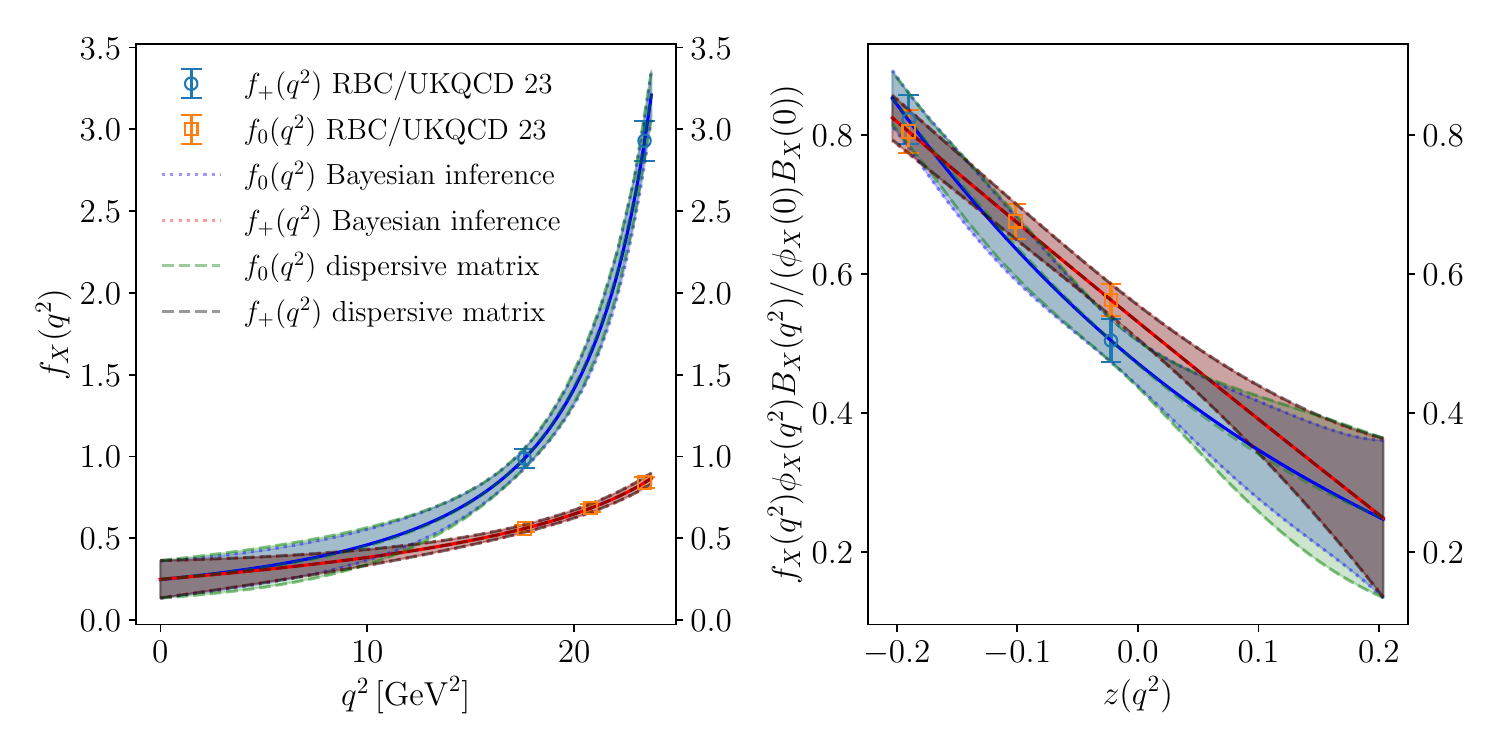}
    \caption{Comparison of Bayesian inference with the dispersive-matrix method for the form factors of exclusive semileptonic $B_s\to K\ell\nu$ decay.}
    \label{fig:BFF vs. DM comparison}
\end{figure}

In this section we compare our results to the dispersive-matrix method~\cite{Lellouch:1995yv}, 
which has recently received renewed attention in Ref.~\cite{DiCarlo:2021dzg},
and which has been applied to exclusive semileptonic $B_s\to K\ell\nu$ decay in Ref.~\cite{Martinelli:2022tte}.   
Fig.~\ref{fig:BFF vs. DM comparison} shows the comparison of both methods for the fit to the data set RBC/UKQCD 23.
The results for the dispersive-matrix method were obtained with our own implementation of the algorithm proposed in Ref.~\cite{DiCarlo:2021dzg}.
We find central values and error bands in excellent agreement.
While the dispersive-matrix computes a distribution of results for every value of the momentum transfer $q^2$, Bayesian inference
predicts the parameters of the BGL expansion and their correlations. Besides the conceptual simplicity of the Bayesian-inference 
fitting strategy, the results for Bayesian inference are hence more 
convenient for use in further processing, \emph{e.g.}~for making predictions for phenomenology as discussed in the next section.
\section{Phenomenological analysis}\label{sec:Phenomenological analysis}
Having parameterised the form factors $f_+(q^2)$ and $f_0(q^2)$ over the full
kinematically allowed phase space $0\le q^2\le q^2_{\rm max}$, various
phenomenologically relevant quantities can be computed. In the following we
provide determinations of the CKM matrix element $|V_{ub}|$, two versions of
the $R$-ratio (the \emph{traditional} one and an improved version which has been
advocated in Ref.~\cite{Flynn:2023nhi}) and the differential decay
rate. Additionally, results and discussion of the forward-backward and
polarisation asymmetries can be found in App.~\ref{app:asymmetries}.
Here we concentrate mainly on results for combined fits over 
 data sets. The results we would obtain from fits to 
individual data sets are summarised in tables in App.~\ref{sec:Bayesian individual data sets}.

\subsection{Determination of $|V_{ub}|$}
By combining experimental measurements of $d\Gamma(B_s\to K \ell\nu)/dq^2$ with
theoretical predictions for the form factors $f_0$ and $f_+$ the CKM matrix
element $|V_{ub}|$ can be determined using
Eq.~\eqref{eq:B_semileptonic_rate}. Currently, the only available measurements
have been performed by LHCb who provide the ratio of branching fractions
$R_{BF}$~\cite{LHCb:2020ist},
\begin{equation}
R_{\rm BF}=\frac{\mathcal{B}(B_s^0\to K^-\mu^+\nu_\mu)}{\mathcal{B}(B_s^0\to D_s^-\mu^+\nu_\mu)}\,.
\end{equation}
These values are given for two integrated $q^2$ bins, which we will refer to as `low' and `high',
\begin{align}
  \text{$q^2\leq 7\gev^2$:} \quad& R^{\rm low}_{BF}=1.66(80)(86)\times10^{-3}\nonumber\,,\\
  \text{$q^2\geq 7\gev^2$:} \quad& R^{\rm high}_{BF}=3.25(21)(^{+18}_{-19})\times10^{-3}\,.
\end{align}
Using the life time of the $B_s^0$ meson
$\tau_{B_s^0}=1.520(5)\,\mathrm{ps}$~\cite{HFLAV:2022pwe,ParticleDataGroup:2022pth} and the branching
ratio~\cite{LHCb:2020cyw}
\begin{equation}
  \mathcal{B}(B_s^0\to D_s^-\mu^+\nu_\mu)= 2.49(12)(21)\times 10^{-2}\,,
\end{equation}
this can be used to determine $|V_{ub}|$ from
\begin{equation}
	|V_{ub}|=\sqrt{\frac{R^{\rm bin}_{BF}\,\mathcal{B}(B_s^0\to D_s^-\mu^+\nu_\mu)}{\tau_{B_s^0}\,\Gamma^{\rm bin}_0(B_s\to K\ell\nu)}}\,,
\end{equation}
where we defined the reduced decay rate $\Gamma^{\rm bin}_0=\Gamma^{\rm
  bin}/|V_{ub}|^2$. Since we have obtained the BGL parameterisation of the form
factors, $\Gamma^{\rm bin}_0$ can be computed by numerically integrating the
right-hand side of Eq.~\eqref{eq:B_semileptonic_rate} over the appropriate $q^2$
bin. After symmetrising the errors on the input data, we generate multivariate
distributions for the aforementioned experimental inputs, assuming the
systematic uncertainties of the branching fractions $R_{BF}$ and the branching
ratio $\mathcal{B}$ to be 100\% correlated and all other uncertainties to be
uncorrelated (\emph{cf.}  Ref.~\cite{Martinelli:2022tte}).  The form factors
$f_0$ and $f_+$ are constructed from the samples for the BGL coefficients that we
have found from our algorithm. Combining these distributions provides a fully
correlated analysis framework to determine $|V_{ub}|$ from either bin as well as
from a weighted average. Numerical values of our results are presented in
Tab.~\ref{tab:BstoK_zfit_observables}.
\begin{table}
  \centering
  \tiny
  \begin{tabular}{l@{\hspace{1mm}}llllllllll}
\hline\hline
$K_+$&$K_0$&\multicolumn{1}{c}{$f(q^2=0)$}&\multicolumn{1}{c}{$R_{B_s\to K}^{\rm impr}$}&\multicolumn{1}{c}{$R_{B_s\to K}$}&\multicolumn{1}{c}{$\frac{\Gamma^\tau}{|V_{ub}|^2}\,[\frac 1{\rm ps}]$}&\multicolumn{1}{c}{$\frac{\Gamma^\mu}{|V_{ub}|^2}\,[\frac 1{\rm ps}]$}&\multicolumn{1}{c}{$V^{\rm low}_{\rm CKM}$}&\multicolumn{1}{c}{$V^{\rm high}_{\rm CKM}$}&\multicolumn{1}{c}{$V^{\rm full}_{\rm CKM}$}&\\
\hline
2&2&0.217(16)&1.544(15)&0.735(15)&4.94(31)&6.72(51)&0.00365(35)&0.00338(30)&0.00349(31)\\
2&3&0.166(25)&1.587(26)&0.809(37)&4.36(36)&5.41(65)&0.00449(64)&0.00365(36)&0.00386(41)\\
3&2&0.234(16)&1.684(38)&0.758(19)&4.41(31)&5.83(51)&0.00367(35)&0.00374(34)&0.00370(33)\\
3&3&0.286(36)&1.682(36)&0.700(37)&4.80(40)&6.89(87)&0.00319(41)&0.00359(34)&0.00343(35)\\
3&4&0.277(53)&1.688(42)&0.715(64)&4.73(46)&6.7(1.2)&0.00333(60)&0.00364(36)&0.00356(40)\\
4&3&0.288(37)&1.689(42)&0.701(39)&4.79(41)&6.87(90)&0.00319(42)&0.00362(35)&0.00344(35)\\
4&4&0.286(93)&1.687(41)&0.709(98)&4.80(54)&7.0(1.6)&0.00335(88)&0.00362(37)&0.00358(41)\\
5&5&0.286(87)&1.686(44)&0.709(94)&4.81(54)&7.0(1.6)&0.00332(86)&0.00360(36)&0.00356(41)\\
6&6&0.282(85)&1.686(45)&0.713(93)&4.78(53)&6.9(1.6)&0.00336(85)&0.00361(37)&0.00357(41)\\
7&7&0.288(85)&1.686(44)&0.706(90)&4.81(54)&7.0(1.6)&0.00330(80)&0.00361(36)&0.00355(40)\\
8&8&0.290(90)&1.686(44)&0.704(96)&4.82(57)&7.1(1.7)&0.00330(89)&0.00361(38)&0.00356(42)\\
9&9&0.297(90)&1.685(43)&0.697(95)&4.87(56)&7.2(1.7)&0.00324(87)&0.00359(37)&0.00353(42)\\
10&10&0.300(93)&1.685(44)&0.694(98)&4.89(59)&7.3(1.8)&0.00322(86)&0.00357(37)&0.00352(42)\\
\hline\hline\\
\end{tabular}

  \begin{tabular}{l@{\hspace{1mm}}llllllllll}
\hline\hline
$K_+$&$K_0$&\multicolumn{1}{c}{$I[\mathcal{A}_{\rm FB}^\tau]\,[\frac 1{\rm ps}]$}&\multicolumn{1}{c}{$I[\mathcal{A}_{\rm FB}^\mu]\,[\frac 1{\rm ps}]$}&\multicolumn{1}{c}{$\mathcal{\bar A}_{\rm FB}^\tau$}&\multicolumn{1}{c}{$\mathcal{\bar A}_{\rm FB}^\mu$}&\multicolumn{1}{c}{$I[\mathcal{A}_{\rm pol}^\tau]\,[\frac 1{\rm ps}]$}&\multicolumn{1}{c}{$I[\mathcal{A}_{\rm pol}^\mu]\,[\frac 1{\rm ps}]$}&\multicolumn{1}{c}{$\mathcal{\bar A}_{\rm pol}^\tau$}&\multicolumn{1}{c}{$\mathcal{\bar A}_{\rm pol}^\mu$}&\\
\hline
2&2&1.345(87)&0.0302(34)&0.2724(18)&0.00448(21)&0.732(72)&6.64(50)&0.148(12)&0.98749(57)\\
2&3&1.18(10)&0.0212(42)&0.2715(18)&0.00388(33)&0.53(10)&5.35(64)&0.121(17)&0.98887(80)\\
3&2&1.243(88)&0.0322(36)&0.2817(22)&0.00551(31)&0.23(11)&5.74(51)&0.052(24)&0.98422(93)\\
3&3&1.37(12)&0.0439(88)&0.2855(32)&0.00632(59)&0.24(12)&6.76(85)&0.050(23)&0.9821(16)\\
3&4&1.35(14)&0.042(12)&0.2851(37)&0.00618(82)&0.23(13)&6.6(1.1)&0.047(25)&0.9825(21)\\
4&3&1.37(12)&0.0443(91)&0.2858(33)&0.00640(63)&0.22(13)&6.75(88)&0.046(26)&0.9819(17)\\
4&4&1.37(17)&0.046(21)&0.2856(58)&0.0063(16)&0.23(13)&6.8(1.6)&0.047(26)&0.9821(41)\\
5&5&1.38(17)&0.046(20)&0.2855(56)&0.0063(15)&0.23(14)&6.9(1.5)&0.048(27)&0.9822(39)\\
6&6&1.37(17)&0.045(19)&0.2852(56)&0.0062(15)&0.23(14)&6.8(1.5)&0.048(28)&0.9823(38)\\
7&7&1.38(17)&0.046(20)&0.2856(55)&0.0063(14)&0.23(14)&6.9(1.6)&0.048(27)&0.9821(37)\\
8&8&1.38(19)&0.047(21)&0.2858(59)&0.0064(15)&0.23(14)&6.9(1.6)&0.048(27)&0.9820(39)\\
9&9&1.40(18)&0.049(21)&0.2862(59)&0.0065(15)&0.23(13)&7.1(1.6)&0.047(27)&0.9818(39)\\
10&10&1.40(19)&0.050(23)&0.2864(61)&0.0065(15)&0.23(14)&7.2(1.8)&0.048(27)&0.9817(40)\\
\hline\hline\\
\end{tabular}

  \caption{Summary of results based on combined fit to HPQCD 14~\cite{Bouchard:2014ypa} and RBC/UKQCD 23~\cite{Flynn:2023nhi}. Definitions for the asymmetries $\mathcal{A}$ can be found in App.~\ref{app:asymmetries}.}\label{tab:BstoK_zfit_observables}
\end{table}
For the combined  fit to HPQCD~14 and RBC/UKQCD~23 we find the results to be
stable for $(K_+,K_0)\ge(5,5)$ and we choose this truncation for our main result
\begin{align}\label{eq:ourVub1}
  |V_{ub}|=&3.56(41)\times10^{-3}\,\,\textrm{\cite{Bouchard:2014ypa,Flynn:2023nhi}}.
\end{align}
As we will see shortly, also other observables that we computed have stable
central values and errors when further increasing the truncation.  We make the
same choice $(K_+,K_0)=(5,5)$ for the combined fit to lattice and sum-rule data
HPQCD~14 and RBC/UKQCD~23 and Khodjamirian~17,
\begin{align}\label{eq:ourVub2}
  |V_{ub}|=&3.13(28)\times10^{-3}\,\,\textrm{\cite{Bouchard:2014ypa,Flynn:2023nhi,Khodjamirian:2017fxg}}.
\end{align}
In both cases, the error on $|V_{ub}|$ is currently dominated by the experimental
uncertainty (we ran the fit again assuming vanishing experimental uncertainties and obtained
$|V_{ub}|=3.67(17)\times 10^{-3}$ and $|V_{ub}|=3.23(14)\times 10^{-3}$,
respectively).
 We note that while the results for $|V_{ub}|$ obtained for the
  `low' and `high' bins agree for the analysis with HPQCD~14 and RBC/UKQCD~23
  (\emph{cf.} Tab.~\ref{tab:BstoK_zfit_observables}), they are at tension in the
  analysis that also includes the sum-rule result Khodjamirian 17
  (\emph{cf.} Tab.~\ref{tab:BstoK_zfit_observables with sum rules}), where
  $|V_{ub}^{\rm low}|=2.84(27)$ and $|V_{ub}^{\rm high}|=3.54(33)$.
For comparison we quote the world averages for exclusive and inclusive determinations of $|V_{ub}|$
\begin{align}
  |V_{ub}|_{\rm exclusive}^{\rm FLAG\, 21}\times 10^{-3}=&\,3.74(17)\,\textrm{
  \cite{Lattice:2015tia,Flynn:2015mha,delAmoSanchez:2010af,Lees:2012vv,Ha:2010rf,Sibidanov:2013rkk,FlavourLatticeAveragingGroupFLAG:2021npn}}\label{eq:Vubexclusive}\,,\\
  |V_{ub}|_{\rm inclusive}^{\rm }\times 10^{-3}	=&\,4.13(26) \,\textrm{\cite{ParticleDataGroup:2022pth, HFLAV:2022pwe, Gambino:2007rp, Lange:2005yw, Andersen:2005mj}}\,\label{eq:Vubinclusive}\,.
\end{align}

\subsection{Differential decay width}
\begin{figure*}
  \includegraphics[width=7.5cm]{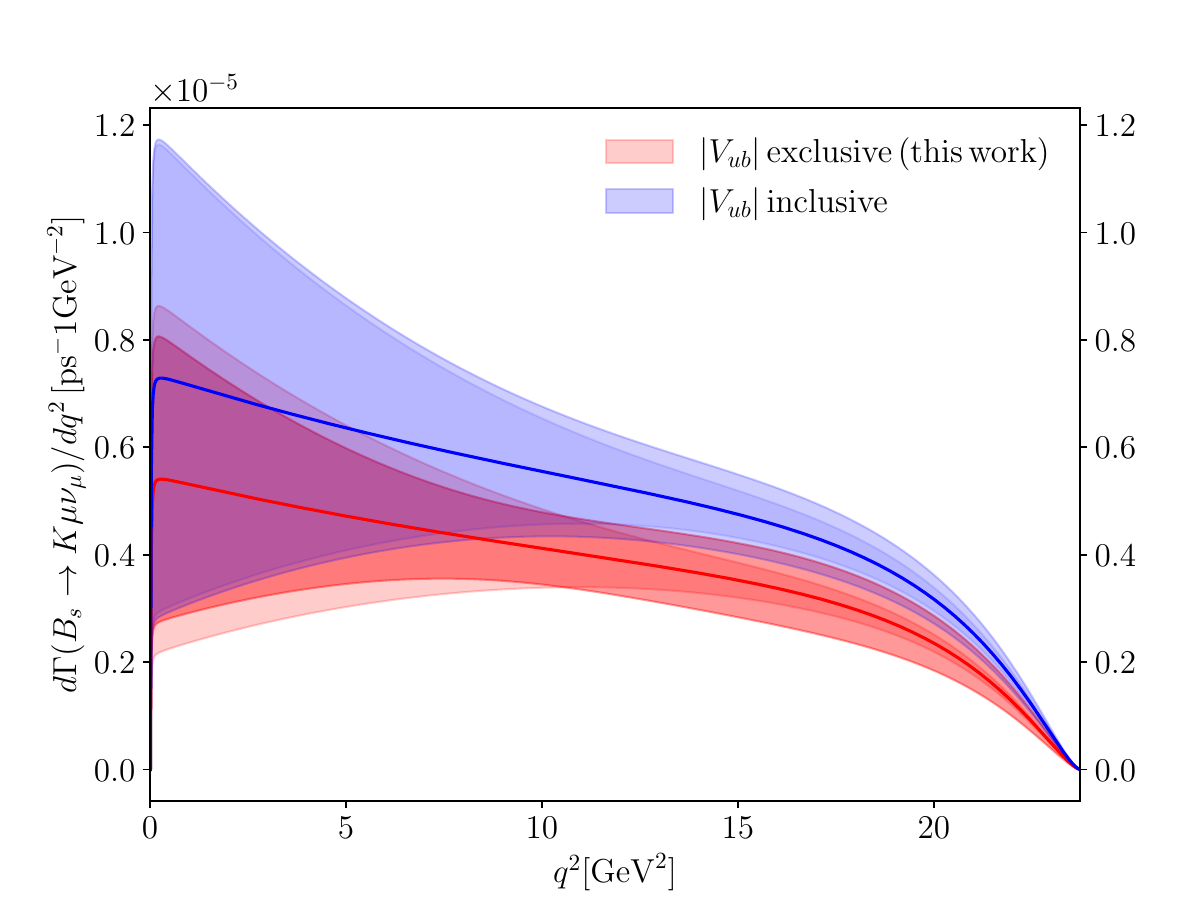}
  \includegraphics[width=7.5cm]{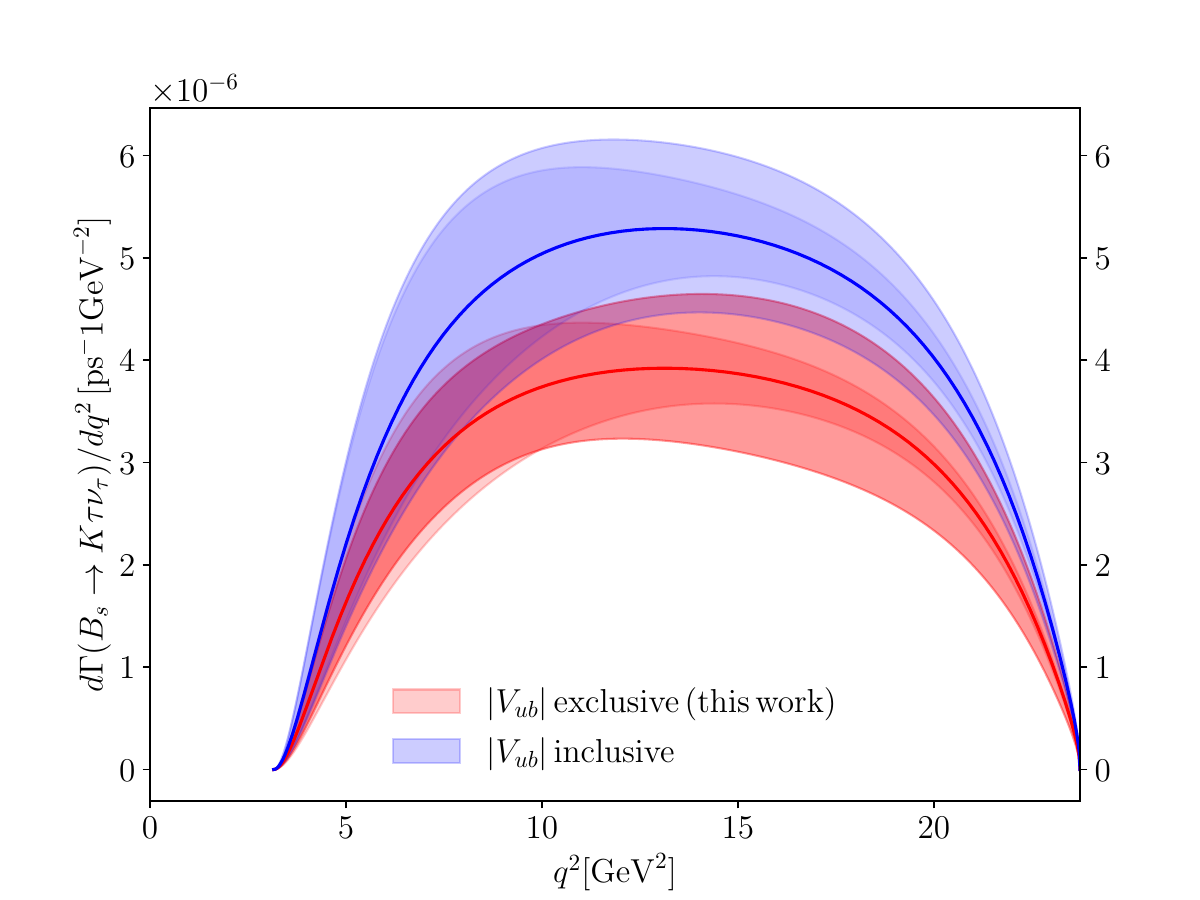}\\
  \includegraphics[width=7.5cm]{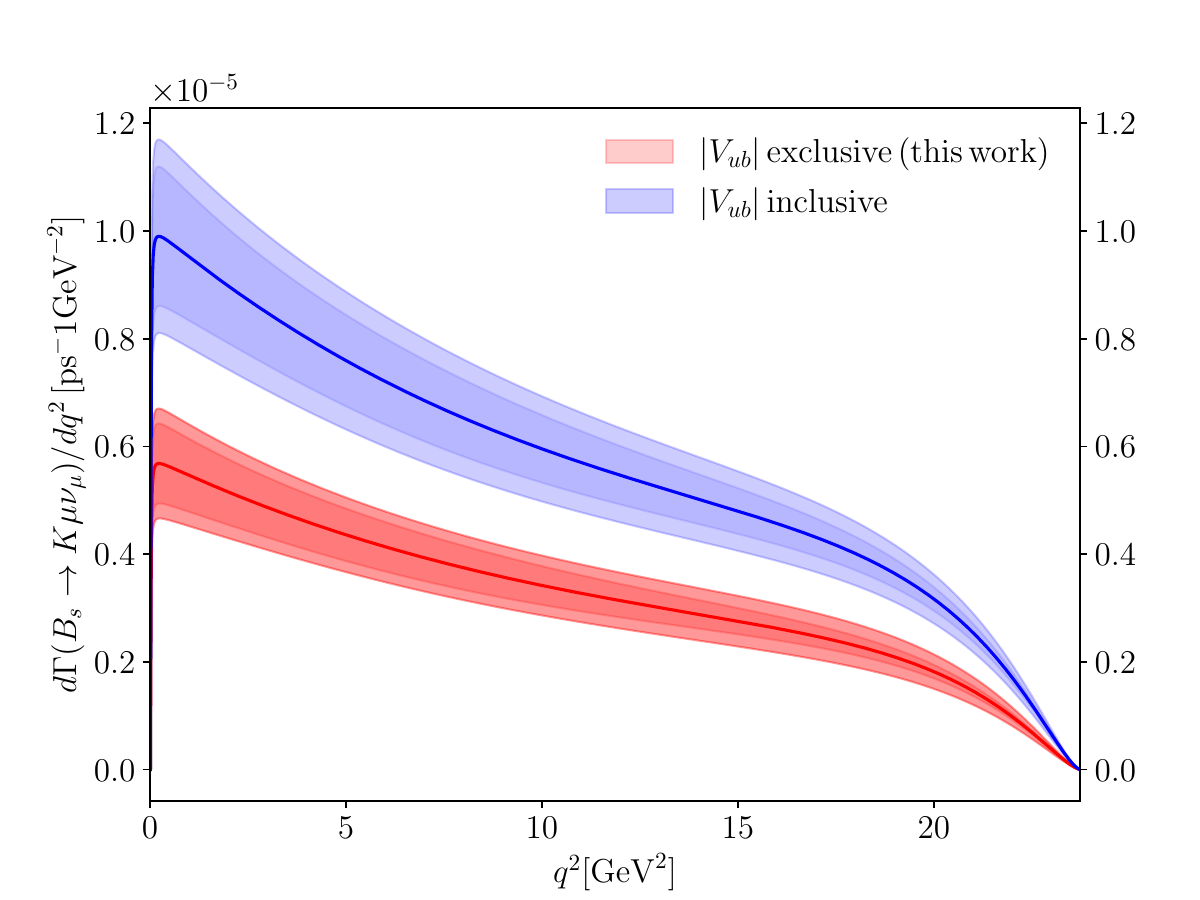}
  \includegraphics[width=7.5cm]{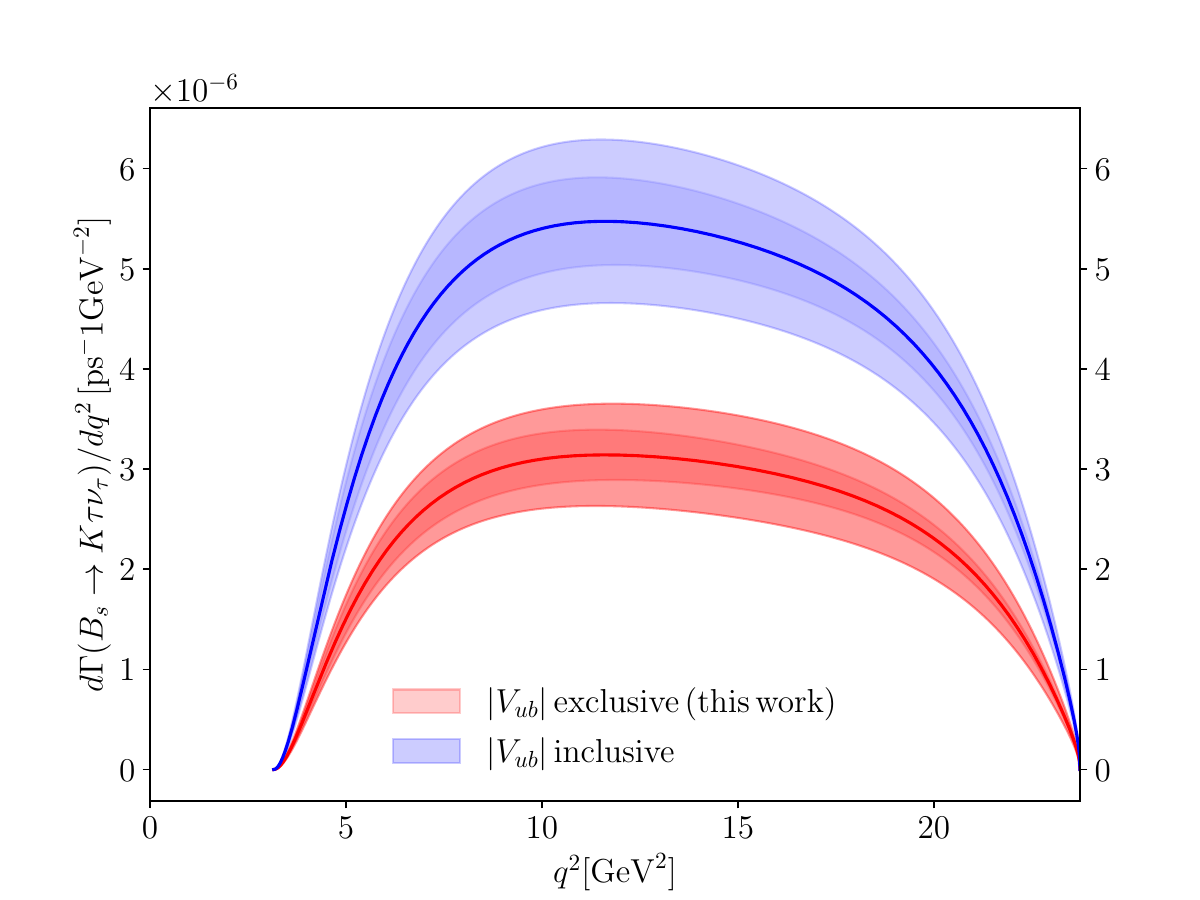}
  \caption{The differential decay width $d\Gamma/d q^2$ for $B_s\to K \mu
    \nu_\mu$ (left) and $B_s\to K \tau \nu_\tau$ (right).  The top row
    shows results from the fit to HPQCD 14 and RBQ/UKQCD 23, while the lower
    panel shows the result where the fit also includes the sum-rule result
    Khodjamirian 17.  The values for $|V_{ub}|$ are taken from
    Eq.~\eqref{eq:Vubinclusive}, \eqref{eq:ourVub1} (top) and
    \eqref{eq:ourVub2} (bottom).  The darker (lighter) shading indicates the
    error without (with) the contribution from the error on $|V_{ub}|$.}
  \label{fig:dGamma/dqsq}
\end{figure*}

From the analysis of lattice and sum-rule data we can make SM predictions for
the shape of the differential decay width $d\Gamma/dq^2$. In
Fig.~\ref{fig:dGamma/dqsq} we illustrate this, assuming our result for
$|V_{ub}|$ from the lattice and lattice+sum rules analyses in
Eq.~\eqref{eq:ourVub1} and \eqref{eq:ourVub2}, and the results from the
inclusive decay analysis in Eqs.~\eqref{eq:Vubinclusive}. The predicted shapes
of the inclusive and exclusive differential decay rates are visibly
different. In particular, after including the sum-rule result, the shapes can be
clearly and statistically significantly distinguished. Such detailed studies of
decay-rate shapes can shed light on the
tension between inclusive and exclusive CKM determinations.

\subsection{$R$ ratios}
Lepton flavour universality (LFU), \emph{i.e.} the identical coupling of leptons
to gauge bosons, is an accidental symmetry of the SM. Testing LFU therefore
provides crucial tests of the SM. One way to perform such tests is by comparing
semileptonic decays with different leptons in the final state. Due to their
different masses, the shapes of the differential decay rates and (partial)
integrals thereof will differ. Of particular interest are ratios which are
independent of the relevant CKM matrix elements (in our case $|V_{ub}|$) since
this eliminates sources of uncertainty. One such observable is the traditional
$R$-ratio, defined by
\begin{equation}
  R_{B_s\to K}= \frac{\int_{m_\tau^2}^{q^2_\text{max}}dq^2\,
    \frac{d\Gamma(B_s\to K\tau\nu_\tau)}{dq^2}}
  {\int_{m_\ell^2}^{q^2_\text{max}}dq^2\,
    \frac{d\Gamma(B_s \to K\ell\nu_\ell)}{dq^2}}\,.
  \label{Eq.RBstoK}
\end{equation}
Here $\ell$ denotes the $e$ or $\mu$, whereas the numerator only contains the
tau lepton. Since $m_e/M_{B_s} \ll m_\mu/M_{B_s} \ll 1$ the contribution
stemming from $f_0$ is negligible in the denominator
(\emph{cf.}~Eq.~\eqref{eq:B_semileptonic_rate}). One immediate consequence of
this is, that the decay into $e$ or $\mu$ does not provide experimental
information on $f_0$, so that this is only accessible via non-perturbative
methods~\cite{ElKhadra:1989iu}.

Ref.~\cite{Flynn:2023nhi} (motivated by Ref.~\cite{Isidori:2020eyd}) advocates
an improved definition of a ratio $R^\mathrm{imp}_{B_s\to K}$ as a more precise
test of LFU. This ratio improves over the traditional $R$-ratio by adjusting the
integration range to be the same in numerator and
denominator~\cite{Isidori:2020eyd, Freytsis:2015qca, Bernlochner:2016bci} and by
constructing it in a way, that form factors in the numerator and denominator
appear with the same weights~\cite{Isidori:2020eyd}.  To do this, they rewrite
the differential decay rate in equation~\eqref{eq:B_semileptonic_rate} with
lepton $\ell$ in the final state in the form
\begin{equation}
    \frac{d\Gamma(B_s\to K\ell\nu)}{dq^2} =
  \Phi\, \omega_\ell(q^2) \big[ F_V^2 + (F_S^\ell)^2\big],
\end{equation}
where 
\begin{align}
  \Phi &= \eta_\text{EW} \frac{G_F^2 |V_{ub}|^2}{24\pi^3}\,,\\
  \omega_\ell(q^2) &= \bigg(1-\frac{m_\ell^2}{q^2}\bigg)^2
               \bigg(1+\frac{m_\ell^2}{2q^2}\bigg)\,,\\
  F_V^2 &= |{\bf p}_K|^3 |f_+(q^2)|^2\,,\\
  \label{eq:FSlcontrib}
  (F_S^\ell)^2 &= \frac{3}{4} \frac{m_\ell^2|{\bf p}_K|}{m_\ell^2+2q^2}
              \frac{(M^2-m^2)^2}{M^2}\, |f_0(q^2)|^2\,.
\end{align}
The notation $\omega_\ell$ and $(F_S^\ell)^2$ is chosen to explicitly indicate
where the lepton mass $m_\ell$ enters. With this, the improved $R$-ratio can now
be defined as
\begin{equation}
  R^\text{imp}_{B_s\to K} =
  \frac{\int_{\qsqmin}^{\qsqmax} dq^2\,
   \frac{d\Gamma(B_s\to K\tau\bar\nu_\tau)}{dq^2}}
  {\int_{\qsqmin}^{\qsqmax} dq^2\,
     \left[\frac{\omega_\tau(q^2)}{\omega_\ell(q^2)}\right]\,
   \frac{d\Gamma(B_s\to K\ell\bar\nu_\ell)}{dq^2}},
  \label{Eq.RKtopi_UN}
\end{equation}
where again $\ell = e, \mu$. This matches the analogous definition for a vector
final state in Ref.~\cite{Isidori:2020eyd} and can be computed for
experimentally measured decay rates. Ref.~\cite{Flynn:2023nhi} proposes this ratio as an improved way to monitor LFU. In the SM (dropping the scalar form factor
for $\ell = e,\mu$) this can be approximated as
\begin{equation}
  R^\text{imp,SM}_{B_s\to K} \approx 1 + \frac{\int_{\qsqmin}^{\qsqmax} dq^2\,
    \omega_\tau(q^2) (F_S^\tau)^2}
  {\int_{\qsqmin}^{\qsqmax} dq^2\,
    \omega_\tau(q^2) F_V^2}\,.
\end{equation}
Table~\ref{tab:BstoK_zfit_observables} lists the values for $R_{B_s\to K}$
and $R_{B_s\to K}^{\rm impr.}$ and several other quantities of phenomenological
interest. As above, only results where the coefficients of the $z$ expansion
have stabilised should be considered in order to be free of truncation errors in
the $z$ expansion. {We note, that the relative uncertainty of the
  improved $R$-ratio is substantially smaller than for the traditional
  one}. For convenience, we provide numerical values of our preferred order for
the Bayesian inference with $(K_+,K_0)=(5,5)$ based on HPQCD 14 and RBC/UKQCD 23:
\begin{align}
R_{B_s\to K}=&\,0.709(94)\,,\\
R_{B_s\to K}^{\rm impr,SM}=&\,1.686(44)\,.
\end{align}

\subsection{Further phenomenological results}
We can also compute forward-backward and polarisation asymmetries. Details are discussed in App.~\ref{app:asymmetries} and 
Tab.~\ref{tab:executive summary} summarises all central fit results.
\begin{table}
    \centering
    \begin{tabular}{l|ll}
    \hline\hline
    \multicolumn{1}{c|}{observable}  &\multicolumn{1}{c}{lattice}&\multicolumn{1}{c}{lattice+sum rules}\\
    \hline
    $|V_{ub}|$&0.00356(41)&0.00313(28)\\
$f_+(0)$&0.286(87)&0.335(22)\\
$R_{B_s\to K}$&0.709(94)&0.653(26)\\
$R_{B_s\to K}^{\rm impr}$&1.686(44)&1.688(44)\\
$\mathcal{\bar A}_{\rm FB}^\mu$&0.0063(15)&0.00718(49)\\
$\mathcal{\bar A}_{\rm FB}^\tau$&0.2855(56)&0.2885(23)\\
$I[\mathcal{A}_{\rm FB}^\mu]/{\rm ps}$&0.046(20)&0.0552(59)\\
$I[\mathcal{A}_{\rm FB}^\tau]/{\rm ps}$&1.38(17)&1.446(97)\\
$\mathcal{\bar A}_{\rm pol}^\mu$&0.9822(39)&0.9799(14)\\
$\mathcal{\bar A}_{\rm pol}^\tau$&0.048(27)&0.043(27)\\
$I[\mathcal{A}_{\rm pol}^\mu]/{\rm ps}$&6.9(1.5)&7.54(66)\\
$I[\mathcal{A}_{\rm pol}^\tau]/{\rm ps}$&0.23(14)&0.22(14)\\
$\Gamma^\mu/ |V_{ub}|^2/{\rm ps}$&7.0(1.6)&7.69(67)\\
$\Gamma^\tau/|V_{ub}|^2/{\rm ps}$&4.81(54)&5.01(34)\\
\hline\hline
%
    \end{tabular}
    \caption{Summary of main results, where `lattice` refers to the combined fit over HPQCD 14 and RBC/UKQCD 23, and where `lattice+sum rules` refers to the fit with the same lattice results plus
    the sum-rule result Khodjamirian 17 for $f_+(0)$.}
    \label{tab:executive summary}
\end{table}

\section{Conclusions and outlook}
The  main results of this paper are:
\begin{itemize}
    \item We have generalised the BGL~\cite{Boyd:1994tt} unitarity constraint towards
    exclusive semileptonic processes for which the flavour-structure of the weak current allows for a
particle-production threshold that lies below the pair-production threshold of 
the asymptotic-state pair of the process. For instance,
for the semileptonic process $B_s\to K\ell\nu$ the $t$-channel  $B\pi$ threshold lies below the $B_s K$ threshold. The 
modified unitarity constraint is restricted to contributions from above
the $B_s K$ threshold. This problem has recently also been addressed in~\cite{Berns:2018vpl,Gubernari:2020eft,Gubernari:2022hxn,Blake:2022vfl}.
While fundamentally equivalent, we find the solution proposed here more elegant and also simpler to implement.
A simple modification of existing fit-codes  allows the 
modified unitarity constraint presented in Eq.~\eqref{eq:modified unitarity constraint} to be imposed.
{\color{black}  The thus-modified unitarity constraint could in the presence of truncation be strengthened accidentally (this also applies to the work of \cite{Berns:2018vpl,Gubernari:2020eft,Gubernari:2022hxn,Blake:2022vfl}). We propose an alternative BGL expansion, which can be used to check whether this is the case.  We also discuss a way to correct the asymptotic behaviour of the BGL expansion of the vector form factor~\cite{Buck:1998kp,Becher:2005bg} and show in each case that the correction, as anticipated in~\cite{Becher:2005bg}, does at the current level of statistics not
impact the form factor in the semileptonic region in a measurable way.}
\item The second central result of this paper is a novel method that allows,
using Bayesian inference,
for \textcolor{black}{model-independent parameterisations of hadronic form factors with controlled truncation errors}.
This is achieved by using quantum-field-theoretical unitarity and analyticity as
regulators, to keep the less or unconstrained higher-order coefficients
in an untruncated BGL expansion under control. We show how kinematical constraints
like $f_+(0)=f_0(0)$ for the vector and scalar form factors at zero 
momentum transfer in pseudo-scalar to pseudo-scalar meson decay,
can be taken into account exactly. The new unitarity constraint of
Eq.~\eqref{eq:modified unitarity constraint}, which within 
Bayesian inference corresponds to a flat prior, is taken into account
in a fully consistent way, leading to meaningful central values 
and errors in the computation of observables based on 
the form-factor parameterisations.
The approach presented here is similar in spirit to the recently revived idea
of the dispersive-matrix method~\cite{DiCarlo:2021dzg,Martinelli:2021frl}. In fact, our results agree very well with
the ones determined in~\cite{Martinelli:2022tte}. The method proposed here
is however conceptually simpler, and besides the exact implementation of 
constraints like $f_+(0)=f_0(0)$, allows for straight-forwardly 
combining different, potentially correlated data sets into a global
fit.
We demonstrate how this works in practice by presenting fits to lattice, sum-rule and experimental data,
and make a range of predictions with relevance for phenomenology. 

We recommend to use the complementary information gained from Bayesian-inference based fits and
frequentist fits to asses how well model and data are compatible, and to obtain 
parameterisations of form factors that are free of any bias originating
from truncations.
\end{itemize}

Looking ahead, we plan to extend our work to other decay channels, for which 
lattice and potentially also experimental data is available (\emph{e.g.}~$B\to\pi \ell\nu$, $B_{(s)}\to D_{(s)}\ell\nu$, 
$B_{(s)}\to D^\ast_{(s)}\ell\nu$, $\Lambda_b\to (p,\Lambda_c^{(\ast)})\ell\nu, \dots$, in order to make truncation-independent predictions for a wider set of SM parameters and observables.

\acknowledgments We thank our RBC/UKQCD collaborators, and Greg Ciezarek, Gilberto Colangelo, Luigi Del Debbio, Danny van Dyk, Sasha Zhiboedov and M\'eril Reboud for fruitful discussions. This project has received funding from Marie
Sklodowska-Curie grant 894103 (EU Horizon 2020).

\appendix

\section{The BGL parameterisation and unitarity}\label{app:The BGL parameterisation}

The discussion in this section reviews work by Boyd, Grinstein and
Lebed~\cite{Boyd:1994tt,Boyd:1995sq,Boyd:1997kz}, used also by  Arnesen
et al.~\cite{Arnesen:2005ez}. For convenience we first recall the $z$ transformation from Eq.~\eqref{eq:z-fn-defn}, but written using $t=q^2$,
\begin{equation}
  z(t;t_*,t_0) = \frac{\sqrt{t_*-t} - \sqrt{t_*-t_0}}%
                      {\sqrt{t_*-t} + \sqrt{t_*-t_0}}\,.
\end{equation}
As noted in Sec.~\ref{Sec:BGL modified}, $t_*$ denotes the start of the cut in the $t$-channel, which for the decay $B_s\to K\ell\nu$ is $t_*=(M_B+M_\pi)^2$. As before we set $t_\pm =
(M_{B_s}\pm M_K)^2$, with $t_- = q^2_\text{max}$ the upper end of the kinematical
range for physical semileptonic decay. We choose $t_0$ to symmetrise the range of $z$ corresponding to $0\leq t\leq q^2_\text{max}$.

Continuing the discussion started in Sec.~\ref{Sec:BGL modified}, the idea is that the product $B_X\phi_X f_X$ is analytic inside the unit circle in $z$ and hence has a power series expansion in $z$. When there is a single sub-threshold pole, the Blaschke factor is given by:
\begin{equation}
  \label{eq:blaschke}
  B_X(q^2) = \frac{z(q^2;t_*,t_0)-z(\mpole^2;t_*,t_0)}
                  {1-z(q^2;t_*,t_0)z(\mpole^2;t_*,t_0)}
                  = z(q^2;t_*,\mpole^2).
\end{equation}
Here $\mpole$ is the mass of a pole sitting between $t_-=q^2_\mathrm{max}$ and $t_*$. If there is no such pole, then we set $B_X(q^2) = 1$. If there are $n$ sub-threshold poles at positions $z_i$ with masses $m_i$, then the Blaschke factor is the product
\begin{equation}
  \label{eq:Blaschke}
  B(q^2)=\prod_{i=0}^{n-1} \frac{z-z_i}{1-z_i z}=\prod_{i=0}^{n-1} z(q^2;t_*,m_i^2).
\end{equation}
It has the property that $|B(z)|=1$ for $z$ on the unit circle, a fact used in deriving the analyticity/unitarity bounds.

For $f_+$, the $1^-$ $B^*$ vector-meson mass lies above $q^2_{\rm max}$ and below the $B\pi$ threshold at $t_*$ and we include 
this single pole in the expression~\eqref{eq:Blaschke} for the Blaschke factor. In the $0^+$ channel, the theoretically predicted mass $M_{B^\ast(0^+)}=5.63 \gev$~\cite{Bardeen:2003kt} sits above the $B\pi$ threshold. For $f_0$ we therefore do not need to include a pole mass, and set $B=1$ in this case. 

The outer functions are given by
\begin{align}
  \label{eq:phiplus}
  \phi_+(q^2,t_0) &= \sqrt{\frac{\eta_I}{48\pi\chi_{1^-}(0)}}\,
                    \frac{r_q^{1/2}}{r_0^{1/2}}\,(r_q+r_0)
                    \big(r_q+\sqrt{t_*}\,\big)^{-5}
                    (t_+-q^2)^{3/4}
                    (r_q+r_-)^{3/2}\,, \\
  \label{eq:phizero} 
  \phi_0(q^2,t_0) &= \sqrt{\frac{\eta_It_+t_-}{16\pi\chi_{0^+}(0)}}\,
                    \frac{r_q^{1/2}}{r_0^{1/2}}\,(r_q+r_0)
                    \big(r_q+\sqrt{t_*}\,\big)^{-4}
                    (t_+-q^2)^{1/4}
                    (r_q+r_-)^{1/2}\,,
\end{align}
where we have set $r_q=\sqrt{t_*-q^2}$, $r_-=\sqrt{t_*-t_-}$ and
$r_0=\sqrt{t_*-t_0}$.

Let us now discuss our choice for $\chi_{1^-}(0)$ and $\chi_{0^+}(0)$. We first recall some steps in the derivation of the unitarity bounds~\cite{Boyd:1994tt,Boyd:1995sq,Boyd:1997kz}:
\begin{enumerate}
  \item Compute the vacuum polarisation function of two currents
    $J_\mu = \bar u \gamma_\mu b$,
    \begin{equation}
      \label{eq:vacpol}
      \Pi_{\mu\nu}(q) = i \int d^4x\, e^{iq\cdot x}
      \langle0| \mathrm{T} J_\mu(x)J^\dagger_\nu(0) |0\rangle
      = (q_\mu q_\nu-q^2 g_{\mu\nu})\Pi_\mathrm{T}(q^2) + q_\mu q_\nu
      \Pi_\mathrm{L}(q^2)\,.
    \end{equation}
  \item The $\Pi_\mathrm{T,L}$ defined in Eq.~\eqref{eq:vacpol}
    satisfy once- or twice-subtracted dispersion relations
    \begin{align}
      \label{eq:chi-defn_T}
      \chi_\mathrm{T}(Q^2) &=
      \frac12\left.\frac{\partial^2\big(q^2 \Pi_\mathrm{T}(q^2)\big)}{\partial
        (q^2)^2}\right|_{q^2=-Q^2}
      = \frac1\pi \int_0^\infty dt\,
      \frac{t\, \mathrm{Im}\,\Pi_\mathrm{T}(t)}{(t+Q^2)^3},\\
      \label{eq:chi-defn_L}
      \chi_\mathrm{L}(Q^2) &=
      \left.\frac{\partial\big(q^2\Pi_\mathrm{L}(q^2)\big)}{\partial
        q^2}\right|_{q^2=-Q^2}
      = \frac1\pi \int_0^\infty dt\,
      \frac{t\, \mathrm{Im}\,\Pi_\mathrm{L}(t)}{(t+Q^2)^2}\,,
    \end{align}
    where we follow the notation of Refs.~\cite{Lellouch:1995yv,DiCarlo:2021dzg}.
   \item The absorptive parts $\mathrm{Im}\,\Pi_\mathrm{T,L}(t)$ are
     found by inserting real intermediate states between the two
     currents in Eq.~\eqref{eq:vacpol}. For a judicious choice of $\mu$
     and $\nu$ this is a sum of positive definite terms. One can then
     obtain inequalities (bounds) by concentrating on intermediate 
     $\bar B_sK$ pairs. By analyticity
     and crossing symmetry, this constrains the shape in $t=q^2$ of
     the form factors in the physical region $0 \leq t \leq t_-$.
   \item The $\chi$'s come from evaluating the current-current correlator and
     depend on the ratio $u = m_u/m_b$. $\chi(0)$ corresponds to the lowest moment of
     $\Pi(t)$ computed with an OPE up to some number of loops and with
     condensate contributions. Detailed expressions with the
     perturbative parts to two loops are given in Ref.~\cite{Boyd:1997kz};
     three-loop perturbative contributions were calculated by Grigo et
     al. (GHMS) in Ref.~\cite{Grigo:2012ji}. 

     For the decay of interest in this paper, $B_s\to K\ell\nu$, we can approximate the ratio $u$ by zero. Using two-loop perturbative expressions from
     BGL~\cite{Boyd:1997kz}, with $m_b = m_b^\text{pole}$, $\chi_\text{T,L}$ are given by
     \begin{align}
       \chi_\text{T}(0)_{u=0}=
       \chi_{1^-}(0) &= \frac{3[1+1.140\alpha_s(m_b)]}{32\pi^2m_b^2}
       - \frac{\bar m_b \langle\bar uu\rangle}{m_b^6}
       - \frac{\langle\alpha_s G^2\rangle}{12\pi m_b^6}, \\
       \chi_\text{L}(0)_{u=0}=
       \chi_{0^+}(0) &= \frac{[1+0.751\alpha_s(m_b)]}{8\pi^2}
       + \frac{\bar m_b \langle\bar uu\rangle}{m_b^4}
       + \frac{\langle\alpha_s G^2\rangle}{12\pi m_b^4}\,.
     \end{align}
     The expressions in Grigo et al.~\cite{Grigo:2012ji} use the
     $\overline{\text{MS}}$ $b$ mass evaluated at its own scale, $\bar m_b(\bar m_b)$, instead of
     $m_b^\text{pole}$. Applying the relation
     \begin{equation}
       m_b^\text{pole} = \bar m_b
       \left(1+\frac43 \frac{\alpha_s(\bar m_b)}\pi\right)
       + O(\alpha_s^2)
     \end{equation}
     shows
     agreement of the perturbative terms above with the terms up to two loops in~\cite{Grigo:2012ji}. We use the 3-loop results for our numerical values for $\chi_{1^-,0^+}$
with $m_b=\bar m_b(\bar m_b)$, taking $\bar m_b(\bar
m_b) = 4.163\gev$ and $\alpha_s^{(5)}(\bar m_b) = 0.2268$
from Ref.~\cite{Chetyrkin:2009fv}. In the quark-condensate term, $\bar m_b$ and $\langle\bar
     uu\rangle$ should both be evaluated in the same scheme with the
     same scale, for example $\overline{\text{MS}}$ at scale
     $\mu=1\gev$ or $2\gev$. We ran the mass to $\bar m_b(2\gev) =
4.95\gev$ using the \texttt{RunDec}
package~\cite{Chetyrkin:2000yt,Schmidt:2012az,Herren:2017osy} and
combined it with $\langle\bar u u\rangle = -(274\mev)^3$, using a
weighted mean of $2+1+1$ and $2+1$ flavour estimates for
$\Sigma^{1/3}$ in SU(2) in the 2021 FLAG
review~\cite{FlavourLatticeAveragingGroupFLAG:2021npn,Bazavov:2010yq,Cichy:2013gja,Alexandrou:2017bzk,Borsanyi:2012zv,Durr:2013goa,Boyle:2015exm,Cossu:2016eqs,Aoki:2017paw}).
We took $\langle\alpha_s G^2\rangle = 0.0635(35)\gev^4$ from a
sum rules average~\cite{Narison:2018dcr}. The condensate terms are
small compared to the perturbative parts. We obtain:
\begin{equation}
\label{eq:our-chis}
\begin{aligned}
\chi_{1^-}(0)&=6.03\times 10^{-4}\,{\rm GeV}^{-2}\,,\\
\chi_{0^+}(0)&=1.48\times 10^{-2}\,.
\end{aligned}
\end{equation}
\end{enumerate}
With the above considerations and the proposed modification in Sec.~\ref{Sec:BGL modified} we obtain the unitarity bound 
in Eq.~\eqref{eq:modified unitarity constraint}.

In~\cite{Martinelli:2022tte} the two susceptibilities were computed nonperturbatively,
\begin{equation}
\begin{aligned}
\chi_{1^-}(0)&=4.45(1.16)\times 10^{-4}\,{\rm GeV}^{-2}\,,\\
\chi_{0^+}(0)&=2.04(0.20)\times 10^{-2}\,.
\end{aligned}
\end{equation}
We checked that our results for observables do not change significantly when using 
these values instead of the ones in Eq.~\eqref{eq:our-chis}.

\section{Comment on the generalised BGL unitarity constraint and relation to Refs.~\cite{Gubernari:2020eft,Gubernari:2022hxn,Blake:2022vfl}}
\label{app:Modified unitarity constraint}
The authors of Refs.~\cite{Gubernari:2020eft,Gubernari:2022hxn,Blake:2022vfl} introduce a modified BGL expansion
\begin{equation}\label{eq:modified BGL expansion}
f(z) = \frac{1}{B(q^2)\phi(q^2,t_0)}\sum_{i=0}^{K-1} b_i\, p_i(z)\,,
\end{equation}
 in terms of  a complete set of orthogonal polynomials $p_i(z)$ with
 \begin{equation}
     \langle p_i|p_j\rangle_\alpha=\delta_{ij}\,,
 \end{equation}
 where the inner product is as defined in Eq.~\eqref{eq:Blake inner product}, restricted
 to the arc $[-\alpha,\alpha]$ of the unit circle.
 The unitarity constraint in Eq.~\eqref{eq:vanilla unitarity constraint} 
 then takes the simple form
\begin{equation}
\sum\limits_{i=0}^{K-1} |b_i|^2 \leq 1.
\end{equation} 
We now show that the approach of Refs.~\cite{Gubernari:2020eft,Gubernari:2022hxn,Blake:2022vfl} is equivalent to 
the original BGL expansion in terms of a polynomial in $\{1,z,z^2,\dots\}$ up to the
modified unitarity constraint in Eq.~\eqref{eq:modified unitarity constraint}. 
By the construction of Refs.~\cite{Gubernari:2020eft,Gubernari:2022hxn,Blake:2022vfl}, $K-1$ is the maximum order of $z$ in 
both the original BGL expansion in Eq.~\eqref{eq:BGLparametrisation} and the 
one in Eq.~\eqref{eq:modified BGL expansion}. Therefore, the coefficients $a_i$ and $b_i$
are related by a linear transformation, or, in other words,
\begin{equation}
\sum\limits_{i=0}^{K-1} b_i p_i(z) = \sum\limits_{i=0}^{K-1}  a_i z^i.
\end{equation}
Using the inner product defined in Eq.~\eqref{eq:Blake inner product} we now
project on the orthonormal polynomials $p_j(z)$,
\begin{equation}
 \sum\limits_{i=0}^{K-1} b_i \langle p_i|p_j\rangle_{\alpha} = \sum\limits_{i=0}^{K-1} a_i \langle z^i|p_j\rangle_{\alpha}\,.
\end{equation}
Using the orthonormality of the $p_i(z)$ we get
\begin{equation}
    b_j = \sum\limits_{i=0}^{K-1} a_i \langle z^i|p_j\rangle_{\alpha}\,,
\end{equation}
which defines the linear transformation between the $a_i$ and $b_i$.
Using this result we can rewrite the unitarity constraint of Refs.~\cite{Gubernari:2020eft,Gubernari:2022hxn,Blake:2022vfl} as
\begin{equation}\label{eq:modified unitarity constraint app}
\sum\limits_{i=0}^{K-1} |b_i|^2 =\sum_{j,k,l=0}^{K-1} a_k^* \langle z^k|p_j\rangle_\alpha \langle p_j|z^l\rangle_{\alpha}a_l
=\sum_{k,l=0}^{K-1} a_k^* \langle z^k|z^l\rangle_{\alpha}\, a_l\le 1\,,
\end{equation}
which follows from the completeness $\sum_i |p_i\rangle\langle p_j| = 1$ of the $p_i(z)$.
This modified BGL unitarity constraint can be computed immediately to any desired order $K-1$,  recalling that $\langle z^i|z^j \rangle_\alpha$ is known from Eq.~\eqref{eq:metric}.
Thus, the modified BGL expansion in Refs.~\cite{Gubernari:2020eft,Gubernari:2022hxn,Blake:2022vfl} and the original BGL 
expansion~\cite{Boyd:1994tt} with the modified unitarity constraint Eq.~\eqref{eq:modified unitarity constraint app} agree exactly. While the implementation in Refs.~\cite{Gubernari:2020eft,Gubernari:2022hxn,Blake:2022vfl}
requires the computation of the polynomials $p_i(z)$ via recursion relations, the
proposal made here allows the continued use of BGL-fit implementations. Only 
the unitarity constraint needs to be modified according to Eq.~\eqref{eq:modified unitarity constraint app}.

We make another observation in this context. By construction,
\begin{equation}
    \langle p_i|p_j\rangle = \gamma_{ik}\langle z^k|z^l\rangle_\alpha \gamma_{jl}=\delta_{jl}\,,
\end{equation}
where $\gamma_{ik}$ is the polynomial coefficient multiplying $z^k$ in the expansion of $p_i(z)$ in the basis $\{1,z,z^2,\dots\}$. In Refs.~\cite{Gubernari:2020eft,Gubernari:2022hxn,Blake:2022vfl} these coefficients are computed using recursion relations based on the work in Refs.~\cite{Szego:1939,Simon:2004}. An alternative way to compute the coefficients is as follows. Define the matrix $M_{kl}=\langle z^k|z^l\rangle_\alpha$ and rewrite the previous equation as
\begin{equation}
\label{eq:mchol}
    \gamma\, M\,\gamma^T=\identity\;\;\leftrightarrow \;\;M=\gamma^{-1}(\gamma^T)^{-1}\,.
\end{equation}
Since $\gamma$ is lower triangular, Eq.~\eqref{eq:mchol} provides a Cholesky decomposition of $M$. $M$ is symmetric positive definite for $0 < \alpha \leq \pi$, making the decomposition unique. We know $M$ analytically and hence we can compute $\gamma$, and the polynomials $p_i(z)$, by Cholesky decomposition.

 \section{Constraining the asymptotic behaviour of the BGL expansion}\label{app:modified BGL}
The asymptotic behaviour of the BGL ansatz for $f_+(t=q^2)$ for large $t$ with
the choice of outer function as detailed in App.~\ref{app:The BGL
  parameterisation} is
\begin{equation}    
f_+(t)=\frac{1}{B_+(t)\phi_+(t)}\sum\limits_k a_{+,k} z(t)^k \stackrel{z\approx 1}{=}{\rm polynomial \,in\,}\{t^{1/4},t^{-1/4},t^{-3/4},t^{-5/4},\dots\}\,.\label{eq:BGLasympt}
\end{equation}
This expression could potentially allow for a diverging form factor,
incompatible with the expectation from perturbation theory.

In principle, the dispersion relation Eq.~\eqref{eq:vanilla unitarity constraint}, written in terms of the $t$ integral,
\begin{equation}
\frac 1\pi\int_{t_+}^\infty dt \left|\frac {dz(t)}{dt}\right||B_+(t)\phi_+(t)f_+(t)|^2\le 1\,,\label{eq:uconstraint}
\end{equation}
has constraining power. Let us analyse the integral kernel as follows:
The Jacobian has the asymptotic behaviour
\begin{equation}\abs{\frac {dz(t)}{dt}}=\frac{\sqrt{{t_\ast}-{t_0}}}{
    \sqrt{{t_\ast}-t}(\sqrt{t_\ast-t}+\sqrt{t_\ast-t_0})^2 }\stackrel{z\approx 1}\sim t^{-3/2}\,.
\end{equation}
Together with the asymptotic behaviour
\begin{equation}|	B_+(t)\phi_+(t)|^2 \stackrel{z\approx 1}{\sim}{\rm polynomial \,in\,}\{t^{-1/2},t^{-1},t^{-3/2},\dots\}\,,
\end{equation}
we see that the asymptotic behaviour of the vector form factor is not sufficiently
constrained. In particular,  by Eq.~\eqref{eq:uconstraint} the form factor is 
only constrained to \(f_+(t)\lesssim t^{1/2}\).

In the following we propose a modified BGL expansion, which is
constrained such that the leading three powers in the asymptotic
behaviour of Eq.~\eqref{eq:BGLasympt} are suppressed in the
large-\(t\) limit:  Let us first observe that

\begin{equation}\sum_{k=0}^{\infty}a_{+,k} z^k\stackrel{z\approx 1}{=}\sum_{k=0}^\infty a_{+,k}\left(1+k\,\alpha\, t^{-1/2}+k^2\,\beta\,t^{-1}+\dots\right)\,,
\end{equation}
where \(\alpha\) and \(\beta\) depend on \(t_0\), \(t_+\) and
\(t_\ast\). We find that the first three derivatives (\(n=0,1,2\)) of
the same sum with respect to \(z\) provide the same polynomial structure
in \(k^n\), and setting the derivatives to zero will therefore remove
the contribution of the leading three powers in Eq.~\eqref{eq:BGLasympt} in the limit \(z\to 1\):
\begin{align}
\sum_{k=0}^\infty a_{+,k} z^k &\stackrel{z=1}{=}0\,,\\
\frac d{dz}\sum_{k=0}^\infty a_{+,k} z^k &\stackrel{z=1}{=}\sum_{k=1}^\infty k\,a_{+,k}=0\,,\\
\frac {d^2}{dz^2}\sum_{k=0}^\infty a_{+,k} z^k &\stackrel{z=1}{=}\sum_{k=2}^\infty k(k-1)\,a_{+,k}=0\,.
\end{align}
These sum rules have first been proposed in Ref.~\cite{Buck:1998kp,Becher:2005bg}. We can solve the above system for the three coefficients \(a_{+,j}\),
\(a_{+,j+1}\), \(a_{+,j+2}\):

\begin{align}
    a_{+,j\hphantom{+0}}&=-\frac 12\sum\limits_{\substack{k=0\\k \neq j+\{0,1,2\}}}^\infty\left[(j-k)^2+3(j-k)+2\right]a_{+,k}&\hspace{-1ex}\equiv&\hspace{-2ex}\sum\limits_{\substack{k=0\\k \neq j+\{0,1,2\}}}^\infty\rho_{k,j}a_{+,k}\,,\\
a_{+,j+1}&=\;\;\;\;\;\;\sum\limits\limits_{\substack{k=0\\k \neq j+\{0,1,2\}}}^\infty\left[(j-k)^2+2(j-k)\right]a_{+,k}&\hspace{-1ex}\equiv&\hspace{-2ex}\sum\limits_{\substack{k=0\\k \neq j+\{0,1,2\}}}^\infty\sigma_{k,j}a_{+,k}\,,\\
a_{+,j+2}&=-\frac 12\sum\limits\limits_{\substack{k=0\\k \neq j+\{0,1,2\}}}^\infty\left[(j-k)^2+(j-k)\right]a_{+,k}&\hspace{-1ex}\equiv&\hspace{-2ex}\sum\limits_{\substack{k=0\\k \neq j+\{0,1,2\}}}^\infty\tau_{k,j}a_{+,k}.
\end{align}
The correspondingly modified BGL expansion for the vector form factor
then reads
\begin{align}
f_+(t)
&=\frac{1}{B_+(t)\phi_+(t)}\sum\limits_{\substack{k=0\\k \neq j+\{0,1,2\}}} a_{+,k} \left(z(t)^k
+\rho_{k,j}z(t)^{j}+\sigma_{k,j}z(t)^{j+1}+\tau_{k,j}z(t)^{j+2}\right)\,,\label{eq:BGLmodified}
\end{align}
with the associated unitarity constraint (here for the case $\alpha=\pi$
and noting that $\alpha\neq \pi$ can
be implemented straight-forwardly)
\begin{equation}
  \begin{aligned}
    \sum_{k=0}^\infty|a_{+,k}|^2 &= \abs{a_j}^2 + \abs{a_{j+1}}^2 +\abs{a_{j+2}}^2 + \!\!\!\!\!\!\sum\limits_{\substack{k=0\\k \neq j+\{0,1,2\}}} \!\!\!\!\!\!\abs{a_{+,k}}^2\\
    &= \Big|\!\!\!\!\!\!\!\!\sum\limits_{\substack{k=0\\k \neq j+\{0,1,2\}}}\!\!\!\!\!\! \rho_{k,j}a_{+,k}\Big|^2
    +\Big|\!\!\!\!\!\!\!\!\sum\limits_{\substack{k=0\\k \neq j+\{0,1,2\}}} \!\!\!\!\!\!\sigma_{k,j}a_{+,k}\Big|^2
    +\Big|\!\!\!\!\!\!\!\!\sum\limits_{\substack{k=0\\k \neq j+\{0,1,2\}}} \!\!\!\!\!\!\tau_{k,j}a_{+,k}\Big|^2
    + \!\!\!\!\!\!\sum\limits_{\substack{k=0\\k \neq j+\{0,1,2\}}}\!\!\!\!\!\! \abs{a_{+,k}}^2 \leq 1\,.
  \end{aligned}
\end{equation}

\section{Implementation details for the algorithm}\label{app:prior metric}
The choice of prior metric $M$ in Eq.~\eqref{eq:prior} has to ensure
that ${\bf a}^TM{\bf a}\le 2$, in order for the
accept-reject step to be well defined. Since we have used
the kinematical constraint $f_0(0)=f_+(0)$ to eliminate the 
BGL parameter $a_{0,0}$, $M$ is a $(K_++K_0-1)\times (K_++K_0-1)$ matrix. 
A naive choice could then be
\begin{equation}
    M=\left(\begin{array}{cc}\mathcal{M}^{++}&0\\0& \mathcal{M}^{00}
\end{array}\right)\,,\label{eq:not a good metric}
\end{equation}
where $\mathcal{M}^{XX}_{ij}=\langle z^i|z^j\rangle$. Clearly, ${\bf a}_+^T\mathcal{M}^{++}{\bf a}_+\le 1$. However, since the parameter
$a_{0,0}$ has been eliminated, the reduced norm 
$\bar {\bf a}_0^T\mathcal{M}^{00}\bar {\bf a}_0\equiv \sum_{i,j=1}^{K_0-1}a_{0,i} \mathcal{M}^{00}_{ij}a_{0,j}$ can be larger than 1. This metric is therefore not suitable  in view of the accept-reject step.

Let us instead start with the parameter vectors before eliminating
the $a_{0,0}$ component, for which ${\bf a}_+^T \mathcal{M}^{++}{\bf a}_+\le 1$ and  ${\bf a}_0^T \mathcal{M}^{00}{\bf a}_0\le 1$. Then,
\begin{equation}
\label{eq:two bilinears}
\begin{aligned}
    {\bf a}_+^T \mathcal{M}^{++}{\bf a}_+ + {\bf a}_0^T \mathcal{M}^{00}{\bf a}_0
    &=  { a}_{+,\mu} \mathcal{M}^{++}_{\mu,\nu}{a}_{+,\nu}
    + a_{0,i} \mathcal{M}^{00}_{i,j} a_{0,j}\\
    &\phantom{=}
    {} + a_{0,0}\left(\mathcal{M}^{00}_{0,0}a_{0,0}+2\mathcal{M}^{0,0}_{0,i}a_{0,i}\right)\\
    &\le 2\,,
\end{aligned}
\end{equation}
where Greek indices are summed starting from $0$ and Latin indices
 starting from $1$. Using the  kinematical
constraint $f_+(0)=f_0(0)$ we can now eliminate $a_{0,0}$ using (\emph{cf.} Eq.~\eqref{eq:kinematical constraint})
\begin{align}
    a_{0,0}=\frac{B_0(0)\phi_0(0,t_0)}{B_+(0)\phi_+(0,t_0)}
    \sum\limits_{k=0}^{K_+-1}a_{+,k}z^k(0)-\sum\limits_{k=1}^{K_0-1}a_{0,k} z^k(0)\,.
\end{align}
Eq.~\eqref{eq:two bilinears} can then be rewritten in the 
compact form
\begin{align}
{\bf a}^T M{\bf a}={\bf a}^T\left(
    \begin{array}{cc}
    M^{++}&M^{+0}\\
    M^{0+}& M^{00}
    \end{array}\right){\bf a}\le 2\,,
\end{align}
where, defining $z_{\rm max}=z(0)$,
\begin{equation}
\begin{aligned}
M^{++}_{\mu,\nu}&=\,
\mathcal{M}^{++}_{\mu,\nu}+\left(\frac{B_0(0)\phi_0(0,t_0)}{B_+(0)\phi_+(0,t_0)}\right)^2\mathcal{M}_{0,0}^{00}z^\mu_{\rm max}z^\nu_{\rm max}\,,\\
M^{0+}_{\mu, i}&=\,-\left(\frac{B_0(0)\phi_0(0,t_0)}{B_+(0)\phi_+(0,t_0)}\right)
z^\mu_{\rm max}\left(\mathcal{M}_{0,0}^{00}z^i_{\rm max}-\mathcal{M}_{0,i}^{00}\right)\,,\\
M^{+0}_{i,\mu}&=\,M^{0+}_{\mu, i}\,,\\
M^{00}_{i,j}&=\mathcal{M}^{00}_{i,j}+\mathcal{M}^{00}_{0,0} z^i_{\rm max}z^j_{\rm max}
-\mathcal{M}^{00}_{0,i}z^j_{\rm max}-\mathcal{M}^{00}_{0,j}z^i_{\rm max}\,.
\end{aligned}
\end{equation}
We choose $M$ as the metric for the prior term in Eq.~\eqref{eq:full_PD}.
\textcolor{black}{The modifications of $M$ required to represent the modified BGL expansion 
defined in App.~\ref{app:modified BGL} are straight forward.}

\section{Results for forward-backward and polarisation asymmetries}\label{app:asymmetries}
Here we will present the underlying formulae for two more phenomenologically
relevant quantities that can be computed from the form-factor parameterisation:
the forward-backward and polarisation asymmetries.

\begin{figure*}
  \includegraphics[width=8cm]{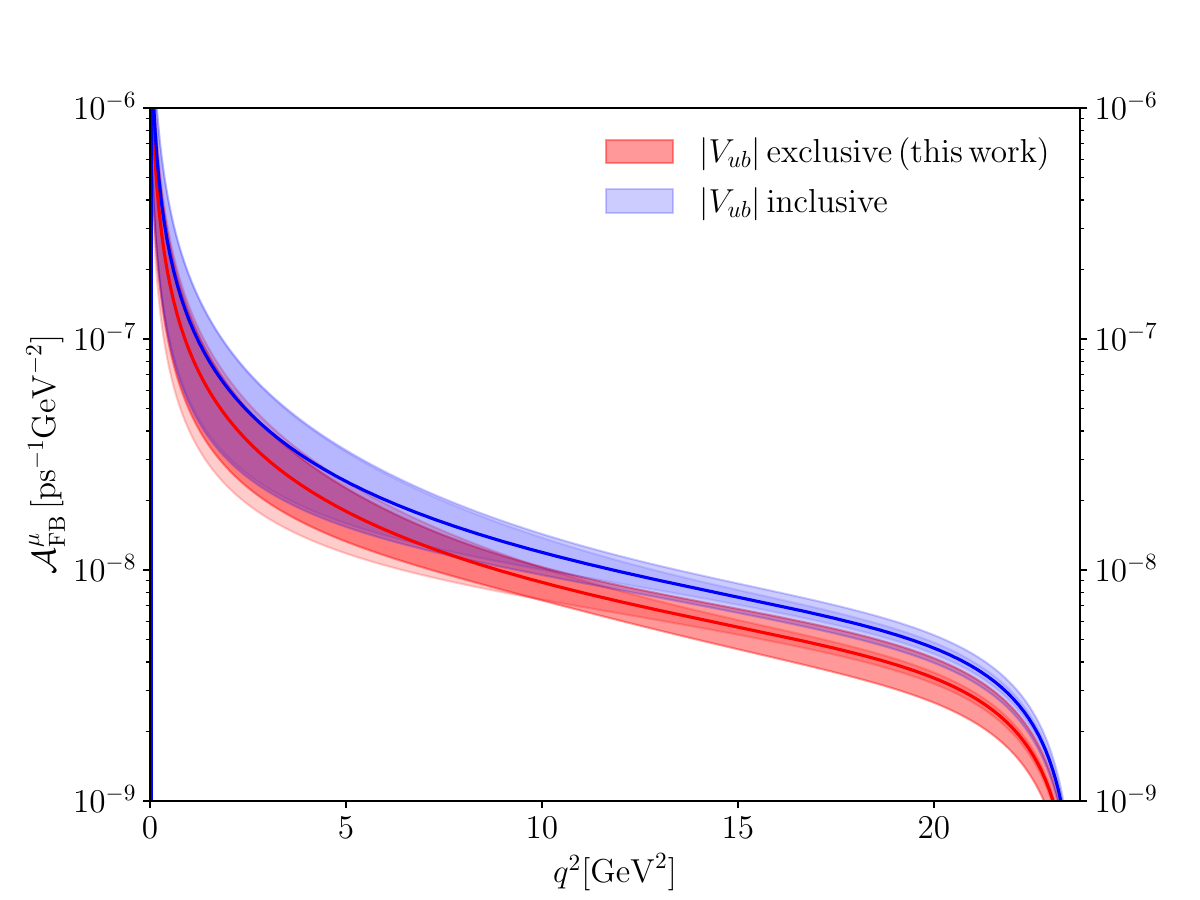}
  \includegraphics[width=8cm]{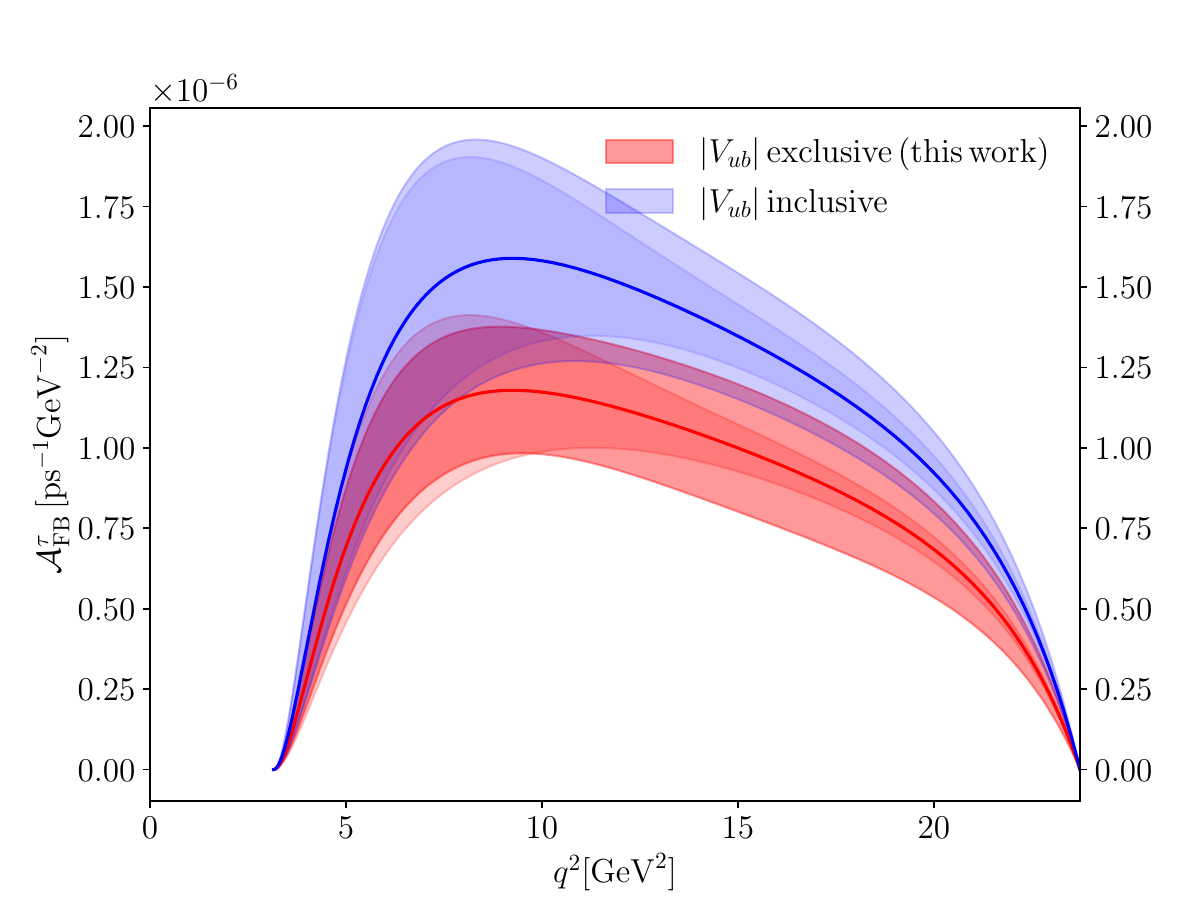}
  \caption{Forward-backward asymmetries $\mathcal{A}_{\rm FB}^{\mu}$ (left) and
    $\mathcal{A}_{\rm FB}^{\tau}$ (right). For $|V_{ub}|$ exclusive we take our
    determination (\emph{cf.} Eq.~\eqref{eq:ourVub1}). The value for $|V_{ub}|$
    inclusive is taken from Eq.~\eqref{eq:Vubinclusive}. The inner shading does
    not include the uncertainty contributions from $|V_{ub}|$.}
  \label{fig:AFB}
\end{figure*}

The forward-backward asymmetry is defined as
\begin{equation}
  \mathcal{A}^\ell_{\rm FB}(q^2)\equiv\left[\int\limits_0^1-\int\limits_{-1}^0\right]
  d\cos\theta_\ell\frac{d^2\Gamma( B_s\to K\ell\nu)}{dq^2d\cos\theta_\ell}\,,
\end{equation}
where $\theta_\ell$ is the angle between the $B_s$ momentum and the lepton
$\ell$ in the rest frame of the $\ell$--$\nu$ system. In the SM this can be
expressed as~\cite{Meissner:2013pba}
\begin{equation}
  \mathcal{A}^\ell_{\rm FB}(q^2)=\frac{\eta_{EW}G_F^2 |V_{ub}|^2}{32\pi^3 M_{B_s}}\left(1-\frac {m_\ell^2}{q^2}\right)^2|{\bf p}_K|^2 \frac{m_\ell^2}{q^2}\left(M_{B_s}^2-M_K^2\right)f_+(q^2)f_0(q^2)\,.
\end{equation}
Our results for the combined Bayesian inference of the HPQCD~14 and the
RBC/UKQCD~23 datasets are shown in Fig.~\ref{fig:AFB} for the cases $\ell=\mu$ on the left 
and $\tau$ on the right.
Furthermore, we define the integrated forward-backward asymmetry
$I[\mathcal{A}_{\rm FB}]$ and the average forward-backward asymmetry
$\bar{\mathcal{A}}_{\rm FB}$ as
\begin{equation}
  I[\mathcal{A}^\ell_{\rm FB}]=\int\limits_{m_\ell^2}^{q^2_{\rm max}}dq^2\mathcal{A}^\ell_{\rm FB}(q^2)/|V_{ub}|^2\,,\label{eq:IAFB}
\end{equation}
and
\begin{equation}
  \mathcal{\bar A}^\ell_{\rm FB}=\frac{\int_{m_\ell^2}^{q_{\rm max}^2}dq^2\mathcal{A}^\ell_{\rm FB}(q^2)}
			 {\int_{m_\ell^2}^{q_{\rm max}^2}dq^2d\Gamma(B_s\to K\ell\nu)/dq^2}\,.
          \label{eq:AbarFB}
\end{equation}
Numerical results for these values are provided in Tab.~\ref{tab:BstoK_zfit_observables}.

\begin{figure*}
  \includegraphics[width=8cm]{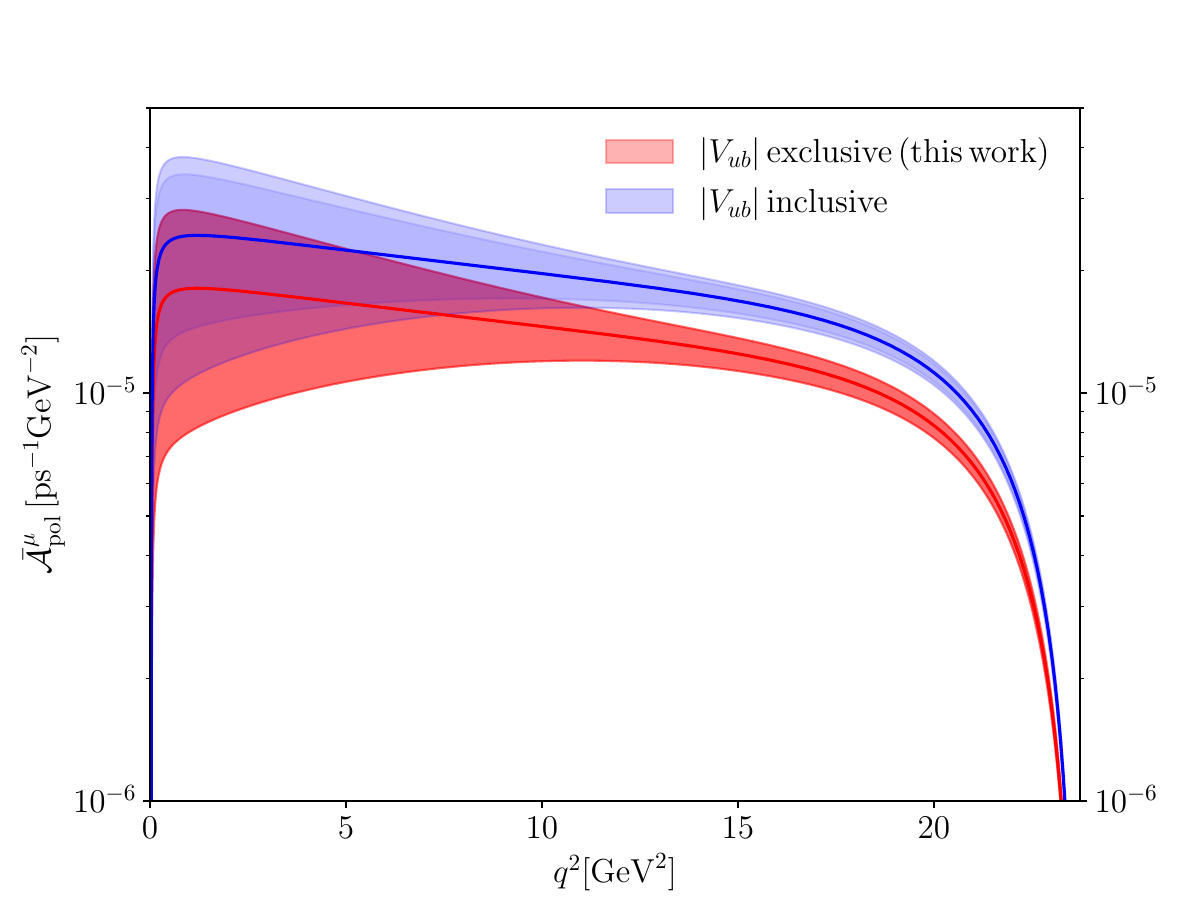}
  \includegraphics[width=8cm]{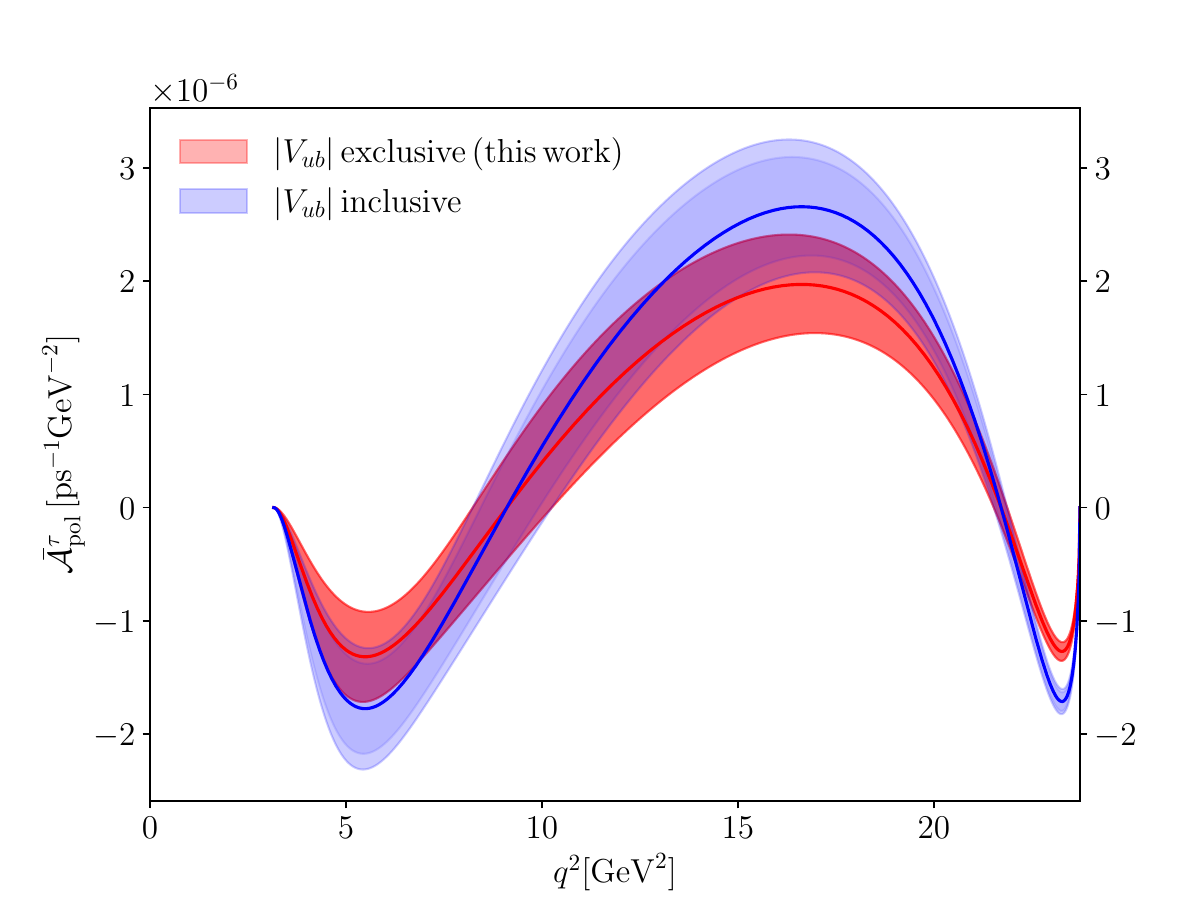}
  \caption{Polarisation asymmetries $\mathcal{A}_{\rm pol}^{\mu}$ (left) and
    $\mathcal{A}_{\rm pol}^{\tau}$ (right). We again take the value for
    $|V_{ub}|$ exclusive from Eq.~\eqref{eq:ourVub1} and the value for
    $|V_{ub}|$ inclusive from Eq.~\eqref{eq:Vubinclusive}. The inner shading
    does not include the uncertainties of the CKM matrix element.}
  \label{fig:Apol}
\end{figure*}

Another observable which can be computed from the form factor parameterisation
is the polarisation asymmetry $\mathcal{A}^\ell_{\rm pol}$. This is defined to
be the difference between the left-handed and the right-handed contributions to
the decay rate~\cite{Meissner:2013pba},
\begin{equation}
  \mathcal{A}^\ell_{\rm pol}(q^2)=\frac{d\Gamma(\ell,{\rm LH})}{dq^2}-\frac{d\Gamma (\ell,{\rm RH})}{dq^2}\,,
\end{equation}
and can be used to probe for helicity-violating interactions. In the SM this
takes the form
\begin{align}
  \frac {d \Gamma(\ell,{\rm LH})}{dq^2}=&\frac{\eta_{EW}G_F^2|V_{ub}|^2|{\bf p}_K|^3}{24\pi^3}\left(1-\frac {m_\ell^2}{q^2}\right)^2f_+^2(q^2)\,,\nonumber\\
  \\[-1ex]
  \frac {d \Gamma(\ell,{\rm RH})}{dq^2}=&\frac{\eta_{EW}G_F^2|V_{ub}|^2|{\bf p}_K|}{24\pi^3}\frac{m_\ell^2}{q^2}\left(1-\frac {m_\ell^2}{q^2}\right)^2 \left(\frac 38 \frac{(M_{B_s}^2-M_K^2)^2}{M_{B_s}^2}f_0^2(q^2)+\frac{|{\bf p}_K|^2}{2} f_+^2(q^2)\right)\,.\quad\nonumber
\end{align}
The polarisation distribution is shown in Fig.~\ref{fig:Apol}. Finally, using
analogous definitions to Eqs.~\eqref{eq:IAFB} and \eqref{eq:AbarFB} we define
$I[\mathcal{A}^\ell_{\rm pol}]$ and $\mathcal{\bar A}^\ell_{\rm pol}$ and
provide numerical values in Tab.~\ref{tab:BstoK_zfit_observables}.

\section{Further numerical results}\label{app:Further numerical results}

\subsection{Results for observables from Bayesian fits to individual lattice data sets}\label{sec:Bayesian individual data sets}
The results for observables computed from the Bayesian-inference fit to RBC/UKQCD 23 can be found in Tab.~\ref{tab:observables RBCUKQCD 23}, the ones for FNAL/MILC 19 in Tab.~\ref{tab:observables FNALMILC 19}, and the ones for HPQCD 14 in Tab.~\ref{tab:observables HPQCD 14}. The corresponding BGL coefficients are listed in Tabs.~\ref{tab:BI results for BGL coefficients a+} and~\ref{tab:BI results for BGL coefficients a0}.
\begin{table*}[bt!]
\begin{center}
\tiny
\begin{tabular}{l@{\hspace{1mm}}llllllllll}
\hline\hline
$K_+$&$K_0$&\multicolumn{1}{c}{$f(q^2=0)$}&\multicolumn{1}{c}{$R_{B_s\to K}^{\rm impr}$}&\multicolumn{1}{c}{$R_{B_s\to K}$}&\multicolumn{1}{c}{$\frac{\Gamma^\tau}{|V_{ub}|^2}\,[\frac 1{\rm ps}]$}&\multicolumn{1}{c}{$\frac{\Gamma^\mu}{|V_{ub}|^2}\,[\frac 1{\rm ps}]$}&\multicolumn{1}{c}{$V^{\rm low}_{\rm CKM}$}&\multicolumn{1}{c}{$V^{\rm high}_{\rm CKM}$}&\multicolumn{1}{c}{$V^{\rm full}_{\rm CKM}$}&\\
\hline
2&2&0.222(21)&1.545(17)&0.741(19)&5.37(43)&7.25(70)&0.00356(39)&0.00325(30)&0.00336(32)\\
2&3&0.087(39)&1.657(46)&0.954(75)&3.70(50)&3.94(81)&0.0070(22)&0.00408(46)&0.00420(52)\\
3&2&0.231(21)&1.721(57)&0.774(27)&4.34(45)&5.62(72)&0.00375(42)&0.00382(41)&0.00379(39)\\
3&3&0.248(88)&1.721(56)&0.76(10)&4.48(72)&6.1(1.7)&0.0039(14)&0.00381(46)&0.00381(52)\\
3&4&0.25(12)&1.722(64)&0.77(15)&4.51(84)&6.2(2.3)&0.0042(22)&0.00380(48)&0.00382(53)\\
4&3&0.249(86)&1.72(12)&0.76(12)&4.55(82)&6.3(2.0)&0.0039(16)&0.00378(53)&0.00379(59)\\
4&4&0.25(12)&1.72(12)&0.78(17)&4.53(89)&6.3(2.4)&0.0043(29)&0.00381(57)&0.00383(62)\\
5&5&0.25(11)&1.72(11)&0.77(16)&4.57(90)&6.4(2.4)&0.0041(24)&0.00376(55)&0.00378(61)\\
6&6&0.26(11)&1.71(11)&0.76(16)&4.63(88)&6.5(2.4)&0.0040(26)&0.00375(54)&0.00376(58)\\
7&7&0.26(11)&1.71(11)&0.75(15)&4.67(90)&6.7(2.4)&0.0038(19)&0.00373(56)&0.00374(62)\\
8&8&0.26(11)&1.70(12)&0.74(15)&4.71(94)&6.8(2.6)&0.0038(19)&0.00371(55)&0.00372(62)\\
9&9&0.27(11)&1.70(12)&0.74(16)&4.76(98)&7.0(2.7)&0.0038(20)&0.00370(59)&0.00371(66)\\
10&10&0.28(11)&1.71(13)&0.73(16)&4.80(99)&7.1(2.8)&0.0037(31)&0.00368(58)&0.00368(62)\\
\hline\hline\\
\end{tabular}
\\
\begin{tabular}{l@{\hspace{1mm}}llllllllll}
\hline\hline
$K_+$&$K_0$&\multicolumn{1}{c}{$I[\mathcal{A}_{\rm FB}^\tau]\,[\frac 1{\rm ps}]$}&\multicolumn{1}{c}{$I[\mathcal{A}_{\rm FB}^\mu]\,[\frac 1{\rm ps}]$}&\multicolumn{1}{c}{$\mathcal{\bar A}_{\rm FB}^\tau$}&\multicolumn{1}{c}{$\mathcal{\bar A}_{\rm FB}^\mu$}&\multicolumn{1}{c}{$I[\mathcal{A}_{\rm pol}^\tau]\,[\frac 1{\rm ps}]$}&\multicolumn{1}{c}{$I[\mathcal{A}_{\rm pol}^\mu]\,[\frac 1{\rm ps}]$}&\multicolumn{1}{c}{$\mathcal{\bar A}_{\rm pol}^\tau$}&\multicolumn{1}{c}{$\mathcal{\bar A}_{\rm pol}^\mu$}&\\
\hline
2&2&1.46(12)&0.0320(46)&0.2720(21)&0.00440(27)&0.794(92)&7.16(68)&0.148(13)&0.98768(73)\\
2&3&0.99(14)&0.0115(41)&0.2679(27)&0.00284(46)&0.31(13)&3.90(80)&0.082(27)&0.9912(11)\\
3&2&1.23(13)&0.0315(46)&0.2825(28)&0.00560(44)&0.14(15)&5.53(71)&0.031(34)&0.9838(13)\\
3&3&1.27(23)&0.038(19)&0.2836(77)&0.0058(15)&0.13(16)&6.0(1.7)&0.030(35)&0.9833(39)\\
3&4&1.28(27)&0.040(26)&0.2833(91)&0.0057(19)&0.14(17)&6.1(2.2)&0.030(38)&0.9834(49)\\
4&3&1.29(26)&0.038(19)&0.2820(80)&0.0058(16)&0.18(31)&6.2(2.0)&0.034(65)&0.9832(45)\\
4&4&1.28(28)&0.039(25)&0.2817(93)&0.0058(20)&0.16(31)&6.2(2.4)&0.031(64)&0.9833(52)\\
5&5&1.30(28)&0.040(24)&0.2821(89)&0.0057(18)&0.18(29)&6.3(2.3)&0.035(60)&0.9834(49)\\
6&6&1.31(28)&0.041(24)&0.2826(88)&0.0058(18)&0.19(29)&6.4(2.3)&0.036(58)&0.9832(48)\\
7&7&1.33(28)&0.043(24)&0.2831(85)&0.0060(18)&0.20(31)&6.6(2.4)&0.037(62)&0.9829(47)\\
8&8&1.34(29)&0.043(25)&0.2827(86)&0.0059(18)&0.23(32)&6.7(2.5)&0.042(64)&0.9831(47)\\
9&9&1.35(31)&0.045(27)&0.2830(90)&0.0060(18)&0.23(34)&6.8(2.6)&0.041(67)&0.9827(49)\\
10&10&1.37(31)&0.047(27)&0.2832(93)&0.0062(18)&0.23(36)&7.0(2.7)&0.040(69)&0.9823(49)\\
\hline\hline\\
\end{tabular}

\end{center}
\caption{Results for observables from Bayesian-inference fit to RBC/UKQCD 23~\cite{Flynn:2023nhi}.}\label{tab:observables RBCUKQCD 23}
\end{table*}
\begin{table*}[bt!]
\begin{center}
\tiny
\begin{tabular}{l@{\hspace{1mm}}llllllllll}
\hline\hline
$K_+$&$K_0$&\multicolumn{1}{c}{$f(q^2=0)$}&\multicolumn{1}{c}{$R_{B_s\to K}^{\rm impr}$}&\multicolumn{1}{c}{$R_{B_s\to K}$}&\multicolumn{1}{c}{$\frac{\Gamma^\tau}{|V_{ub}|^2}\,[\frac 1{\rm ps}]$}&\multicolumn{1}{c}{$\frac{\Gamma^\mu}{|V_{ub}|^2}\,[\frac 1{\rm ps}]$}&\multicolumn{1}{c}{$V^{\rm low}_{\rm CKM}$}&\multicolumn{1}{c}{$V^{\rm high}_{\rm CKM}$}&\multicolumn{1}{c}{$V^{\rm full}_{\rm CKM}$}&\\
\hline
2&2&0.120(26)&1.476(20)&0.802(31)&3.41(31)&4.27(54)&0.00565(93)&0.00398(38)&0.00422(43)\\
2&3&0.180(30)&1.435(20)&0.712(32)&3.92(37)&5.54(75)&0.00429(64)&0.00367(35)&0.00381(40)\\
3&2&0.119(27)&1.517(29)&0.828(37)&3.22(31)&3.92(55)&0.0059(10)&0.00415(40)&0.00439(46)\\
3&3&0.177(31)&1.460(29)&0.728(37)&3.76(39)&5.20(78)&0.00446(73)&0.00379(39)&0.00394(44)\\
3&4&0.108(52)&1.430(33)&0.794(58)&3.26(45)&4.16(87)&0.0064(22)&0.00405(44)&0.00414(49)\\
4&3&0.059(80)&1.427(36)&0.835(71)&3.14(49)&3.84(94)&0.0084(37)&0.00411(46)&0.00418(49)\\
4&4&0.06(11)&1.428(34)&0.821(93)&3.18(53)&4.0(1.2)&0.0083(39)&0.00409(48)&0.00415(51)\\
5&5&0.07(11)&1.428(36)&0.823(91)&3.17(53)&4.0(1.1)&0.0083(41)&0.00410(48)&0.00416(51)\\
6&6&0.07(10)&1.429(36)&0.817(90)&3.20(52)&4.0(1.1)&0.0080(38)&0.00407(48)&0.00413(51)\\
7&7&0.08(10)&1.431(36)&0.814(92)&3.21(53)&4.1(1.2)&0.0079(39)&0.00410(48)&0.00415(51)\\
8&8&0.09(10)&1.433(36)&0.808(95)&3.23(53)&4.1(1.2)&0.0077(39)&0.00406(46)&0.00411(49)\\
9&9&0.10(10)&1.432(36)&0.798(97)&3.27(56)&4.2(1.3)&0.0073(37)&0.00404(47)&0.00409(50)\\
10&10&0.11(10)&1.435(35)&0.79(10)&3.32(55)&4.4(1.3)&0.0070(37)&0.00401(47)&0.00406(50)\\
\hline\hline\\
\end{tabular}
\\
\begin{tabular}{l@{\hspace{1mm}}llllllllll}
\hline\hline
$K_+$&$K_0$&\multicolumn{1}{c}{$I[\mathcal{A}_{\rm FB}^\tau]\,[\frac 1{\rm ps}]$}&\multicolumn{1}{c}{$I[\mathcal{A}_{\rm FB}^\mu]\,[\frac 1{\rm ps}]$}&\multicolumn{1}{c}{$\mathcal{\bar A}_{\rm FB}^\tau$}&\multicolumn{1}{c}{$\mathcal{\bar A}_{\rm FB}^\mu$}&\multicolumn{1}{c}{$I[\mathcal{A}_{\rm pol}^\tau]\,[\frac 1{\rm ps}]$}&\multicolumn{1}{c}{$I[\mathcal{A}_{\rm pol}^\mu]\,[\frac 1{\rm ps}]$}&\multicolumn{1}{c}{$\mathcal{\bar A}_{\rm pol}^\tau$}&\multicolumn{1}{c}{$\mathcal{\bar A}_{\rm pol}^\mu$}&\\
\hline
2&2&0.884(89)&0.0132(34)&0.2589(36)&0.00305(41)&0.715(83)&4.24(54)&0.210(17)&0.9914(11)\\
2&3&1.02(10)&0.0217(52)&0.2592(36)&0.00387(44)&0.94(11)&5.48(73)&0.239(17)&0.9895(12)\\
3&2&0.844(89)&0.0128(34)&0.2619(38)&0.00322(45)&0.578(97)&3.88(54)&0.179(22)&0.9908(12)\\
3&3&0.98(11)&0.0209(53)&0.2615(40)&0.00396(48)&0.82(13)&5.15(76)&0.218(23)&0.9891(13)\\
3&4&0.82(13)&0.0122(60)&0.2510(80)&0.00278(82)&0.81(13)&4.13(85)&0.250(31)&0.9923(22)\\
4&3&0.78(15)&0.0098(64)&0.2476(99)&0.00236(92)&0.80(13)&3.81(93)&0.257(35)&0.9934(25)\\
4&4&0.79(16)&0.0121(97)&0.248(11)&0.0027(13)&0.80(13)&4.0(1.1)&0.255(34)&0.9924(34)\\
5&5&0.79(16)&0.0117(90)&0.248(10)&0.0027(12)&0.80(13)&3.9(1.1)&0.255(34)&0.9925(33)\\
6&6&0.80(16)&0.0121(92)&0.249(10)&0.0027(13)&0.80(14)&4.0(1.1)&0.253(34)&0.9924(33)\\
7&7&0.80(16)&0.0125(98)&0.249(10)&0.0028(13)&0.80(14)&4.0(1.1)&0.252(35)&0.9923(34)\\
8&8&0.81(16)&0.013(10)&0.250(10)&0.0029(13)&0.80(13)&4.1(1.2)&0.250(35)&0.9920(35)\\
9&9&0.83(17)&0.014(11)&0.251(10)&0.0029(13)&0.81(14)&4.2(1.2)&0.250(34)&0.9918(36)\\
10&10&0.84(17)&0.015(12)&0.252(10)&0.0031(14)&0.81(14)&4.3(1.3)&0.246(34)&0.9914(38)\\
\hline\hline\\
\end{tabular}

\end{center}
\caption{Results for observables from Bayesian-inference fit to FNAL/MILC 19~\cite{Bazavov:2019aom}.}\label{tab:observables FNALMILC 19}
\end{table*}
\begin{table*}[bt!]
\begin{center}
\tiny
\begin{tabular}{l@{\hspace{1mm}}llllllllll}
\hline\hline
$K_+$&$K_0$&\multicolumn{1}{c}{$f(q^2=0)$}&\multicolumn{1}{c}{$R_{B_s\to K}^{\rm impr}$}&\multicolumn{1}{c}{$R_{B_s\to K}$}&\multicolumn{1}{c}{$\frac{\Gamma^\tau}{|V_{ub}|^2}\,[\frac 1{\rm ps}]$}&\multicolumn{1}{c}{$\frac{\Gamma^\mu}{|V_{ub}|^2}\,[\frac 1{\rm ps}]$}&\multicolumn{1}{c}{$V^{\rm low}_{\rm CKM}$}&\multicolumn{1}{c}{$V^{\rm high}_{\rm CKM}$}&\multicolumn{1}{c}{$V^{\rm full}_{\rm CKM}$}&\\
\hline
2&2&0.208(25)&1.524(37)&0.727(25)&4.51(45)&6.23(76)&0.00383(47)&0.00352(35)&0.00363(37)\\
2&3&0.226(34)&1.511(41)&0.704(39)&4.67(49)&6.67(97)&0.00361(53)&0.00344(34)&0.00349(38)\\
3&2&0.233(27)&1.609(58)&0.733(27)&4.44(45)&6.08(77)&0.00368(45)&0.00367(37)&0.00367(38)\\
3&3&0.293(41)&1.592(57)&0.664(40)&4.84(51)&7.3(1.1)&0.00310(44)&0.00349(35)&0.00333(36)\\
3&4&0.293(56)&1.593(60)&0.667(59)&4.85(58)&7.4(1.4)&0.00313(55)&0.00349(37)&0.00338(40)\\
4&3&0.294(42)&1.594(60)&0.663(40)&4.85(52)&7.4(1.1)&0.00309(44)&0.00348(36)&0.00332(36)\\
4&4&0.285(92)&1.593(60)&0.677(88)&4.83(62)&7.3(1.7)&0.00328(86)&0.00350(38)&0.00346(42)\\
5&5&0.277(88)&1.595(62)&0.685(85)&4.81(62)&7.2(1.7)&0.00333(85)&0.00351(38)&0.00348(42)\\
6&6&0.277(88)&1.592(63)&0.685(86)&4.79(63)&7.2(1.7)&0.00335(88)&0.00350(38)&0.00348(43)\\
7&7&0.282(89)&1.592(60)&0.680(87)&4.82(64)&7.3(1.7)&0.00332(89)&0.00350(38)&0.00347(43)\\
8&8&0.283(88)&1.594(61)&0.679(85)&4.83(64)&7.3(1.7)&0.00330(85)&0.00351(37)&0.00347(41)\\
9&9&0.289(91)&1.594(62)&0.674(88)&4.85(64)&7.4(1.8)&0.00327(89)&0.00350(38)&0.00347(42)\\
10&10&0.293(95)&1.593(60)&0.670(91)&4.87(67)&7.5(1.9)&0.00325(92)&0.00349(38)&0.00346(42)\\
\hline\hline\\
\end{tabular}
\\
\begin{tabular}{l@{\hspace{1mm}}llllllllll}
\hline\hline
$K_+$&$K_0$&\multicolumn{1}{c}{$I[\mathcal{A}_{\rm FB}^\tau]\,[\frac 1{\rm ps}]$}&\multicolumn{1}{c}{$I[\mathcal{A}_{\rm FB}^\mu]\,[\frac 1{\rm ps}]$}&\multicolumn{1}{c}{$\mathcal{\bar A}_{\rm FB}^\tau$}&\multicolumn{1}{c}{$\mathcal{\bar A}_{\rm FB}^\mu$}&\multicolumn{1}{c}{$I[\mathcal{A}_{\rm pol}^\tau]\,[\frac 1{\rm ps}]$}&\multicolumn{1}{c}{$I[\mathcal{A}_{\rm pol}^\mu]\,[\frac 1{\rm ps}]$}&\multicolumn{1}{c}{$\mathcal{\bar A}_{\rm pol}^\tau$}&\multicolumn{1}{c}{$\mathcal{\bar A}_{\rm pol}^\mu$}&\\
\hline
2&2&1.22(13)&0.0278(51)&0.2708(37)&0.00443(34)&0.74(15)&6.15(75)&0.164(29)&0.98767(96)\\
2&3&1.26(14)&0.0314(70)&0.2709(38)&0.00465(44)&0.81(18)&6.59(96)&0.173(31)&0.9872(12)\\
3&2&1.23(13)&0.0319(59)&0.2780(43)&0.00524(51)&0.46(19)&5.99(76)&0.103(40)&0.9852(15)\\
3&3&1.36(15)&0.045(10)&0.2814(48)&0.00612(66)&0.53(20)&7.2(1.1)&0.110(40)&0.9830(18)\\
3&4&1.37(17)&0.046(14)&0.2814(50)&0.00611(83)&0.53(22)&7.3(1.3)&0.109(41)&0.9830(22)\\
4&3&1.37(15)&0.046(10)&0.2815(50)&0.00616(71)&0.53(22)&7.2(1.1)&0.109(42)&0.9829(20)\\
4&4&1.36(19)&0.046(21)&0.2810(69)&0.0060(15)&0.53(21)&7.2(1.7)&0.109(42)&0.9834(41)\\
5&5&1.35(19)&0.044(20)&0.2806(67)&0.0058(15)&0.53(22)&7.1(1.6)&0.109(44)&0.9837(39)\\
6&6&1.35(20)&0.044(20)&0.2803(69)&0.0058(15)&0.53(22)&7.1(1.7)&0.111(44)&0.9838(39)\\
7&7&1.35(20)&0.045(20)&0.2806(69)&0.0059(15)&0.53(21)&7.2(1.7)&0.111(43)&0.9835(39)\\
8&8&1.36(20)&0.045(20)&0.2808(69)&0.0059(15)&0.53(22)&7.2(1.7)&0.109(44)&0.9835(39)\\
9&9&1.36(20)&0.047(21)&0.2812(71)&0.0060(15)&0.53(22)&7.3(1.7)&0.109(44)&0.9832(40)\\
10&10&1.37(21)&0.048(23)&0.2815(72)&0.0061(15)&0.53(22)&7.4(1.8)&0.109(43)&0.9831(41)\\
\hline\hline\\
\end{tabular}

\end{center}
\caption{Results for observables from  Bayesian-inference fit to HPQCD 14~\cite{Bouchard:2014ypa}.}\label{tab:observables HPQCD 14}
\end{table*}
\subsection{Combined frequentist fit to HPQCD 14, FNAL/MILC 19 and RBC/UKQCD 23}\label{sec:all lattice data results}
The results for the BGL coefficients from the combined frequentist fit to HPQCD 14, FNAL/MILC 19 and RBC/UKQCD 23 
can be found in Tab.~\ref{tab:all lattice data frequentist}. Judging from the $p$-value
no acceptable combined fit over the three data sets is possible.
\begin{table*}[bt!]
\begin{center}
\tiny
\begin{tabular}{l@{\hspace{1mm}}llllllllllllllllllllllllllllllllllllllllllllllllll}
\hline\hline
$K_+$&$K_0$&\multicolumn{1}{c}{$a_{+,0}$}&\multicolumn{1}{c}{$a_{+,1}$}&\multicolumn{1}{c}{$a_{+,2}$}&\multicolumn{1}{c}{$a_{+,3}$}&\multicolumn{1}{c}{$a_{+,4}$}&\multicolumn{1}{c}{$a_{+,5}$}&\multicolumn{1}{c}{$a_{+,6}$}&\multicolumn{1}{c}{$a_{+,7}$}&$p$&$\chi^2/N_{\rm dof}$&$N_{\rm dof}$\\
\hline
2&2&0.02641(58)&-0.0824(26)&- &- &- &- &- &-& 0.00& 5.15&14&\\
2&3&0.02668(68)&-0.0811(31)&- &- &- &- &- &-& 0.00& 5.50&13&\\
3&2&0.02477(68)&-0.0829(26)&0.054(12)&- &- &- &- &-& 0.00& 3.95&13&\\
3&3&0.02534(73)&-0.0792(31)&0.062(12)&- &- &- &- &-& 0.00& 3.89&12&\\
3&4&0.02534(73)&-0.0781(34)&0.067(14)&- &- &- &- &-& 0.00& 4.19&11&\\
4&3&0.02535(73)&-0.0776(38)&0.074(20)&0.023(30)&- &- &- &-& 0.00& 4.19&11&\\
4&4&0.02592(97)&-0.033(50)&0.69(69)&2.1(2.3)&- &- &- &-& 0.00& 4.53&10&\\
5&5&0.0266(10)&0.052(65)&2.21(97)&11.1(5.6)&17.2(15.1)&- &- &-& 0.00& 5.04&8&\\
\hline\hline\\
\end{tabular}
\\
\begin{tabular}{l@{\hspace{1mm}}llllllllllllllllllllllllllllllllllllllllllllllllll}
\hline\hline
$K_+$&$K_0$&\multicolumn{1}{c}{$a_{0,0}$}&\multicolumn{1}{c}{$a_{0,1}$}&\multicolumn{1}{c}{$a_{0,2}$}&\multicolumn{1}{c}{$a_{0,3}$}&\multicolumn{1}{c}{$a_{0,4}$}&\multicolumn{1}{c}{$a_{0,5}$}&\multicolumn{1}{c}{$a_{0,6}$}&\multicolumn{1}{c}{$a_{0,7}$}&$p$&$\chi^2/N_{\rm dof}$&$N_{\rm dof}$\\
\hline
2&2&0.0854(17)&-0.2565(75)&- &- &- &- &- &-& 0.00& 5.15&14&\\
2&3&0.0856(18)&-0.2527(91)&0.021(27)&- &- &- &- &-& 0.00& 5.50&13&\\
3&2&0.0858(18)&-0.2501(77)&- &- &- &- &- &-& 0.00& 3.95&13&\\
3&3&0.0864(18)&-0.2379(95)&0.061(28)&- &- &- &- &-& 0.00& 3.89&12&\\
3&4&0.0869(19)&-0.231(13)&0.067(29)&-0.08(10)&- &- &- &-& 0.00& 4.19&11&\\
4&3&0.0869(19)&-0.229(15)&0.091(48)&- &- &- &- &-& 0.00& 4.19&11&\\
4&4&0.0887(27)&-0.08(17)&2.2(2.4)&7.0(7.9)&- &- &- &-& 0.00& 4.53&10&\\
5&5&0.0887(28)&0.07(20)&6.1(3.3)&41.5(19.0)&93.3(44.0)&- &- &-& 0.00& 5.04&8&\\
\hline\hline\\
\end{tabular}

\end{center}
\caption{Results for the frequentist BGL fit to HPQCD 14~\cite{Bouchard:2014ypa}, FNAL/MILC 19~\cite{Bazavov:2019aom} and RBC/UKQCD 23~\cite{Flynn:2023nhi}.
     The tables show the results for BGL coefficients for different orders of the fit.
    Results for higher truncations are possible in principle (\emph{i.e.} up to ($K_+,K_0)=(8,8)$), 
    but higher-order fluctuate wildly -- we removed these results from the tables.}\label{tab:all lattice data frequentist}
\end{table*}
\subsection{Combined Bayesian fit to RBC/UKQCD 23, HPQCD 14 and Khodjamirian 17}\label{app:results RBC/UKQCD 23, HPQCD 14 and Khodjamirian 17}\label{sec:all lattice data with sum rules}
Results for the BGL coefficients of the combined frequentist fit over lattice results 
by RBC/UKQCD 23, HPQCD 14 and sum-rule results by Khodjamirian 17 can be found in 
Tab.~\ref{tab:combined BGL frequ with sum rule}, the corresponding results for the Bayesian-inference fit in 
Tab.~\ref{tab:combined BGL Bayesian with sum rules}, and results for phenomenology from the Bayesian fit in 
Tab.~\ref{tab:BstoK_zfit_observables with sum rules}.
\begin{table*}[bt!]
\begin{center}
\tiny
\begin{tabular}{l@{\hspace{1mm}}llllllllllllllllllllllllllllllllllllllllllllllllll}
\hline\hline
$K_+$&$K_0$&\multicolumn{1}{c}{$a_{+,0}$}&\multicolumn{1}{c}{$a_{+,1}$}&\multicolumn{1}{c}{$a_{+,2}$}&\multicolumn{1}{c}{$a_{+,3}$}&\multicolumn{1}{c}{$a_{+,4}$}&\multicolumn{1}{c}{$a_{+,5}$}&$p$&$\chi^2/N_{\rm dof}$&$N_{\rm dof}$\\
\hline
2&2&0.02936(75)&-0.0786(32)&- &- &- &-& 0.00& 5.60&9&\\
2&3&0.02950(81)&-0.0780(35)&- &- &- &-& 0.00& 6.27&8&\\
3&2&0.02580(99)&-0.0762(32)&0.090(16)&- &- &-& 0.01& 2.47&8&\\
3&3&0.02567(99)&-0.0691(37)&0.126(19)&- &- &-& 0.63& 0.76&7&\\
3&4&0.02564(99)&-0.0685(39)&0.130(20)&- &- &-& 0.55& 0.83&6&\\
4&3&0.0256(10)&-0.0702(48)&0.127(19)&0.035(88)&- &-& 0.53& 0.85&6&\\
4&4&0.0253(10)&-0.0717(49)&0.141(23)&0.12(12)&- &-& 0.56& 0.78&5&\\
5&5&0.0256(13)&-0.051(56)&0.33(51)&-0.4(1.3)&-4.9(13.0)&-& 0.29& 1.25&3&\\
6&6&0.0300(32)&0.33(26)&6.4(4.1)&15.1(10.8)&-152.1(100.2)&-596.6(407.5)& 0.31& 1.04&1&\\
\hline\hline\\
\end{tabular}
\\
\begin{tabular}{l@{\hspace{1mm}}llllllllllllllllllllllllllllllllllllllllllllllllll}
\hline\hline
$K_+$&$K_0$&\multicolumn{1}{c}{$a_{0,0}$}&\multicolumn{1}{c}{$a_{0,1}$}&\multicolumn{1}{c}{$a_{0,2}$}&\multicolumn{1}{c}{$a_{0,3}$}&\multicolumn{1}{c}{$a_{0,4}$}&\multicolumn{1}{c}{$a_{0,5}$}&$p$&$\chi^2/N_{\rm dof}$&$N_{\rm dof}$\\
\hline
2&2&0.0985(25)&-0.259(10)&- &- &- &-& 0.00& 5.60&9&\\
2&3&0.0984(25)&-0.256(11)&0.015(31)&- &- &-& 0.00& 6.27&8&\\
3&2&0.0982(25)&-0.246(10)&- &- &- &-& 0.01& 2.47&8&\\
3&3&0.0970(25)&-0.220(12)&0.139(37)&- &- &-& 0.63& 0.76&7&\\
3&4&0.0966(27)&-0.223(13)&0.159(51)&0.13(23)&- &-& 0.55& 0.83&6&\\
4&3&0.0970(25)&-0.220(12)&0.140(37)&- &- &-& 0.53& 0.85&6&\\
4&4&0.0956(28)&-0.226(14)&0.194(61)&0.34(30)&- &-& 0.56& 0.78&5&\\
5&5&0.0956(33)&-0.22(13)&0.2(1.2)&0.2(3.0)&-1.0(29.7)&-& 0.29& 1.25&3&\\
6&6&0.0951(35)&-0.12(19)&1.7(2.2)&3.3(4.7)&-36.7(55.0)&-132.6(164.5)& 0.31& 1.04&1&\\
\hline\hline\\
\end{tabular}

\end{center}
\caption{Results for the frequentist BGL fit to HPQCD 14~\cite{Bouchard:2014ypa}, RBC/UKQCD 23~\cite{Flynn:2023nhi}
     and Khodjamirian 17~\cite{Khodjamirian:2017fxg}. The tables show the results for BGL coefficients for different orders of the fit.}\label{tab:combined BGL frequ with sum rule}
\end{table*}
\begin{table*}[bt!]
\begin{center}
\tiny
\begin{tabular}{l@{\hspace{1mm}}llllllllllllllllllllllllllllllllllllllllllllllllll}
\hline\hline
$K_+$&$K_0$&\multicolumn{1}{c}{$a_{+,0}$}&\multicolumn{1}{c}{$a_{+,1}$}&\multicolumn{1}{c}{$a_{+,2}$}&\multicolumn{1}{c}{$a_{+,3}$}&\multicolumn{1}{c}{$a_{+,4}$}&\multicolumn{1}{c}{$a_{+,5}$}&\multicolumn{1}{c}{$a_{+,6}$}&\multicolumn{1}{c}{$a_{+,7}$}&\multicolumn{1}{c}{$a_{+,8}$}&\multicolumn{1}{c}{$a_{+,9}$}&\\
\hline
2&2&0.02935(74)&-0.0786(31)&- &- &- &- &- &- &- &-&\\
2&3&0.02948(80)&-0.0779(34)&- &- &- &- &- &- &- &-&\\
3&2&0.02577(98)&-0.0761(32)&0.090(16)&- &- &- &- &- &- &-&\\
3&3&0.02569(97)&-0.0692(37)&0.126(18)&- &- &- &- &- &- &-&\\
3&4&0.02561(100)&-0.0686(39)&0.130(19)&- &- &- &- &- &- &-&\\
4&3&0.0256(10)&-0.0704(47)&0.128(19)&0.038(88)&- &- &- &- &- &-&\\
4&4&0.0253(11)&-0.0717(51)&0.140(23)&0.12(12)&- &- &- &- &- &-&\\
5&5&0.0253(11)&-0.0714(57)&0.141(35)&0.11(13)&-0.03(68)&- &- &- &- &-&\\
6&6&0.0253(10)&-0.0712(54)&0.141(33)&0.10(13)&-0.06(63)&0.11(65)&- &- &- &-&\\
7&7&0.0254(10)&-0.0710(54)&0.142(35)&0.09(13)&-0.10(64)&0.13(72)&-0.12(67)&- &- &-&\\
8&8&0.0253(10)&-0.0709(55)&0.145(34)&0.08(14)&-0.15(65)&0.21(83)&-0.21(87)&0.10(71)&- &-&\\
9&9&0.0254(10)&-0.0707(57)&0.145(36)&0.08(14)&-0.16(66)&0.3(1.0)&-0.3(1.2)&0.2(1.1)&-0.11(77)&-&\\
10&10&0.0253(10)&-0.0704(59)&0.150(38)&0.06(16)&-0.26(68)&0.4(1.2)&-0.5(1.7)&0.5(1.7)&-0.3(1.4)&0.14(86)&\\
\hline\hline\\
\end{tabular}
\\
\begin{tabular}{l@{\hspace{1mm}}llllllllllllllllllllllllllllllllllllllllllllllllll}
\hline\hline
$K_+$&$K_0$&\multicolumn{1}{c}{$a_{0,0}$}&\multicolumn{1}{c}{$a_{0,1}$}&\multicolumn{1}{c}{$a_{0,2}$}&\multicolumn{1}{c}{$a_{0,3}$}&\multicolumn{1}{c}{$a_{0,4}$}&\multicolumn{1}{c}{$a_{0,5}$}&\multicolumn{1}{c}{$a_{0,6}$}&\multicolumn{1}{c}{$a_{0,7}$}&\multicolumn{1}{c}{$a_{0,8}$}&\multicolumn{1}{c}{$a_{0,9}$}&\\
\hline
2&2&0.0985(25)&-0.258(10)&- &- &- &- &- &- &- &-&\\
2&3&0.0983(25)&-0.256(11)&0.014(31)&- &- &- &- &- &- &-&\\
3&2&0.0982(25)&-0.245(10)&- &- &- &- &- &- &- &-&\\
3&3&0.0970(25)&-0.220(12)&0.140(36)&- &- &- &- &- &- &-&\\
3&4&0.0965(27)&-0.224(13)&0.157(50)&0.13(23)&- &- &- &- &- &-&\\
4&3&0.0970(25)&-0.220(12)&0.140(36)&- &- &- &- &- &- &-&\\
4&4&0.0955(28)&-0.226(14)&0.191(60)&0.33(29)&- &- &- &- &- &-&\\
5&5&0.0958(28)&-0.225(13)&0.193(66)&0.28(28)&-0.15(63)&- &- &- &- &-&\\
6&6&0.0958(28)&-0.225(13)&0.191(68)&0.26(27)&-0.19(61)&0.19(64)&- &- &- &-&\\
7&7&0.0958(28)&-0.225(14)&0.197(70)&0.24(26)&-0.29(65)&0.32(71)&-0.23(66)&- &- &-&\\
8&8&0.0958(28)&-0.224(13)&0.200(72)&0.23(26)&-0.38(68)&0.48(88)&-0.42(90)&0.25(72)&- &-&\\
9&9&0.0959(28)&-0.224(13)&0.205(75)&0.21(25)&-0.46(72)&0.7(1.1)&-0.7(1.2)&0.5(1.1)&-0.24(77)&-&\\
10&10&0.0959(27)&-0.223(14)&0.210(79)&0.19(25)&-0.56(80)&0.9(1.3)&-1.1(1.7)&0.9(1.8)&-0.6(1.4)&0.25(84)&\\
\hline\hline\\
\end{tabular}

\caption{Results for the Bayesian-inference BGL fit to HPQCD 14~\cite{Bouchard:2014ypa}, RBC/UKQCD 23~\cite{Flynn:2023nhi}
     and Khodjamirian 17~\cite{Khodjamirian:2017fxg}. The tables show the results for BGL coefficients for different orders of the fit.}\label{tab:combined BGL Bayesian with sum rules}
\end{center}
\end{table*}
\begin{table}[bt!]
     \centering
     \tiny
    \begin{tabular}{l@{\hspace{1mm}}llllllllll}
\hline\hline
$K_+$&$K_0$&\multicolumn{1}{c}{$f(q^2=0)$}&\multicolumn{1}{c}{$R_{B_s\to K}^{\rm impr}$}&\multicolumn{1}{c}{$R_{B_s\to K}$}&\multicolumn{1}{c}{$\frac{\Gamma^\tau}{|V_{ub}|^2}\,[\frac 1{\rm ps}]$}&\multicolumn{1}{c}{$\frac{\Gamma^\mu}{|V_{ub}|^2}\,[\frac 1{\rm ps}]$}&\multicolumn{1}{c}{$V^{\rm low}_{\rm CKM}$}&\multicolumn{1}{c}{$V^{\rm high}_{\rm CKM}$}&\multicolumn{1}{c}{$V^{\rm full}_{\rm CKM}$}&\\
\hline
2&2&0.255(13)&1.547(15)&0.705(11)&5.58(29)&7.92(48)&0.00321(27)&0.00318(27)&0.00319(26)\\
2&3&0.261(17)&1.542(19)&0.699(18)&5.64(32)&8.08(59)&0.00315(29)&0.00315(27)&0.00315(27)\\
3&2&0.268(13)&1.693(37)&0.728(14)&4.88(29)&6.70(49)&0.00328(28)&0.00356(31)&0.00341(28)\\
3&3&0.322(19)&1.683(35)&0.665(20)&5.12(31)&7.71(59)&0.00287(26)&0.00348(31)&0.00313(27)\\
3&4&0.326(21)&1.677(39)&0.659(24)&5.09(32)&7.75(62)&0.00285(27)&0.00349(33)&0.00312(28)\\
4&3&0.323(20)&1.692(41)&0.668(21)&5.08(32)&7.62(62)&0.00289(26)&0.00351(31)&0.00315(27)\\
4&4&0.335(23)&1.687(40)&0.652(25)&4.98(33)&7.65(61)&0.00284(27)&0.00354(33)&0.00312(27)\\
5&5&0.335(22)&1.688(44)&0.653(26)&5.01(34)&7.69(67)&0.00284(27)&0.00354(33)&0.00313(28)\\
6&6&0.333(22)&1.688(42)&0.654(26)&5.01(33)&7.67(64)&0.00284(27)&0.00353(33)&0.00311(28)\\
7&7&0.333(22)&1.685(43)&0.653(26)&5.02(33)&7.70(65)&0.00284(27)&0.00353(33)&0.00312(28)\\
8&8&0.333(22)&1.687(43)&0.653(26)&5.02(33)&7.70(65)&0.00283(27)&0.00352(31)&0.00312(27)\\
9&9&0.334(22)&1.685(43)&0.653(26)&5.04(33)&7.74(66)&0.00283(27)&0.00351(33)&0.00311(28)\\
10&10&0.334(22)&1.686(43)&0.652(26)&5.05(32)&7.76(64)&0.00282(27)&0.00352(32)&0.00310(27)\\
\hline\hline\\
\end{tabular}

    \begin{tabular}{l@{\hspace{1mm}}llllllllll}
\hline\hline
$K_+$&$K_0$&\multicolumn{1}{c}{$I[\mathcal{A}_{\rm FB}^\tau]\,[\frac 1{\rm ps}]$}&\multicolumn{1}{c}{$I[\mathcal{A}_{\rm FB}^\mu]\,[\frac 1{\rm ps}]$}&\multicolumn{1}{c}{$\mathcal{\bar A}_{\rm FB}^\tau$}&\multicolumn{1}{c}{$\mathcal{\bar A}_{\rm FB}^\mu$}&\multicolumn{1}{c}{$I[\mathcal{A}_{\rm pol}^\tau]\,[\frac 1{\rm ps}]$}&\multicolumn{1}{c}{$I[\mathcal{A}_{\rm pol}^\mu]\,[\frac 1{\rm ps}]$}&\multicolumn{1}{c}{$\mathcal{\bar A}_{\rm pol}^\tau$}&\multicolumn{1}{c}{$\mathcal{\bar A}_{\rm pol}^\mu$}&\\
\hline
2&2&1.537(82)&0.0390(32)&0.2753(15)&0.00492(16)&0.800(75)&7.81(47)&0.143(11)&0.98633(45)\\
2&3&1.552(88)&0.0402(41)&0.2752(15)&0.00497(19)&0.83(10)&7.97(58)&0.147(14)&0.98623(49)\\
3&2&1.388(83)&0.0400(33)&0.2846(19)&0.00597(28)&0.22(12)&6.59(48)&0.044(23)&0.98300(84)\\
3&3&1.474(87)&0.0528(52)&0.2882(22)&0.00685(35)&0.24(12)&7.56(58)&0.047(22)&0.98077(99)\\
3&4&1.467(90)&0.0535(56)&0.2880(23)&0.00690(37)&0.26(13)&7.60(61)&0.050(24)&0.9807(10)\\
4&3&1.465(91)&0.0527(53)&0.2883(22)&0.00691(39)&0.22(13)&7.47(61)&0.042(25)&0.9806(11)\\
4&4&1.436(92)&0.0550(59)&0.2884(23)&0.00720(49)&0.22(13)&7.49(60)&0.044(25)&0.9798(14)\\
5&5&1.446(97)&0.0552(59)&0.2885(23)&0.00718(49)&0.22(14)&7.54(66)&0.043(27)&0.9799(14)\\
6&6&1.444(93)&0.0548(58)&0.2885(23)&0.00715(49)&0.22(14)&7.51(63)&0.043(26)&0.9799(14)\\
7&7&1.449(93)&0.0549(59)&0.2884(23)&0.00713(48)&0.23(14)&7.55(64)&0.045(26)&0.9800(14)\\
8&8&1.449(92)&0.0550(58)&0.2885(23)&0.00714(48)&0.23(14)&7.55(63)&0.044(26)&0.9800(14)\\
9&9&1.455(94)&0.0552(59)&0.2885(23)&0.00713(48)&0.23(14)&7.59(65)&0.045(26)&0.9800(14)\\
10&10&1.456(90)&0.0553(59)&0.2887(24)&0.00713(48)&0.23(14)&7.60(63)&0.045(27)&0.9800(14)\\
\hline\hline\\
\end{tabular}

     \caption{Summary of results based on combined fit to HPQCD 14~\cite{Bouchard:2014ypa}, RBC/UKQCD 23~\cite{Flynn:2023nhi}
     and Khodjamirian 17~\cite{Khodjamirian:2017fxg}. Definitions for the asymmetries $\mathcal{A}$ can be found in App.~\ref{app:asymmetries}.}\label{tab:BstoK_zfit_observables with sum rules}
 \end{table}

\bibliographystyle{jhep}
\bibliography{fit}

\providecommand{\href}[2]{#2}\begingroup\raggedright\begin{thebibliography}{10}

\bibitem{Gubernari:2020eft}
N.~Gubernari, D.~van Dyk, and J.~Virto, {\it {Non-local matrix elements in
  $B_{(s)}\to \{K^{(*)},\phi\}\ell^+\ell^-$}},  {\em JHEP} {\bf 02} (2021) 088,
  [\href{http://arxiv.org/abs/2011.09813}{{\tt arXiv:2011.09813}}].

\bibitem{Gubernari:2022hxn}
N.~Gubernari, M.~Reboud, D.~van Dyk, and J.~Virto, {\it {Improved theory
  predictions and global analysis of exclusive $b\to s{\mu}^+{\mu}^-$
  processes}},  {\em JHEP} {\bf 09} (2022) 133,
  [\href{http://arxiv.org/abs/2206.03797}{{\tt arXiv:2206.03797}}].

\bibitem{Blake:2022vfl}
T.~Blake, S.~Meinel, M.~Rahimi, and D.~van Dyk, {\it {Dispersive bounds for
  local form factors in $\Lambda_b\to\Lambda$ transitions}},  {\em Phys. Rev.
  D} {\bf 108} (2023), no.~9 094509,
  [\href{http://arxiv.org/abs/2205.06041}{{\tt arXiv:2205.06041}}].

\bibitem{Workman:2022ynf}
{\bf Particle Data Group} Collaboration, R.~L. Workman and Others, {\it {Review
  of Particle Physics}},  {\em PTEP} {\bf 2022} (2022) 083C01.

\bibitem{FlavourLatticeAveragingGroupFLAG:2021npn}
{\bf Flavour Lattice Averaging Group (FLAG)} Collaboration, Y.~Aoki et~al.,
  {\it {FLAG Review 2021}},  {\em Eur. Phys. J. C} {\bf 82} (2022), no.~10 869,
  [\href{http://arxiv.org/abs/2111.09849}{{\tt arXiv:2111.09849}}].

\bibitem{Colangelo:2000dp}
P.~Colangelo and A.~Khodjamirian, {\it {QCD sum rules, a modern perspective}},
  \href{http://arxiv.org/abs/hep-ph/0010175}{{\tt hep-ph/0010175}}.

\bibitem{Khodjamirian:2020btr}
A.~Khodjamirian, {\em {Hadron Form Factors}: {From Basic Phenomenology to QCD
  Sum Rules}}.
\newblock CRC Press, Taylor \& Francis Group, Boca Raton, FL, USA, 2020.

\bibitem{Boyd:1994tt}
C.~G. Boyd, B.~Grinstein, and R.~F. Lebed, {\it Constraints on form-factors for
  exclusive semileptonic heavy to light meson decays},  {\em Phys. Rev. Lett.}
  {\bf 74} (1995) 4603--4606, [\href{http://arxiv.org/abs/hep-ph/9412324}{{\tt
  hep-ph/9412324}}].

\bibitem{Caprini:1997mu}
I.~Caprini, L.~Lellouch, and M.~Neubert, {\it Dispersive bounds on the shape of
  {$\bar B\to D^{(*)} \ell\bar\nu$} form-factors},  {\em Nucl. Phys.} {\bf
  B530} (1998) 153--181, [\href{http://arxiv.org/abs/hep-ph/9712417}{{\tt
  hep-ph/9712417}}].

\bibitem{Bourrely:2008za}
C.~Bourrely, I.~Caprini, and L.~Lellouch, {\it Model-independent description of
  {$B\to\pi l\nu$} decays and a determination of {$|V_{ub}|$}},  {\em Phys.
  Rev.} {\bf D79} (2009) 013008, [\href{http://arxiv.org/abs/0807.2722}{{\tt
  arXiv:0807.2722}}]. erratum:
  \href{https://doi.org/10.1103/PhysRevD.82.099902}{Phys. Rev. {\bfseries D82}
  (2010) 099902}.

\bibitem{Buck:1998kp}
W.~W. Buck and R.~F. Lebed, {\it {New constraints on dispersive form-factor
  parameterizations from the timelike region}},  {\em Phys. Rev. D} {\bf 58}
  (1998) 056001, [\href{http://arxiv.org/abs/hep-ph/9802369}{{\tt
  hep-ph/9802369}}].

\bibitem{Becher:2005bg}
T.~Becher and R.~J. Hill, {\it Comment on form-factor shape and extraction of
  {$|V_{ub}|$} from {$B\to\pi l \nu$}},  {\em Phys. Lett.} {\bf B633} (2006)
  61--69, [\href{http://arxiv.org/abs/hep-ph/0509090}{{\tt hep-ph/0509090}}].

\bibitem{Bourrely:1980gp}
C.~Bourrely, B.~Machet, and E.~de~Rafael, {\it Semileptonic decays of
  pseudoscalar particles ({$M \to M' \ell \nu_\ell$}) and short distance
  behaviour of quantum chromodynamics},  {\em Nucl. Phys.} {\bf B189} (1981)
  157--181.

\bibitem{Lellouch:1995yv}
L.~Lellouch, {\it Lattice constrained unitarity bounds for {$\bar B^0\to\pi^+
  l^-\bar\nu$} decays},  {\em Nucl. Phys.} {\bf B479} (1996) 353--391,
  [\href{http://arxiv.org/abs/hep-ph/9509358}{{\tt hep-ph/9509358}}].

\bibitem{DiCarlo:2021dzg}
M.~Di~Carlo, G.~Martinelli, M.~Naviglio, F.~Sanfilippo, S.~Simula, and
  L.~Vittorio, {\it Unitarity bounds for semileptonic decays in lattice {QCD}},
   {\em Phys. Rev. D} {\bf 104} (2021), no.~5 054502,
  [\href{http://arxiv.org/abs/2105.02497}{{\tt arXiv:2105.02497}}].

\bibitem{fittingPaperCode}
A.~J{\"u}ttner, {\it {BFF -- Bayesian Form factor Fit code
  \href{https://github.com/andreasjuettner/BFF}{https://github.com/andreasjuettner/BFF},
  \href{https://doi.org/10.5281/zenodo.7799543}{https://doi.org/10.5281/zenodo.7799543}}},
  .

\bibitem{DelDebbio:2021whr}
L.~Del~Debbio, T.~Giani, and M.~Wilson, {\it {Bayesian approach to inverse
  problems: an application to NNPDF closure testing}},  {\em Eur. Phys. J. C}
  {\bf 82} (2022), no.~4 330, [\href{http://arxiv.org/abs/2111.05787}{{\tt
  arXiv:2111.05787}}].

\bibitem{Neil:2022joj}
E.~T. Neil and J.~W. Sitison, {\it {Improved information criteria for Bayesian
  model averaging in lattice field theory}},  {\em Phys. Rev. D} {\bf 109}
  (2024), no.~1 014510, [\href{http://arxiv.org/abs/2208.14983}{{\tt
  arXiv:2208.14983}}].

\bibitem{Jay:2020jkz}
W.~I. Jay and E.~T. Neil, {\it {Bayesian model averaging for analysis of
  lattice field theory results}},  {\em Phys. Rev. D} {\bf 103} (2021) 114502,
  [\href{http://arxiv.org/abs/2008.01069}{{\tt arXiv:2008.01069}}].

\bibitem{Frison:2023lwb}
J.~Frison, {\it {Towards fully bayesian analyses in Lattice QCD}},
  \href{http://arxiv.org/abs/2302.06550}{{\tt arXiv:2302.06550}}.

\bibitem{Duhr:2021mfd}
C.~Duhr, A.~Huss, A.~Mazeliauskas, and R.~Szafron, {\it {An analysis of
  Bayesian estimates for missing higher orders in perturbative calculations}},
  {\em JHEP} {\bf 09} (2021) 122, [\href{http://arxiv.org/abs/2106.04585}{{\tt
  arXiv:2106.04585}}].

\bibitem{Bouchard:2014ypa}
C.~Bouchard, G.~P. Lepage, C.~Monahan, H.~Na, and J.~Shigemitsu, {\it {$B_s \to
  K \ell \nu$} form factors from lattice {QCD}},  {\em Phys.Rev.} {\bf D90}
  (2014), no.~5 054506, [\href{http://arxiv.org/abs/1406.2279}{{\tt
  arXiv:1406.2279}}].

\bibitem{Bazavov:2019aom}
{\bf Fermilab Lattice/MILC} Collaboration, A.~Bazavov et~al., {\it {$B_s\to
  K\ell\nu$} decay from lattice {QCD}},  {\em Phys. Rev.} {\bf D100} (2019),
  no.~3 034501, [\href{http://arxiv.org/abs/1901.02561}{{\tt
  arXiv:1901.02561}}].

\bibitem{Flynn:2023nhi}
{\bf RBC/UKQCD} Collaboration, J.~M. Flynn, R.~C. Hill, A.~J\"uttner, A.~Soni,
  J.~T. Tsang, and O.~Witzel, {\it {Exclusive semileptonic $B_s\to K\ell\nu$
  decays on the lattice}},  {\em Phys. Rev. D} {\bf 107} (2023), no.~11 114512,
  [\href{http://arxiv.org/abs/2303.11280}{{\tt arXiv:2303.11280}}].

\bibitem{Khodjamirian:2017fxg}
A.~Khodjamirian and A.~V. Rusov, {\it {$B_{s}\to K \ell \nu_\ell$ and $B_{(s)}
  \to \pi (K) \ell^+\ell^-$ decays at large recoil and CKM matrix elements}},
  {\em JHEP} {\bf 08} (2017) 112, [\href{http://arxiv.org/abs/1703.04765}{{\tt
  arXiv:1703.04765}}].

\bibitem{LHCb:2020ist}
{\bf LHCb} Collaboration, R.~Aaij et~al., {\it {First observation of the decay
  $B_s^0 \to K^-\mu^+\nu_\mu$ and Measurement of $|V_{ub}|/|V_{cb}|$}},  {\em
  Phys. Rev. Lett.} {\bf 126} (2021), no.~8 081804,
  [\href{http://arxiv.org/abs/2012.05143}{{\tt arXiv:2012.05143}}].

\bibitem{LHCb:2020cyw}
{\bf LHCb} Collaboration, R.~Aaij et~al., {\it {Measurement of $|V_{cb}|$ with
  $B_s^0 \to D_s^{(*)-} \mu^+ \nu_{\mu}$ decays}},  {\em Phys. Rev. D} {\bf
  101} (2020), no.~7 072004, [\href{http://arxiv.org/abs/2001.03225}{{\tt
  arXiv:2001.03225}}].

\bibitem{Na:2015kha}
{\bf HPQCD} Collaboration, H.~Na, C.~M. Bouchard, G.~P. Lepage, C.~Monahan, and
  J.~Shigemitsu, {\it {$B \to D l \nu$} form factors at nonzero recoil and
  extraction of {$|V_{cb}|$}},  {\em Phys. Rev.} {\bf D92} (2015), no.~5
  054510, [\href{http://arxiv.org/abs/1505.03925}{{\tt arXiv:1505.03925}}].
  erratum: \href{https://doi.org/10.1103/PhysRevD.93.119906}{Phys. Rev.
  {\bfseries D93} (2016) 119906}.

\bibitem{Sirlin:1981ie}
A.~Sirlin, {\it Large {$m(W)$}, {$m(Z)$} behavior of the {$O(\alpha)$}
  corrections to semileptonic processes mediated by {$W$}},  {\em Nucl. Phys.}
  {\bf B196} (1982) 83--92.

\bibitem{Flynn:2015mha}
J.~M. Flynn, T.~Izubuchi, T.~Kawanai, C.~Lehner, A.~Soni, R.~S. Van~de Water,
  and O.~Witzel, {\it {$B \to \pi \ell \nu$} and {$B_s \to K \ell\nu$} form
  factors and {$|V_{ub}|$} from $2+1$-flavor lattice {QCD} with domain-wall
  light quarks and relativistic heavy quarks},  {\em Phys. Rev.} {\bf D91}
  (2015), no.~7 074510, [\href{http://arxiv.org/abs/1501.05373}{{\tt
  arXiv:1501.05373}}].

\bibitem{Boyd:1995sq}
C.~G. Boyd, B.~Grinstein, and R.~F. Lebed, {\it Model independent
  determinations of {$\bar B \to D$ (lepton), $D^*$ (lepton) anti-neutrino}
  form-factors},  {\em Nucl. Phys.} {\bf B461} (1996) 493--511,
  [\href{http://arxiv.org/abs/hep-ph/9508211}{{\tt hep-ph/9508211}}].

\bibitem{Boyd:1997qw}
C.~G. Boyd and M.~J. Savage, {\it {Analyticity, shapes of semileptonic
  form-factors, and $\bar B \to \pi \ell\bar\nu$}},  {\em Phys. Rev. D} {\bf
  56} (1997) 303--311, [\href{http://arxiv.org/abs/hep-ph/9702300}{{\tt
  hep-ph/9702300}}].

\bibitem{Arnesen:2005ez}
M.~C. Arnesen, B.~Grinstein, I.~Z. Rothstein, and I.~W. Stewart, {\it A
  precision model independent determination of {$|V_{ub}|$} from {$B\to\pi
  e\nu$}},  {\em Phys. Rev. Lett.} {\bf 95} (2005) 071802,
  [\href{http://arxiv.org/abs/hep-ph/0504209}{{\tt hep-ph/0504209}}].

\bibitem{Bardeen:2003kt}
W.~A. Bardeen, E.~J. Eichten, and C.~T. Hill, {\it Chiral multiplets of
  heavy-light mesons},  {\em Phys. Rev.} {\bf D68} (2003) 054024,
  [\href{http://arxiv.org/abs/hep-ph/0305049}{{\tt hep-ph/0305049}}].

\bibitem{Berns:2018vpl}
A.~Berns and H.~Lamm, {\it Model-independent prediction of {$R(\eta_c)$}},
  {\em JHEP} {\bf 12} (2018) 114, [\href{http://arxiv.org/abs/1808.07360}{{\tt
  arXiv:1808.07360}}].

\bibitem{Szego:1939}
G.~Szeg\"o, {\em Orthogonal Polynomials}, vol.~23.
\newblock American Mathematical Society, 1939.

\bibitem{Simon:2004}
B.~Simon, {\it Orthogonal polynomials on the unit circle: New results},
  \href{http://arxiv.org/abs/math/0405111}{{\tt math/0405111}}.

\bibitem{Lepage:1980fj}
G.~P. Lepage and S.~J. Brodsky, {\it Exclusive processes in perturbative
  quantum chromodynamics},  {\em Phys. Rev. D} {\bf 22} (1980) 2157.

\bibitem{Akhoury:1994tnu}
R.~Akhoury, G.~F. Sterman, and Y.~P. Yao, {\it Exclusive semileptonic decays of
  {$B$} mesons into light mesons},  {\em Phys. Rev. D} {\bf 50} (1994)
  358--372.

\bibitem{Jeffreys:1939xee}
H.~Jeffreys, {\em {The Theory of Probability}}.
\newblock Oxford Classic Texts in the Physical Sciences. 1939.

\bibitem{Cossu:2020yeg}
G.~Cossu, L.~Del~Debbio, A.~J{\"u}ttner, B.~Kitching-Morley, J.~K.~L. Lee,
  A.~Portelli, H.~B. Rocha, and K.~Skenderis, {\it {Nonperturbative Infrared
  Finiteness in a Superrenormalizable Scalar Quantum Field Theory}},  {\em
  Phys. Rev. Lett.} {\bf 126} (2021), no.~22 221601,
  [\href{http://arxiv.org/abs/2009.14768}{{\tt arXiv:2009.14768}}].

\bibitem{Duplancic:2008tk}
G.~Duplancic and B.~Melic, {\it {$B, B_s \to K$ form factors: An Update of
  light-cone sum rule results}},  {\em Phys.Rev.} {\bf D78} (2008) 054015,
  [\href{http://arxiv.org/abs/0805.4170}{{\tt arXiv:0805.4170}}].

\bibitem{Faustov:2013ima}
R.~Faustov and V.~Galkin, {\it {Charmless weak $B_s$ decays in the relativistic
  quark model}},  {\em Phys.Rev.} {\bf D87} (2013), no.~9 094028,
  [\href{http://arxiv.org/abs/1304.3255}{{\tt arXiv:1304.3255}}].

\bibitem{Wang:2012ab}
W.-F. Wang and Z.-J. Xiao, {\it {The semileptonic decays $B/B_s \to (\pi,
  K)(\ell^+\ell^-,\ell\nu,\nu\bar{\nu})$ in the perturbative QCD approach
  beyond the leading-order}},  {\em Phys.Rev.} {\bf D86} (2012) 114025,
  [\href{http://arxiv.org/abs/1207.0265}{{\tt arXiv:1207.0265}}].

\bibitem{Martinelli:2022tte}
G.~Martinelli, S.~Simula, and L.~Vittorio, {\it {Exclusive semileptonic
  $B\to\pi\ell\nu$ and $B_s\to K\ell\nu$ decays through unitarity and lattice
  QCD}},  {\em JHEP} {\bf 08} (2022) 022,
  [\href{http://arxiv.org/abs/2202.10285}{{\tt arXiv:2202.10285}}].

\bibitem{HFLAV:2022pwe}
{\bf HFLAV} Collaboration, Y.~S. Amhis et~al., {\it {Averages of $b$-hadron,
  $c$-hadron, and $\tau$-lepton properties as of 2021}},
  \href{http://arxiv.org/abs/2206.07501}{{\tt arXiv:2206.07501}}.

\bibitem{ParticleDataGroup:2022pth}
{\bf Particle Data Group} Collaboration, R.~L. Workman et~al., {\it {Review of
  Particle Physics}},  {\em PTEP} {\bf 2022} (2022) 083C01.

\bibitem{Lattice:2015tia}
{\bf Fermilab/MILC} Collaboration, J.~A. Bailey et~al., {\it {$|V_{ub}|$ from
  $B\to\pi\ell\nu$ decays and (2+1)-flavor lattice QCD}},  {\em Phys. Rev.}
  {\bf D92} (2015), no.~1 014024, [\href{http://arxiv.org/abs/1503.07839}{{\tt
  arXiv:1503.07839}}].

\bibitem{delAmoSanchez:2010af}
{\bf BaBar} Collaboration, P.~del Amo~Sanchez et~al., {\it {Study of $B \to \pi
  \ell \nu$ and $B \to \rho \ell \nu$ Decays and Determination of $|V_{ub}|$}},
   {\em Phys.Rev.} {\bf D83} (2011) 032007,
  [\href{http://arxiv.org/abs/1005.3288}{{\tt arXiv:1005.3288}}].

\bibitem{Lees:2012vv}
{\bf BaBar} Collaboration, J.~Lees et~al., {\it {Branching fraction and
  form-factor shape measurements of exclusive charmless semileptonic B decays,
  and determination of $|V_{ub}|$}},  {\em Phys.Rev.} {\bf D86} (2012) 092004,
  [\href{http://arxiv.org/abs/1208.1253}{{\tt arXiv:1208.1253}}].

\bibitem{Ha:2010rf}
{\bf Belle} Collaboration, H.~Ha et~al., {\it {Measurement of the decay
  $B^0\to\pi^-\ell^+\nu$ and determination of $|V_{ub}|$}},  {\em Phys.Rev.}
  {\bf D83} (2011) 071101, [\href{http://arxiv.org/abs/1012.0090}{{\tt
  arXiv:1012.0090}}].

\bibitem{Sibidanov:2013rkk}
{\bf Belle} Collaboration, A.~Sibidanov et~al., {\it {Study of Exclusive $B \to
  X_u \ell \nu$ Decays and Extraction of $\|V_{ub}\|$ using Full Reconstruction
  Tagging at the Belle Experiment}},  {\em Phys.Rev.} {\bf D88} (2013), no.~3
  032005, [\href{http://arxiv.org/abs/1306.2781}{{\tt arXiv:1306.2781}}].

\bibitem{Gambino:2007rp}
P.~Gambino, P.~Giordano, G.~Ossola, and N.~Uraltsev, {\it {Inclusive
  semileptonic B decays and the determination of $|V_{ub}|$}},  {\em JHEP} {\bf
  10} (2007) 058, [\href{http://arxiv.org/abs/0707.2493}{{\tt
  arXiv:0707.2493}}].

\bibitem{Lange:2005yw}
B.~O. Lange, M.~Neubert, and G.~Paz, {\it {Theory of charmless inclusive B
  decays and the extraction of $V_{ub}$}},  {\em Phys. Rev. D} {\bf 72} (2005)
  073006, [\href{http://arxiv.org/abs/hep-ph/0504071}{{\tt hep-ph/0504071}}].

\bibitem{Andersen:2005mj}
J.~R. Andersen and E.~Gardi, {\it {Inclusive spectra in charmless semileptonic
  B decays by dressed gluon exponentiation}},  {\em JHEP} {\bf 01} (2006) 097,
  [\href{http://arxiv.org/abs/hep-ph/0509360}{{\tt hep-ph/0509360}}].

\bibitem{ElKhadra:1989iu}
A.~X. El-Khadra, {\em Lattice calculation of meson form-factors for
  semileptonic decays}.
\newblock PhD thesis, UCLA, 1989.

\bibitem{Isidori:2020eyd}
G.~Isidori and O.~Sumensari, {\it {Optimized lepton universality tests in
  $B\rightarrow V \ell {\bar{\nu }}$ decays}},  {\em Eur. Phys. J. C} {\bf 80}
  (2020), no.~11 1078, [\href{http://arxiv.org/abs/2007.08481}{{\tt
  arXiv:2007.08481}}].

\bibitem{Freytsis:2015qca}
M.~Freytsis, Z.~Ligeti, and J.~T. Ruderman, {\it {Flavor models for $\bar{B}
  \to D^{(*)} \tau \bar{\nu}$}},  {\em Phys. Rev. D} {\bf 92} (2015), no.~5
  054018, [\href{http://arxiv.org/abs/1506.08896}{{\tt arXiv:1506.08896}}].

\bibitem{Bernlochner:2016bci}
F.~U. Bernlochner and Z.~Ligeti, {\it {Semileptonic $B_{(s)}$ decays to excited
  charmed mesons with $e,\mu,\tau$ and searching for new physics with
  $R(D^{**})$}},  {\em Phys. Rev. D} {\bf 95} (2017), no.~1 014022,
  [\href{http://arxiv.org/abs/1606.09300}{{\tt arXiv:1606.09300}}].

\bibitem{Martinelli:2021frl}
G.~Martinelli, S.~Simula, and L.~Vittorio, {\it Constraints for the
  semileptonic {$B \to D^{(*)}$} form factors from lattice {QCD} simulations of
  two-point correlation functions},  {\em Phys. Rev. D} {\bf 104} (2021), no.~9
  094512, [\href{http://arxiv.org/abs/2105.07851}{{\tt arXiv:2105.07851}}].

\bibitem{Boyd:1997kz}
C.~G. Boyd, B.~Grinstein, and R.~F. Lebed, {\it Precision corrections to
  dispersive bounds on form-factors},  {\em Phys. Rev.} {\bf D56} (1997)
  6895--6911, [\href{http://arxiv.org/abs/hep-ph/9705252}{{\tt
  hep-ph/9705252}}].

\bibitem{Grigo:2012ji}
J.~Grigo, J.~Hoff, P.~Marquard, and M.~Steinhauser, {\it Moments of heavy quark
  correlators with two masses: exact mass dependence to three loops},  {\em
  Nucl. Phys.} {\bf B864} (2012) 580--596,
  [\href{http://arxiv.org/abs/1206.3418}{{\tt arXiv:1206.3418}}].

\bibitem{Chetyrkin:2009fv}
K.~G. Chetyrkin, J.~H. Kuhn, A.~Maier, P.~Maierhofer, P.~Marquard,
  M.~Steinhauser, and C.~Sturm, {\it Charm and bottom quark masses: An update},
   {\em Phys. Rev.} {\bf D80} (2009) 074010,
  [\href{http://arxiv.org/abs/0907.2110}{{\tt arXiv:0907.2110}}].

\bibitem{Chetyrkin:2000yt}
K.~G. Chetyrkin, J.~H. Kuhn, and M.~Steinhauser, {\it {RunDec}: A {Mathematica}
  package for running and decoupling of the strong coupling and quark masses},
  {\em Comput. Phys. Commun.} {\bf 133} (2000) 43--65,
  [\href{http://arxiv.org/abs/hep-ph/0004189}{{\tt hep-ph/0004189}}].

\bibitem{Schmidt:2012az}
B.~Schmidt and M.~Steinhauser, {\it {CRunDec}: a {C++} package for running and
  decoupling of the strong coupling and quark masses},  {\em Comput. Phys.
  Commun.} {\bf 183} (2012) 1845--1848,
  [\href{http://arxiv.org/abs/1201.6149}{{\tt arXiv:1201.6149}}].

\bibitem{Herren:2017osy}
F.~Herren and M.~Steinhauser, {\it Version 3 of {RunDec} and {CRunDec}},  {\em
  Comput. Phys. Commun.} {\bf 224} (2018) 333--345,
  [\href{http://arxiv.org/abs/1703.03751}{{\tt arXiv:1703.03751}}].

\bibitem{Bazavov:2010yq}
A.~Bazavov et~al., {\it Staggered chiral perturbation theory in the two-flavor
  case and {SU(2)} analysis of the {MILC} data},  {\em PoS} {\bf LATTICE2010}
  (2010) 083, [\href{http://arxiv.org/abs/1011.1792}{{\tt arXiv:1011.1792}}].

\bibitem{Cichy:2013gja}
K.~Cichy, E.~Garcia-Ramos, and K.~Jansen, {\it Chiral condensate from the
  twisted mass {Dirac} operator spectrum},  {\em JHEP} {\bf 10} (2013) 175,
  [\href{http://arxiv.org/abs/1303.1954}{{\tt arXiv:1303.1954}}].

\bibitem{Alexandrou:2017bzk}
C.~Alexandrou, A.~Athenodorou, K.~Cichy, M.~Constantinou, D.~P. Horkel,
  K.~Jansen, G.~Koutsou, and C.~Larkin, {\it Topological susceptibility from
  twisted mass fermions using spectral projectors and the gradient flow},  {\em
  Phys. Rev.} {\bf D97} (2018), no.~7 074503,
  [\href{http://arxiv.org/abs/1709.06596}{{\tt arXiv:1709.06596}}].

\bibitem{Borsanyi:2012zv}
S.~Borsanyi, S.~Durr, Z.~Fodor, S.~Krieg, A.~Schafer, E.~E. Scholz, and K.~K.
  Szabo, {\it {SU(2)} chiral perturbation theory low-energy constants from
  {$2+1$} flavor staggered lattice simulations},  {\em Phys. Rev.} {\bf D88}
  (2013) 014513, [\href{http://arxiv.org/abs/1205.0788}{{\tt
  arXiv:1205.0788}}].

\bibitem{Durr:2013goa}
{\bf Budapest-Marseille-Wuppertal} Collaboration, S.~Dürr et~al., {\it Lattice
  {QCD} at the physical point meets {SU(2)} chiral perturbation theory},  {\em
  Phys. Rev.} {\bf D90} (2014), no.~11 114504,
  [\href{http://arxiv.org/abs/1310.3626}{{\tt arXiv:1310.3626}}].

\bibitem{Boyle:2015exm}
P.~A. Boyle et~al., {\it Low energy constants of {SU(2)} partially quenched
  chiral perturbation theory from {$N_f=2+1$} domain wall {QCD}},  {\em Phys.
  Rev.} {\bf D93} (2016), no.~5 054502,
  [\href{http://arxiv.org/abs/1511.01950}{{\tt arXiv:1511.01950}}].

\bibitem{Cossu:2016eqs}
G.~Cossu, H.~Fukaya, S.~Hashimoto, T.~Kaneko, and J.-I. Noaki, {\it Stochastic
  calculation of the dirac spectrum on the lattice and a determination of
  chiral condensate in {$2+1$}-flavor {QCD}},  {\em PTEP} {\bf 2016} (2016),
  no.~9 093B06, [\href{http://arxiv.org/abs/1607.01099}{{\tt
  arXiv:1607.01099}}].

\bibitem{Aoki:2017paw}
{\bf JLQCD} Collaboration, S.~Aoki, G.~Cossu, H.~Fukaya, S.~Hashimoto, and
  T.~Kaneko, {\it Topological susceptibility of {QCD} with dynamical {M\"obius}
  domain-wall fermions},  {\em PTEP} {\bf 2018} (2018), no.~4 043B07,
  [\href{http://arxiv.org/abs/1705.10906}{{\tt arXiv:1705.10906}}].

\bibitem{Narison:2018dcr}
S.~Narison, {\it {QCD} parameter correlations from heavy quarkonia},  {\em Int.
  J. Mod. Phys.} {\bf A33} (2018), no.~10 1850045,
  [\href{http://arxiv.org/abs/1801.00592}{{\tt arXiv:1801.00592}}]. addendum:
  \href{https://doi.org/10.1142/S0217751X18500458}{Int. J. Mod.
  Phys.A33,no.10,1850045(2018)}.

\bibitem{Meissner:2013pba}
U.-G. Mei\ss{}ner and W.~Wang, {\it {${\bf B_s\to K^{(*)} \ell\bar \nu}$,
  Angular Analysis, S-wave Contributions and ${\bf |V_{ub}|}$}},  {\em JHEP}
  {\bf 01} (2014) 107, [\href{http://arxiv.org/abs/1311.5420}{{\tt
  arXiv:1311.5420}}].

\end{thebibliography}\endgroup
\end{document}